\pdfoutput=1
\documentclass[bibyear]{aa}  
\usepackage{graphicx,url}
\usepackage[varg]{txfonts}     
\usepackage{gensymb} 
\usepackage{pdfcomment,acronym}      
\hypersetup{
  colorlinks=true,   
  urlcolor=blue,     
  linkcolor=red,     
}
\usepackage{natbib,twoopt}
\bibpunct{(}{)}{;}{a}{}{,}    

\makeatletter
\newcommand{\bibnote}[2]{\@namedef{#1note}{#2}}
\newcommand{\biblink}[2]{\@namedef{#1link}{#2}}
\makeatother

\makeatletter
\newcommandtwoopt{\citeads}[3][][]{\href{http://adsabs.harvard.edu/abs/#3}%
{\def\hyper@linkstart##1##2{}%
\let\hyper@linkend\@empty\citealp[#1][#2]{#3}}\biblink{#3}{\href{http://adsabs.harvard.edu/abs/#3}{ADS}}}
\newcommandtwoopt{\citepads}[3][][]{\href{http://adsabs.harvard.edu/abs/#3}%
{\def\hyper@linkstart##1##2{}%
\let\hyper@linkend\@empty\citep[#1][#2]{#3}}\biblink{#3}{\href{http://adsabs.harvard.edu/abs/#3}{ADS}}}
\newcommandtwoopt{\citetads}[3][][]{\href{http://adsabs.harvard.edu/abs/#3}%
{\def\hyper@linkstart##1##2{}%
\let\hyper@linkend\@empty\citet[#1][#2]{#3}}\biblink{#3}{\href{http://adsabs.harvard.edu/abs/#3}{ADS}}}
\newcommandtwoopt{\citeyearads}[3][][]%
{\href{http://adsabs.harvard.edu/abs/#3}
{\def\hyper@linkstart##1##2{}%
\let\hyper@linkend\@empty\citeyear[#1][#2]{#3}}\biblink{#3}{\href{http://adsabs.harvard.edu/abs/#3}{ADS}}}
\makeatother

\newacro{ADS}{Astrophysics Data System}
\newacro{NLTE}{non-local thermodynamic equilibrium}
\newacro{NASA}{National Aeronautics and Space Administration}

\begin{document}


\title{Multiwavelength study of twenty jets emanating from the periphery of active regions}
\subtitle{}

\author{
Sargam M. Mulay \inst{1}
\and
Durgesh Tripathi\inst{2}
\and 
Giulio Del Zanna\inst{1}
\and
Helen Mason\inst{1}
}

\institute{DAMTP, Centre for Mathematical Sciences, University of Cambridge, Wilberforce Road, Cambridge, CB3 0WA, UK\\
\email{smm96@cam.ac.uk} 
\and
Inter-University Centre for Astronomy and Astrophysics, Ganeshkhind, Pune 411007, India.}
\date{Received ; accepted }

 \abstract
{}
{We present a multiwavelength analysis of 20 EUV jets which occurred at the periphery of active regions close to sunspots. We discuss the physical parameters of the jets and their relation with other phenomena such as H$\alpha$ surges, nonthermal type-III radio bursts and hard X-ray emission (HXR).}
{These jets were observed between August 2010 and June 2013 by the Atmospheric Imaging Assembly (AIA) instrument onboard the Solar Dynamic Observatory (SDO). We selected events which were observed on the solar disk within \mbox{+/- 60\degree \space} latitude. Using AIA wavelength channels sensitive to coronal temperatures, we studied the temperature distribution in the jets using the \mbox{\textit{line-of-sight (LOS)}} Differential Emission Measure (DEM) technique. We also investigated the role of the photospheric magnetic field using the LOS magnetogram data from the Helioseismic and Magnetic Imager (HMI) onboard SDO.} 
{It has been observed that most of the jets originated from the western periphery of active regions. Their lifetimes range from \mbox{5 to 39 minutes} with an average of \mbox{18 minutes} and their velocities range from \mbox{87 to 532 km/s} with an average of \mbox{271 km/s}. All the jets are co-temporally associated with H$\alpha$ surges. Most of the jets are co-temporal with nonthermal \mbox{type-III} radio bursts observed by the Wind/WAVES spacecraft in the frequency range from \mbox{20 kHz} to \mbox{13 MHz}. We confirm the source region of these bursts using the Potential Field Source Surface (PFSS) technique. Using Reuven Ramaty High Energy Solar Spectroscopic Imager (RHESSI) observations, we found that half of the jets produced HXR emission and they often shared the same source region as the HXR emission (\mbox{6-12 keV}). 10 out of 20 events showed that the jets originated in a region of flux cancellation and 6 jets in a region of flux emergence. 4 events showed flux emergence and then cancellation during the jet evolution. DEM analyses showed that for most of the spires of the jets, the DEM peaked at around \mbox{log \textit{T} [K] = 6.2/6.3 ($\sim$2 MK)}. In addition, we derived an emission measure and a lower limit of electron density at the location of the spire (jet 1 : \mbox{log EM = 28.6}, \mbox{N$_{e}$ = 1.3$\times$10$^{10}$ cm$^{-3}$}; jet 2 : \mbox{log EM = 28.0}, \mbox{N$_{e}$ = 8.6$\times$10$^{9}$ cm$^{-3}$}) and the footpoint (jet 1 - \mbox{log EM = 28.6}, \mbox{N$_{e}$ = 1.1$\times$10$^{10}$ cm$^{-3}$}; jet 2 : \mbox{log EM = 28.1}, \mbox{N$_{e}$ = 8.4$\times$10$^{9}$ cm$^{-3}$}). These results are in agreement with those obtained earlier by studying individual active region jets.} 
{The observation of flux cancellation, the association with HXR emission and emission of nonthermal type-III radio bursts, suggest that the initiation and therefore, heating is taking place at the base of the jet. This is also supported by the high temperature plasma revealed by the DEM analysis in the jet footpoint (peak in the DEM at \mbox{log \textit{T} [K] = 6.5}). Our results provide substantial constraints for theoretical modelling of the jets and their thermodynamic nature.}

\maketitle

\section{Introduction}
Solar jets are transient phenomena observed in the solar atmosphere. They appear as sharp-edged, impulsive and collimated flows of plasma moving outwards with a bright spot at the footpoint forming an ‘inverted-Y’ topology of magnetic field lines. They are observed throughout the atmosphere i.e. in the photosphere (H$\alpha$, Ca II K surges), chromosphere (UV), transition region (EUV) and corona (X-ray). Jets can occur in different environments such as coronal holes (CHs) (\citeads{2014PASJ...66S..12Y}, \citeads{2014SoPh..289.3313Y}), and active regions (ARs) (\citeads{2011A&A...531L..13I},\citeads{2013ApJ...769...96C}, \citeads{2013A&A...559A...1S}, \citeads{2015MNRAS.446.3741C}), where they have different manifestations.

There have been numerous studies dedicated to the behaviour and different properties of jets such as length, width, lifetime, velocity etc (see e.g.,\citeads{1996PASJ...48..123S}, \citeads{2000ApJ...542.1100S}, \citeads{2013A&A...559A...1S}). The statistical study of X-ray jets \citepads{1996PASJ...48..123S} observed by Soft X-ray Telescope \citepads[SXT; ][]{1991SoPh..136...37T} onboard Yohkoh \citepads{1991SoPh..136....1O} showed that jets most often occur in the western periphery of the leading spot. Other high-resolution observations showed that jets which occur at the periphery of the active regions/sunspots are mostly associated with a nonthermal type-III radio burst which is known to be produced by energetic electrons gyrating along the open magnetic field lines (\citeads{1995ApJ...447L.135K}, \citeads{1996A&A...306..299R}, \citeads{2011A&A...531L..13I}, \citeads{2013ApJ...769...96C}, \citeads{2015MNRAS.446.3741C}). Also, it has been observed that the accelerated electrons which have access to open field lines produce impulsive, electron/ ${^3}$He-rich solar energetic particle (SEP) events in the interplanetary medium (\citeads{2008ApJ...675L.125N}, \citeads{2006ApJ...639..495W}) and the accelerated electrons which are trapped in closed field lines travel downwards, loose their energies due to collision and produce Hard X-ray Emission (HXR) (\citeads{2012ApJ...754....9G}, \citeads{2013ApJ...769...96C}).

Magnetic reconnection (\citeads{1963ApJS....8..177P}, \citeads{1964NASSP..50..425P}, \citeads{1995Natur.375...42Y}) is the fundamental process believed to play an important role in the dynamics of solar jets (\citeads{1992PASJ...44L.173S}, \citeads{1994xspy.conf...29S}, \citeads{2008cosp...37.2858S}, \citeads{2008A&A...481L..57C}, \citeads{2013ApJ...763...24K}, \citeads{2015MNRAS.446.3741C}). It is widely thought that the principle formation mechanism for jets is magnetic reconnection following magnetic flux emergence (\citeads{2007Sci...318.1591S}, \citeads{2008ApJ...673L.211M}). However, the detailed observations of the plasma heating and cooling, and the relationship to the photospheric magnetic field has yet to be understood. It has been observed that jets are associated with emerging field regions, magnetic cancellation \citepads{2008A&A...481L..57C}, and locations of photospheric shearing motions \citepads{1998SoPh..178..379S}. 

A number of authors have studied temperature diagnostics of individual jets (CH and AR) using imaging (TRACE, AIA, XRT) and spectroscopic (SUMER, EIS) observations. The temperatures of jets have been studied by using two techniques based on different approaches - the \textit{filter-ratio method} which assumes a single temperature along the \textit{line-of-sight (LOS)} (e.g., \citeads{2000ApJ...542.1100S}, \citeads{2011AdSpR..48.1490N}, \citeads{2011A&A...526A..19M}, \citeads{2012A&A...545A..67M}, \citeads{2012ApJ...759...15M}, \citeads{2013ApJ...776...16P}, \citeads{2014PASJ...66S..12Y}) and the \textit{Differential Emission Measure (DEM)} method which assumes a multi-thermal plasma along the LOS (e.g., \citetads{2010ApJ...710.1806D}, \citeads{2014A&A...562A..98C}, \citeads{2013ApJ...763...24K}, \citeads{2013ApJ...769...96C}, \citeads{2014ApJ...787L..27S}).

\citetads{2011AdSpR..48.1490N} reported the temperature of CH jets ranging between \mbox{0.8 to 1.3 MK} using the filter ratio method applied to STEREO/SECCHI data; whereas \mbox{\citetads{2013ApJ...776...16P}} found the temperarture of the jet to be \mbox{1.7 MK}. Using a DEM analysis, \citetads{2010ApJ...710.1806D} found the temperature of a CH jet was \mbox{log \textit{T} [K] = 6.15}; whereas \citetads{2014A&A...562A..98C} reported a temperature of about \mbox{log \textit{T} [K] = 5.89}. The presence of \mbox{Fe XIV} and \mbox{Fe XV} emission during the jet event in EIS/Hinode observations confirmed the temperature of CH jets of \mbox{2-3 MK} in an analysis by \citetads{2007PASJ...59S.751C}. Recently, \citetads{2011A&A...526A..19M} showed the presence of high temperature up to \mbox{12 MK} in the footpoints of the jet together with a density of \mbox{4$\times$10$^{10}$ cm$^{-3}$}.

Using AIA/SDO observations, \citetads{2013ApJ...763...24K} studied the emission measure and temperature distribution of a surge which was observed on the edge of an active region. Using the automated method developed by \citetads{2013SoPh..283....5A}, the authors found the average temperature and a density of \mbox{2 MK} and \mbox{4.1$\times$10$^{9}$ cm$^{-3}$} respectively during the maximum rise of the surge. Also, \citetads{2013ApJ...769...96C} studied the temperature structure of an AR jet using the same DEM method and found hot plasma \mbox{$\sim$7 MK} at the footpoint of jet.

\citetads{2014A&A...567A..11Z} investigated the temperature of plasma-blobs observed along the spire of a recurrent AR jet using the DEM method “xrt\_dem\_iterative2.pro” along with Monte Carlo (MC) simulations (details in \citetads{2012ApJ...761...62C}). The authors found the DEM weighted average temperatures of the three blobs to range from \mbox{0.5 to 4 MK} with a median value of \mbox{$\sim$2.3 MK} and the densities of 3.3, 1.9, and \mbox{2.1$\times$10$^{9}$ cm$^{-3}$} respectively. Using simultaneous imaging observation from XRT and spectroscopic observations from EIS, \citetads{2008A&A...481L..57C} observed an AR jet component over a range of temperatures between \mbox{log \textit{T} [K] = 5.4} and 6.4 and the base of the jet was observed to be around \mbox{log \textit{T} [K] = 6.2}. The authors also measured the electon density in the high velocity up-flow component using the Fe XII line ratios to be above \mbox{log N$_{e}$ (cm$^{-3}$) = 11}. \citetads{2012ApJ...759...15M} studied the relationship between the velocity and temperature of an AR jet using EIS and STEREO observations. The authors observed the jet structure over a wide temperature range from \mbox{log \textit{T} [K] = 4.9 to 6.4} and investigated the doppler velocities.

The cooler counterparts of the hot jets (X-ray) are often observed as dark absorbing features \citepads{1973SoPh...28...95R} in the H$\alpha$ observations at chromospheric temperatures named as 'H$\alpha$ surges' (\citeads{1999ApJ...513L..75C}, \citeads{2007A&A...469..331J}, \citeads{1995SoPh..156..245S}, \citeads{1996ApJ...464.1016C}. Also, Yokoyama \& Shibata (\citeyearads{1995Natur.375...42Y}, \citeyearads{1996PASJ...48..353Y}) found both hot and cool jets along the oblique field line in 2D simulations of flux emergence. Almost all the previous studies seem to support an association between X-ray jets, EUV jets and H$\alpha$ surges in regions of evolving magnetic field observed at photospheric heights. These observations seem to indicate that newly emerging flux appearing from below interacts with pre-existing magnetic flux in the different layers of the atmosphere and produces jet-like structures. However, the details of the processes have not yet been determined. This is largely due to inadequate spatial and temporal resolution in the observations. In order to study the local plasma parameters (temperature, density, abundance, flows, magnetic field etc.) involved in the process of jet phenomena, the study would require simultaneous multiwavelength imaging and spectroscopic observations with very high spatial resolution and high-cadence.
 
In this paper, we report multiwavelength observations of 20 EUV jets observed at the periphery of active regions (ARs) during the search period August 2010 to June 2013 \mbox{(See Table \ref{table1})}. We take advantage of the high-cadence (12s) SDO/AIA observations to characterise the jets. Such high-cadence observations are necessary to study these dynamic events. We have carried out the first comprehensive study of multiwavelength observations of active region jets. We used the regularized inversion technique developed by \citetads{2012A&A...539A.146H} to recover differential emission measure (DEM). This method gave us a better representation than the above mentioned methods to investigate the temperature structure in the AR jets. We compute reliable estimates of errors for the DEM over the range of temperatures which further allowed us to investigate the temperature variation of the coronal plasma for different regions in the jet events. We performed a detailed analysis of the temperature distribution for the ‘spire’ and the ‘footpoint’ regions of two jets which has not been done in the previous studies of individual AR jets. In addition, we derived a lower limit of the electron density from these results. We found significant and important differences in the physical parameters for these different regions of the jets. 
 
In \mbox{section 2}, we present information about the instruments we used for this study and the analysis techniques. In \mbox{section 3}, we discuss the characteristic behaviour of active region jets and their association with other phenomena occurring in the solar atmosphere such as nonthermal \mbox{type-III} radio bursts, hard X-ray emission (HXR), soft X-ray flares and H$\alpha$ surges. In section 4, we discuss the DEM analysis of two active region jets and investigate their temperature structure and electron densities at the location of the spire and at the footpoint. The discussion and summary is presented in section 5.

\section{Observations}

\begin{figure*}[!btp]
\begin{center}
\includegraphics[trim=0.1cm 3.4cm 6.7cm 2.0cm,width=0.9\textwidth]{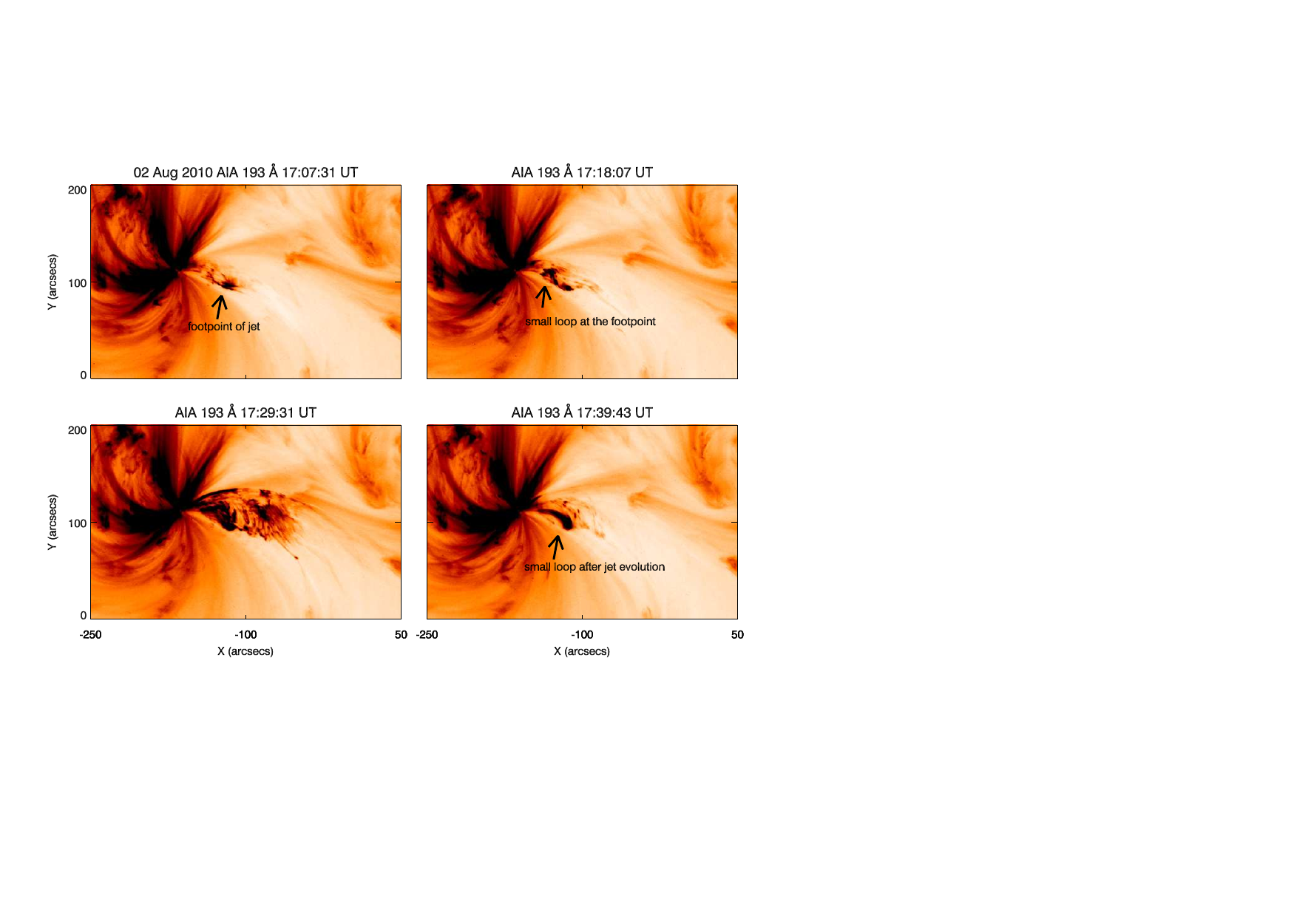}
\caption{The temporal evolution of the jet observed on 2010 August 2 from active region 11092 (N13 E07) in the AIA 193~{\AA} passband. The untwisting nature of the jet plasma is seen with a small loop at the base (reverse color image). The complex, multi-threaded spire appeared to originate from the edge of the sunspot. (See online movie1.mp4) \label{Fig1}}
\end{center}
\end{figure*}

We selected 20 EUV active region jets observed between \mbox{August 2010} and \mbox{June 2013} by the Atmospheric Imaging Assembly \citepads[AIA; ][]{2012SoPh..275...17L} instrument onboard the Solar Dynamic Observatory (SDO). We chose these events using the daily SDO movies \mbox{(http://sdo.gsfc.nasa.gov/data/dailymov.php)} and the automatic solar feature detection tool - Heliophysics Event Knowledgebase \citepads[HEK; ][]{2012SoPh..275...67H} \mbox{(http://www.lmsal.com/isolsearch)}. HEK is a system which catalogs the interesting solar events and features using feature-detection methods and presents a short description of them. 

In this study, we included events satisfying the following conditions :

\begin{itemize}
\item The jets were associated with an active region.
\item The AR jets were observed on the solar disk within \mbox{+/-60\degree} latitude.
\end{itemize}


In the current analysis, we used high temporal (12 sec) and spatial resolution \mbox{(0.6$\arcsec$per pixel)} full disk images of the Sun from AIA/SDO in ten different UV/EUV pass bands. This covers the region from the lower chromosphere to the corona. We also used the LOS magnetogram data (with temporal resolution of \mbox{45 sec} and spatial resolution of \mbox{0.5$\arcsec$} per pixel) from the Helioseismic and Magnetic Imager (HMI) \citepads{2012SoPh..275..207S} onboard the SDO to investigate the possible cause for jets.

To understand the relationship between the nonthermal \mbox{type-III} radio bursts and jets, we used the data from the WAVES \citepads{1995SSRv...71..231B} instrument onboard the Wind satellite. This has two radio receivers - \mbox{RAD 1} \mbox{(20–1040 kHz)} and \mbox{RAD 2} \mbox{(1.075-13.825 MHz)} which provide a dynamic spectrum in radio wavelengths and information about nonthermal bursts observed in the solar atmosphere and interplanetary medium. We also examine the role of Hard X-ray Emission (HXR) associated with the jets using the high-resolution imaging data from the Reuven Ramaty High Energy Solar Spectroscopic Imager (RHESSI) mission \citepads{2002SoPh..210....3L}. This provides us with the detailed information on the positions and structures of thermal and nonthermal HXR sources in different energy bands. In order to investigate the relationship between cool and hot temperature component of jets, we analysed H$\alpha$ data (at 6563~{\AA}) using the ground-based observatories - \textit{Kanzelh\"{o}he Observatory} \citepads{1999ASPC..184..314O}, \textit{Big Bear Solar Observatory (BBSO)} \citepads{1970S&T....39..215Z}, the \textit{Solar Magnetic Activity Research Telescope (SMART)} \citepads{2004ASPC..325..319U} at Hida observatory and the data from \textit{Global Oscillation Network Group (GONG)}. All the data are reduced and prepared using the standard routines available in the SolarSoft libraries and are normalized by the exposure time. We co-aligned HMI magnetograms with AIA 1700~{\AA} channel and then accordingly co-aligned other AIA channels.

We provide a detailed list of events and statistical information about measured physical parameters of the 20 EUV AR jets in Table \ref{table1}. We report the observation of jets in the AIA 193~{\AA} channel and measured their lifetime \textit{(columns 2 and 3)}. We identify the active region locations along with their associated sunspot configurations \textit{(columns 4 and 5)}. We compared jet timings with GOES X-ray flares, RHESSI HXR emission and nonthermal type-III radio burst timings, listed in \textit{columns 6, 7 and 8} respectively. The AIA 171~{\AA} filter images were used to see the fine structure of jets and to calculate the \textit{plane-of-sky} velocities using the time-distance analysis technique. The results are listed in \textit{column 9}. We investigate the relationship of the jets with cool H$\alpha$ surges \textit{(column 10)} and their relationship with photospheric magnetic field topology using HMI magnetogram data \textit{(column 11)}. A \textit{Differential Emission Measure (DEM)} analysis is carried out to obtain the thermal structure of the jets and we present the details in \textit{column 12}. We selected two jets for detailed analysis. We carried out similar analyses (as shown in section 3 and 4) for other jets and summarised additional information and images in the Appendix. 

\section{Data Analysis and Results}

In this section, we discuss the detailed analysis of two active region jets (jet 1 and 13 in Table \ref{table1} which we called as jet 1 and 2 below) and also investigate their relationship to other phenomena.


\subsection{Jet 1 : 2010 August 02}


\begin{figure}[!hbtp]
\begin{center}
\includegraphics[trim=0.7cm 0cm 2.5cm 2.3cm,width=0.5\textwidth]{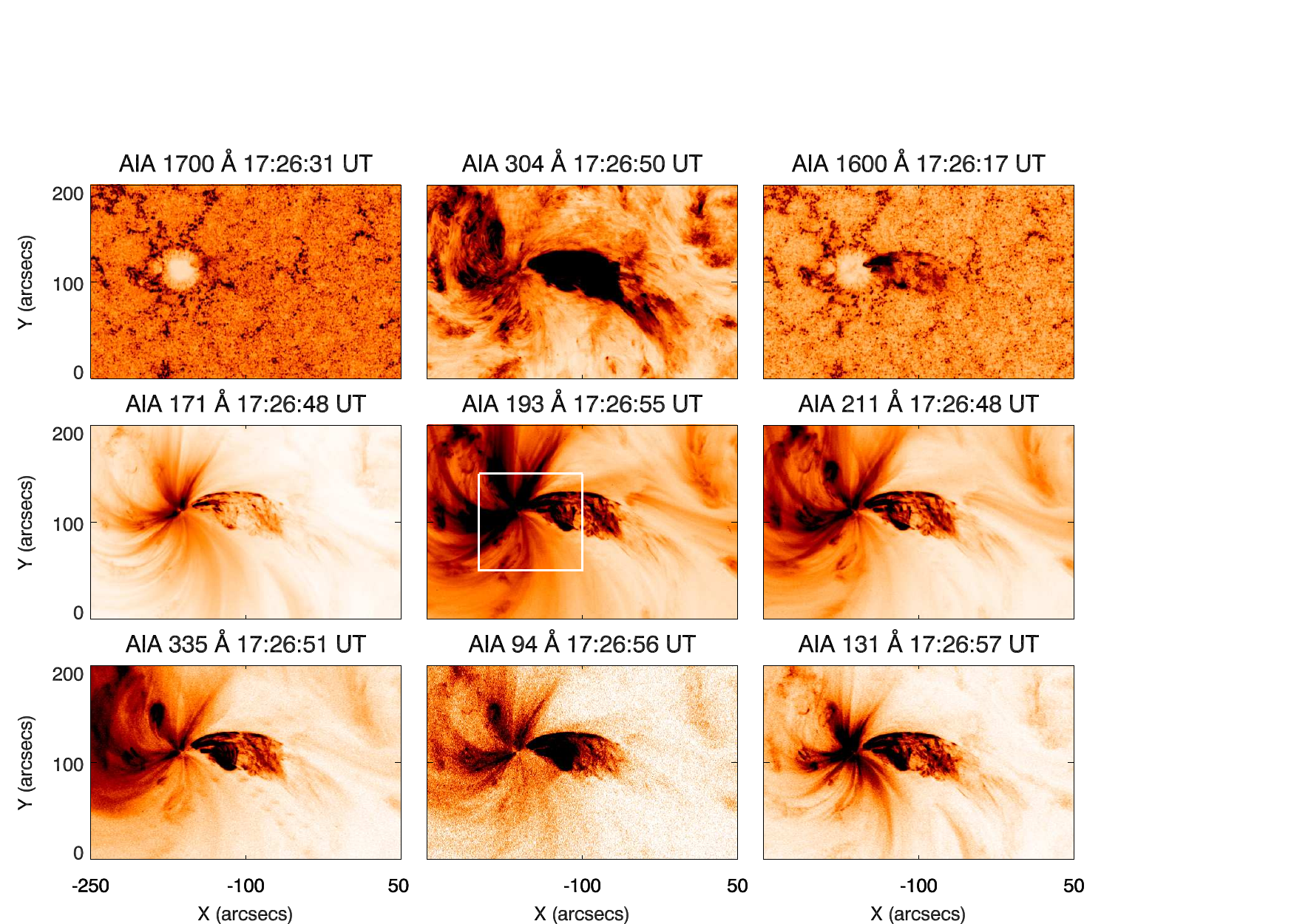}
\caption{The evolution of the jet observed on 2010 August 02 from active region 11092 (N13 E07) at \mbox{17:26 UT} in all AIA passbands. The images show the multi-thermal structure of jet spire originating from the edge of the the sunspot (reverse color image). The white over plotted box shows the field-of-view for the region shown in \mbox{figure \ref{Fig4}}.\label{Fig2}}
\end{center}
\end{figure}

\subsubsection{Overview and Kinematics}
The AIA instrument observed a jet on 2010 August 02 originating from the edge of AR 11092 (N13 E07). Figure \ref{Fig1} shows the temporal evolution of the jet in the AIA 193~{\AA} wavelength channel. The jet appeared to evolve from the western periphery of the active region at  \mbox{17:26 UT} in all wavelength channels of the AIA. \mbox{Figure \ref{Fig2}} shows the multi-thermal structure of the jet spire. The jet activity started at \mbox{17:10 UT} and ended at\mbox{17:35 UT}. The white over plotted box shows the field-of-view for the region shown in  \mbox{figure \ref{Fig4}}. The first brightening at the footpoint was seen at \mbox{17:07 UT} and as time progressed, a small loop starts to develop between the footpoint of the jet and the edge of the umbral-penumbral region of sunspot. At \mbox{17:18:43 UT}, the loop started to expand and became associated with a spire of the jet. The untwisting nature of the complex, multi-threaded spire was clearly visible along with its evolution. The bright loop at the footpoint was clearly observed after the jet evolution at \mbox{17:39 UT}. There was no significant X-ray activity recorded by the Geostationary Operational Environmental Satellite (GOES) during this time.

\begin{figure}[!hbtp]
\begin{center}
\includegraphics[trim=1cm 0cm 1cm 6.7cm, width=0.5\textwidth]{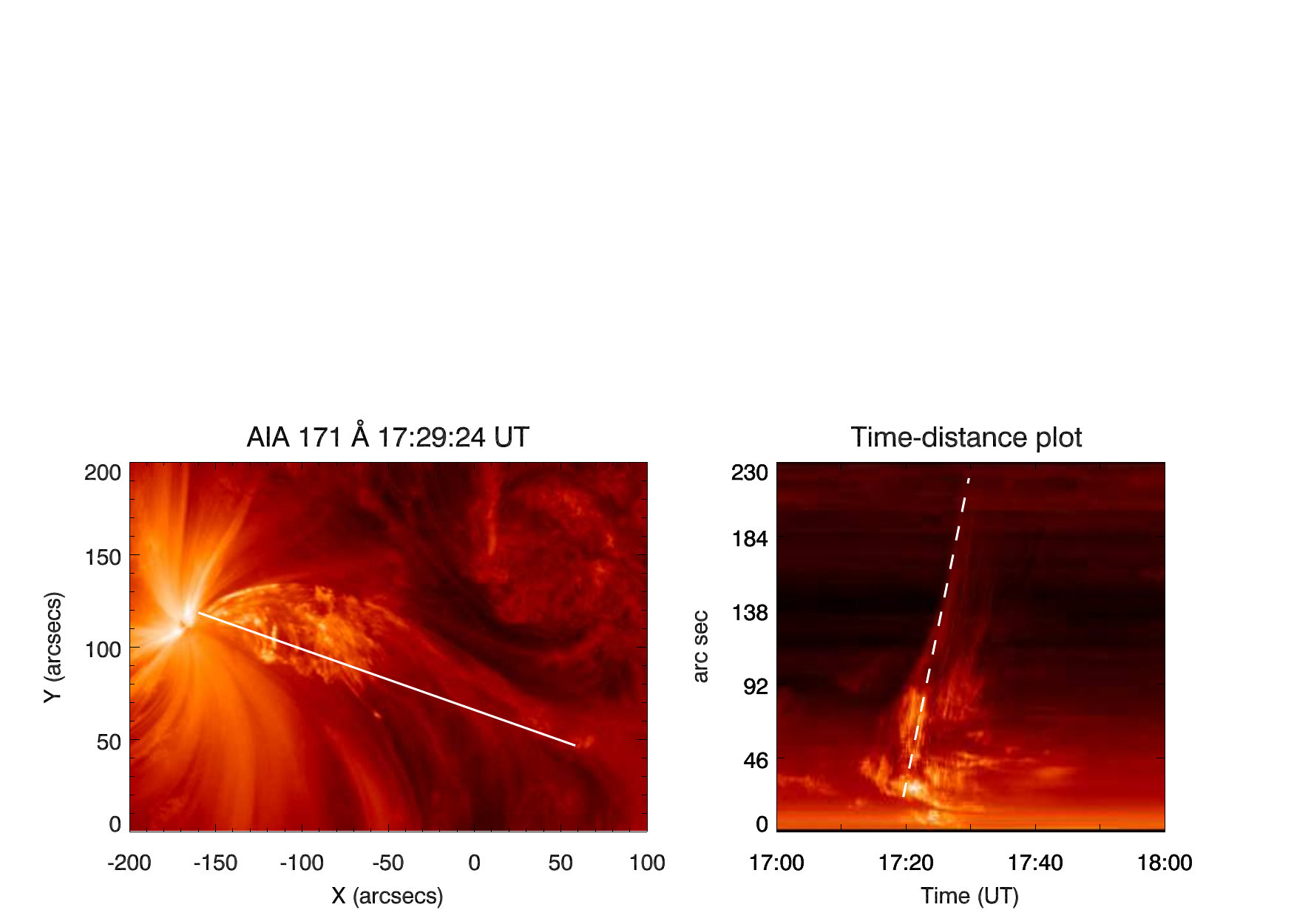}
\caption{Left panel : the jet evolution in the AIA 171 $\AA$ channel. The white line shows the artificial slit which is used to produce a time-distance plot. Right panel : time-distance plot along the jet spire. The white dashed line is used for the velocity calculation. This is found to be  \mbox{236 km/s}.\label{Fig3}}
\end{center}
\end{figure}

In order to estimate the speed of the jet, we performed the time-distance analysis using AIA 171~{\AA} filter images (dominated by Fe IX \mbox{log \textit{T} [K] = 6.0}). We employed an artificial slit along the direction of the jet spire and calculated the \textit{plane-of-sky} velocity of the jet-front as shown in \mbox{Fig. \ref{Fig3}}. The velocity is found to be \mbox{236 km/s}. The complex multi-threaded and untwisting nature of the jet spire is clearly visible in this plot.


\begin{figure}[!hbtp]
\begin{center}
\includegraphics[trim=0.2cm 0cm 8.3cm 4.1cm, width=0.48\textwidth]{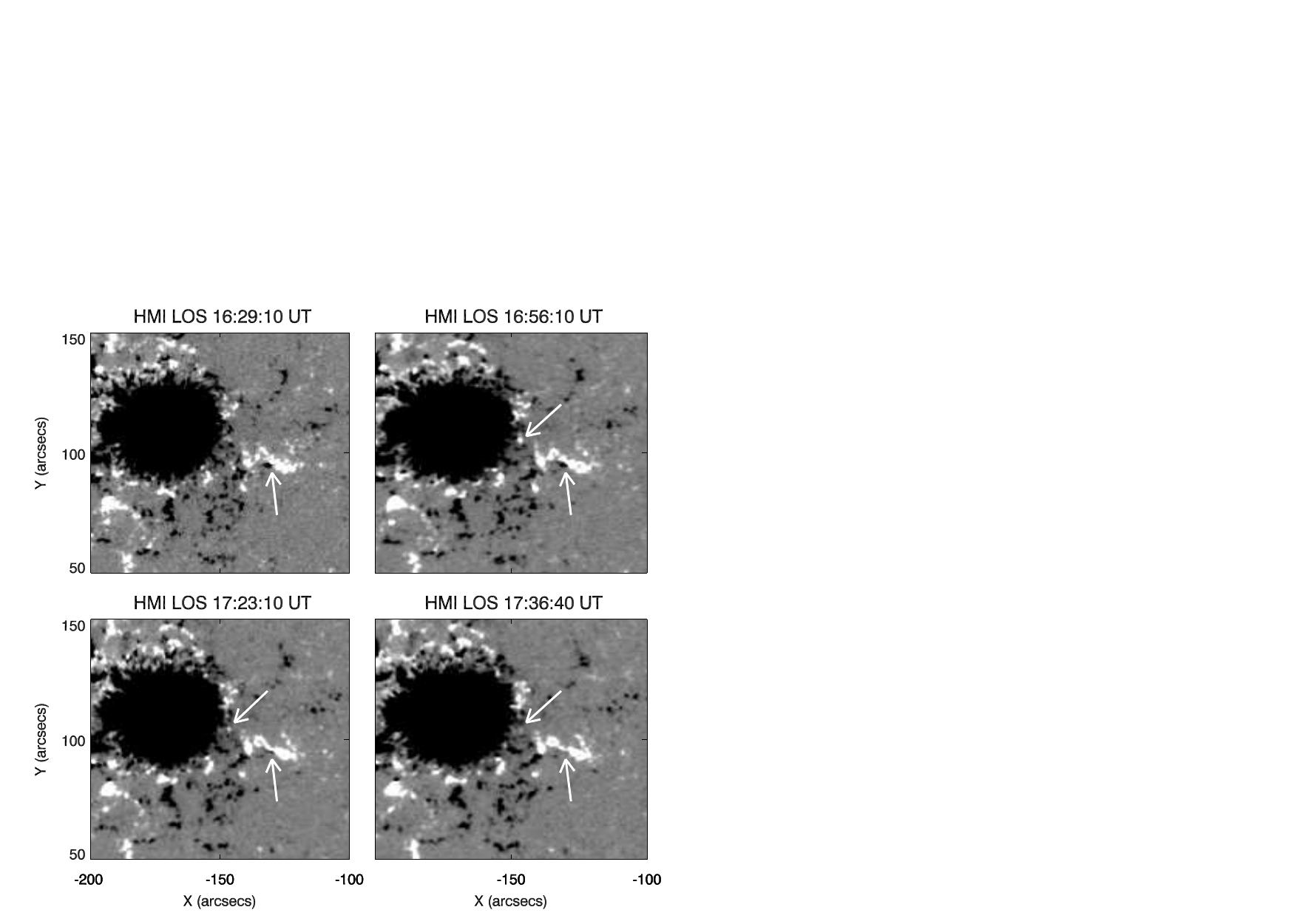}
\caption{HMI LOS magnetograms \mbox{(+/-100 Gauss)} during the jet evolution showing the negative-polarity sunspot (black) surrounded by positive-polarity (white). The white arrows show where there is an emergence of positive-polarity at the edge of the sunspot and negative-polarity cancellation on the western side of sunspot during the jet evolution. (See online \mbox{movie1.mp4}) \label{Fig4}}
\end{center}
\end{figure}


\subsubsection{Photospheric Magnetic Field Evolution}

\mbox{Figure \ref{Fig4}} shows the time evolution of the LOS component of the photospheric magnetic field in the \mbox{+/-100} Gauss range. The active region consists of a negative-polarity sunspot and positive-polarity surrounding it. This shows an anemone structure of the magnetic field (Shibata et al. \citeyearads{1992PASJ...44L.173S}, \citeyearads{1994xspy.conf...29S}). The jet was associated with the edge of a negative-polarity sunspot and a positive-polarity region on the western side of the sunspot as shown by white arrows in \mbox{Fig. \ref{Fig4}}. This activity of magnetic field lasted for more than an hour. The evolution of positive-polarity at the edge of the sunspot indicates flux emergence while the cancellation of the negative-polarity was seen at the western side of the sunspot. This observation is co-spatial with the location of the footpoint of the jet. \mbox{Figure \ref{Fig5}} (bottom panel) shows the AIA 193~{\AA} image at \mbox{17:26 UT} and the  contours are from the HMI LOS at the same time. The negative-polarity sunspot is shown by blue contours and positive-polarity as yellow contours. This image clearly shows the small loop at the footpoint of the jet evolve between the umbral-penumbral boundary of the sunspot and nearby positive-polarity region. The region where the magnetic activity took place is shown by white arrows.


\subsubsection{H$\alpha$ Surge and Hard X-ray Emission}

\begin{figure}[!hbtp]
\begin{center}
\includegraphics[trim=0.5cm 1cm 3cm 8.1cm, width=0.55\textwidth]{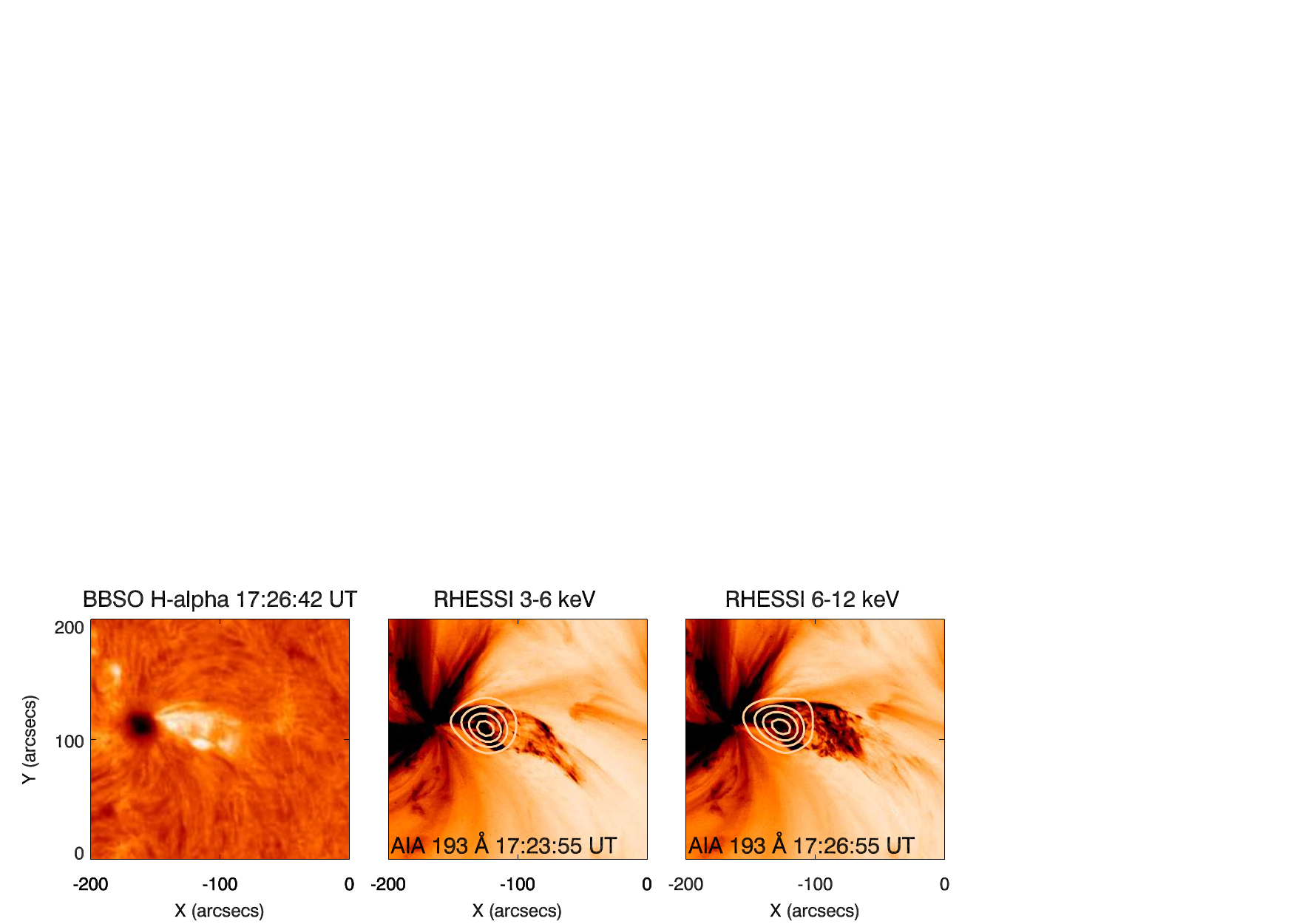}
\includegraphics[trim=3.2cm 0.8cm 3.5cm 0cm,width=0.47\textwidth]{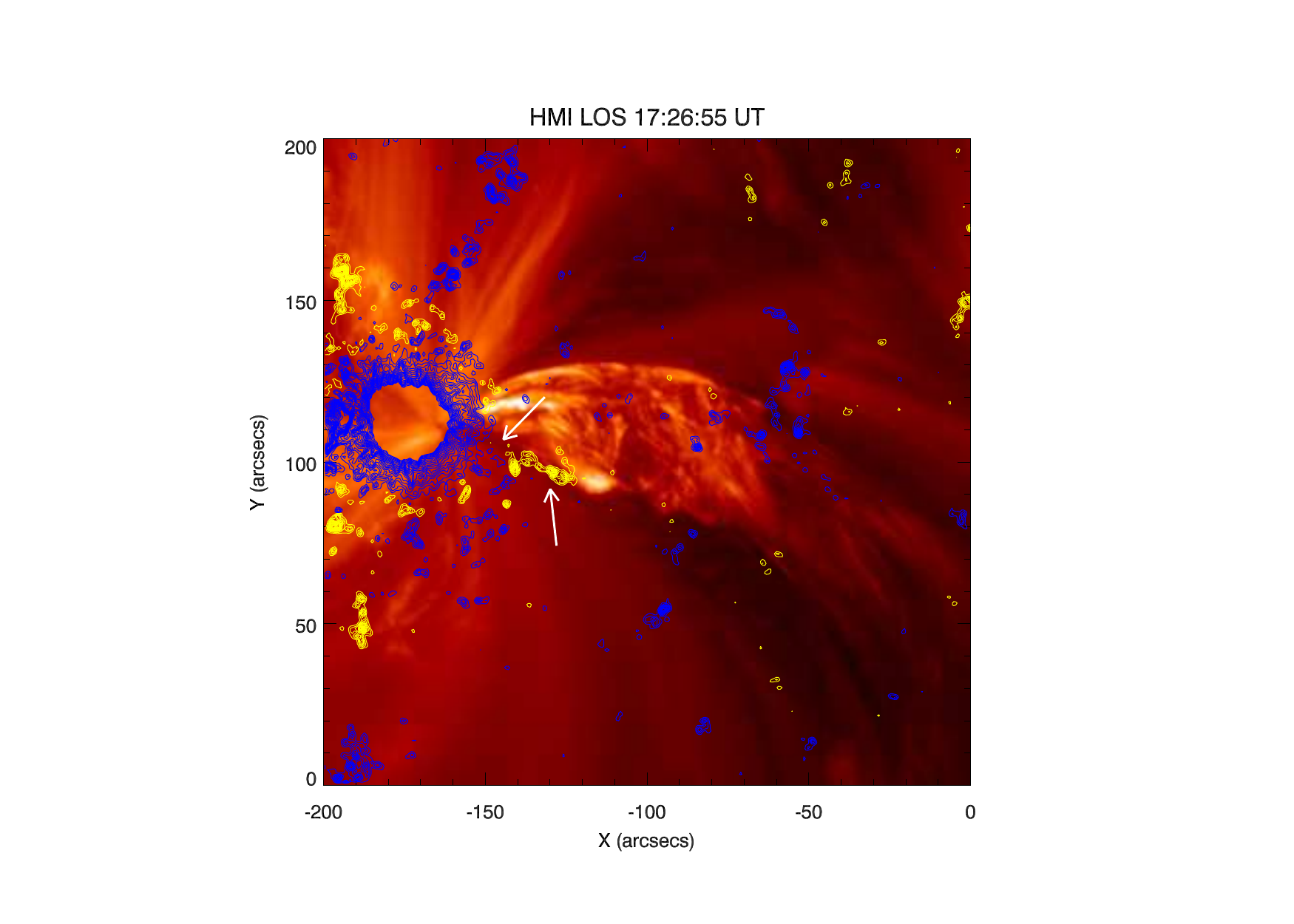} 
\caption{Top left panel : H$\alpha$ image of the jet at \mbox{17:26 UT} observed by the Big Bear Solar Observatory (BBSO). Top middle and right panel : the jet in the AIA 193 $\AA$ channel (reverse color image) with \mbox{3-6 keV} and \mbox{6-12 keV} RHESSI contours  (30, 50 , 70  and 90 \% of peak flux). Bottom panel : AIA 193~{\AA} image and  contours are HMI LOS at \mbox{17:26 UT}. The blue color contours represent the negative-polarity sunspot and the yellow contours show the positive-polarity regions. \label{Fig5}}
\end{center}
\end{figure}


\begin{figure}[!hbtp]
\begin{center}
\includegraphics[trim=0.5cm 0cm 3.5cm 2cm,width=0.45\textwidth]{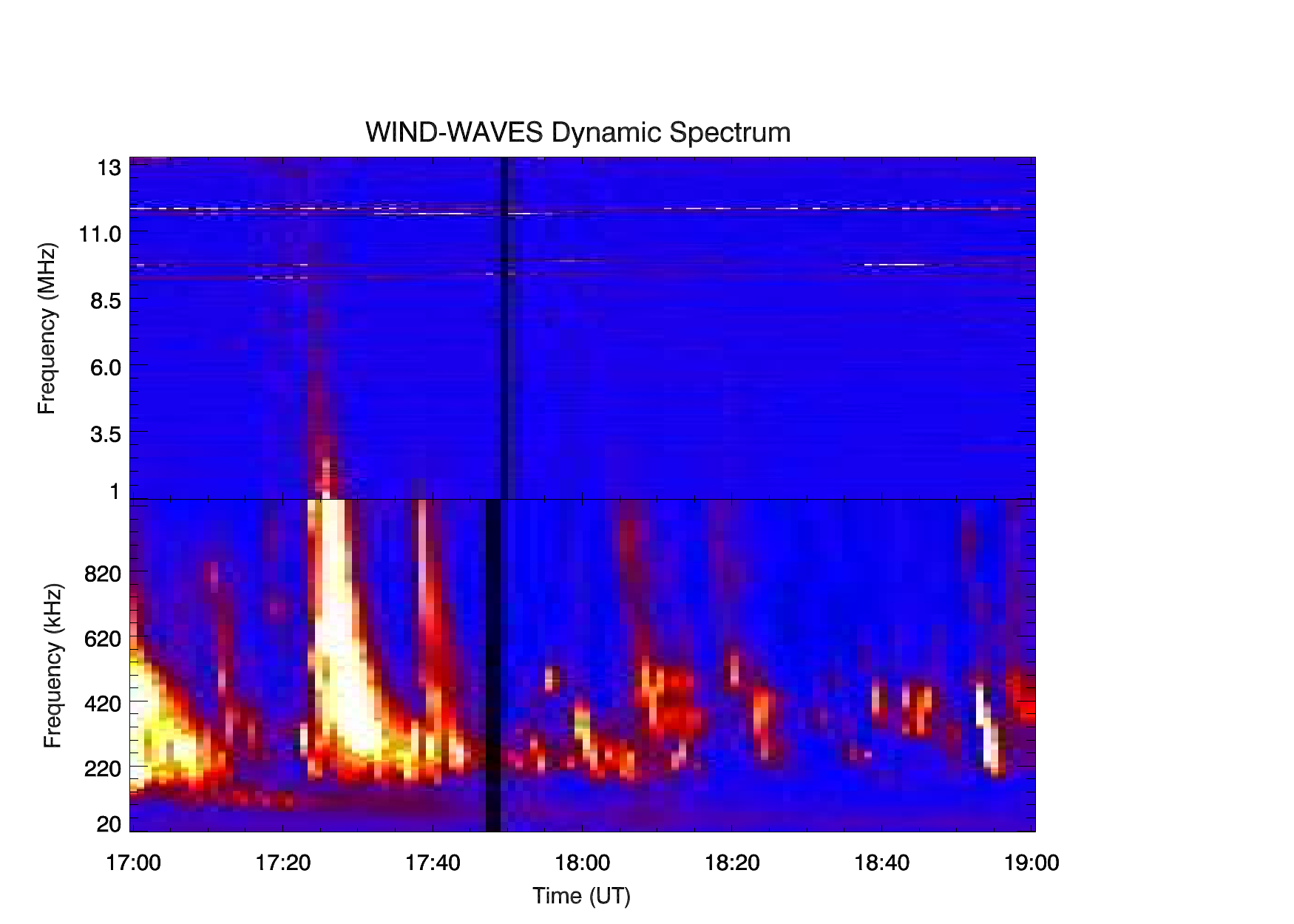}
\includegraphics[trim=1.3cm 0.1cm 1.8cm 0cm,width=0.25\textwidth]{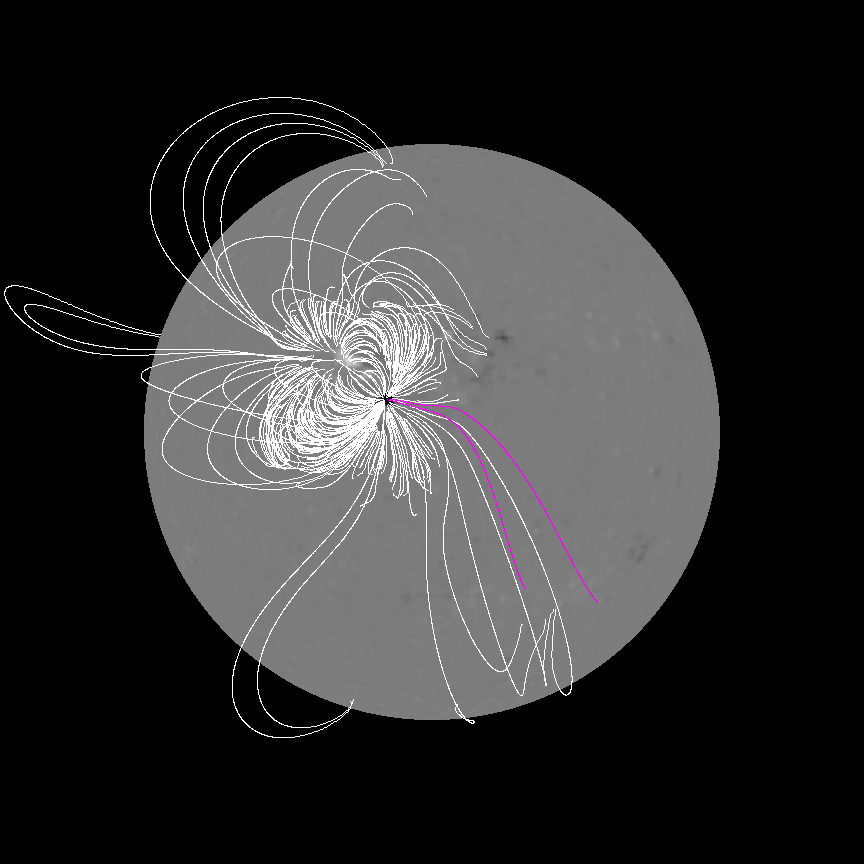}
\caption{Top panel : the dynamic radio spectrum indicates a nonthermal \mbox{type-III} radio burst observed by the \mbox{WIND/WAVES} at \mbox{17:25 UT}. This burst is co-temporal with the jet evolution. The presence of the HXR source and the nonthermal \mbox{type-III} radio burst indicates the signature of particle acceleration. Bottom panel : the PFSS extrapolation of the active region at \mbox{18:04 UT}. The white and pink lines indicate the closed and open magnetic structures in the active region and region nearby respectively. \label{Fig6}}
\end{center}
\end{figure}


We have analysed the H$\alpha$ data from the BBSO to investigate the presence of a low-temperature component during the jet evolution. \mbox{Figure \ref{Fig5}} (top left panel) shows the H$\alpha$ image (reverse colour) of the jet at \mbox{17:26 UT}. The evolution of low-temperature plasma originating from the edge of the sunspot is co-temporal and co-spatial with the hot EUV jet spire plasma as seen in the AIA images. This observation indicates the multi-thermal nature of the jet spire. 

\begin{figure*}[!btp]
\begin{center}
\includegraphics[trim=0.5cm 3.5cm 6cm 5.5cm,width=1.1\textwidth]{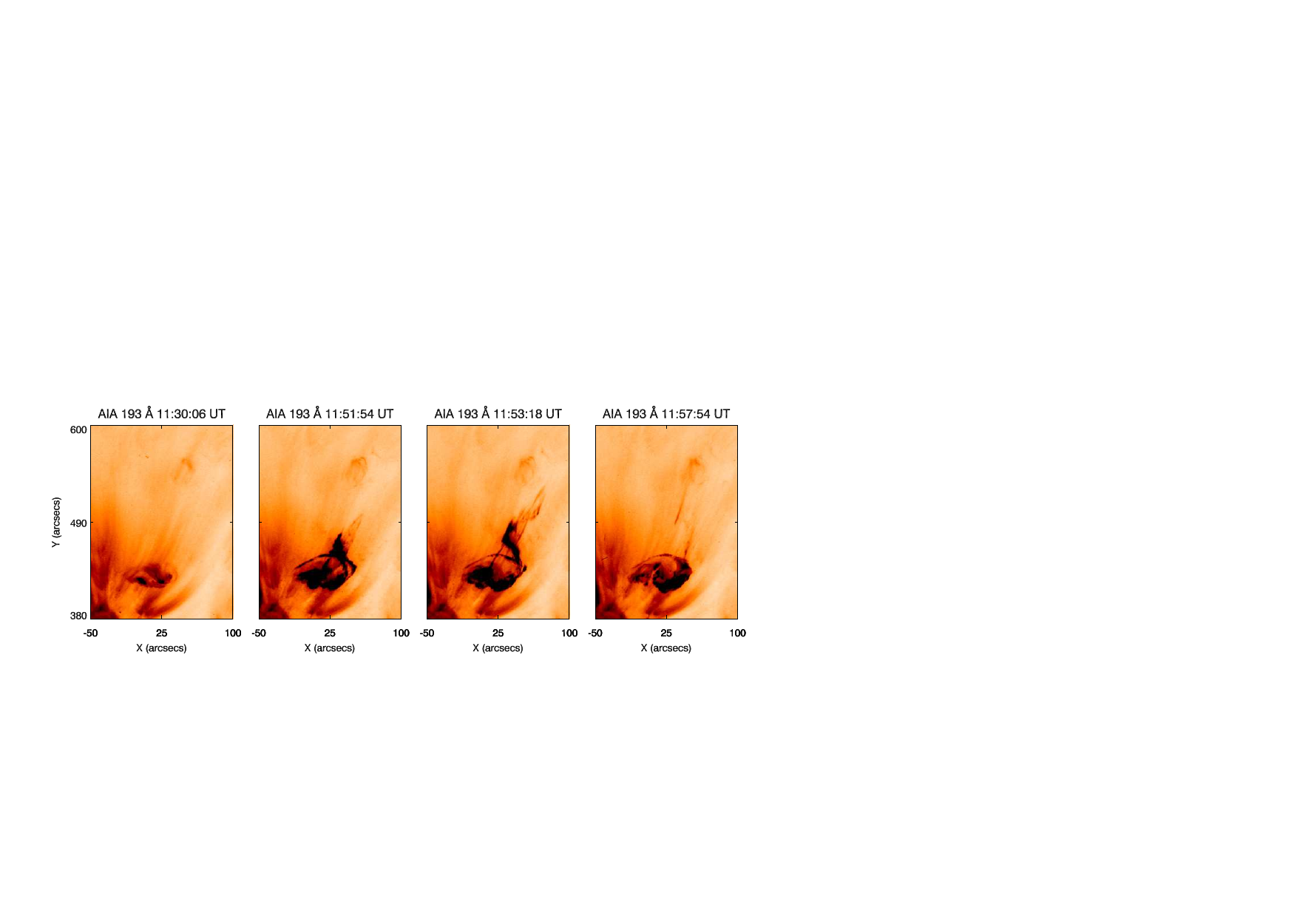}
\caption{The temporal evolution of a jet observed on 2013 March 02 from active region \mbox{AR 11681} (N17 E41) in the AIA 193~{\AA} passband. The complex, multi-threaded spire and a small loop at the footpoint before the jet evolution was clearly observed (reverse color image). The complex spire showed continuous untwisting motion. (See online movie2.mp4) \label{Fig7} }
\end{center}
\end{figure*}

In order to probe the spatial relationship between the jet and HXR sources, we analysed the high-resolution HXR data from the RHESSI satellite. We created images in the \mbox{3-6} and \mbox{6-12 keV} energy band with the CLEAN algorithm \citepads{2002SoPh..210...61H} using \mbox{4-8} sub-collimators. We integrated the HXR fluxes for a few minutes during the jet evolution to get sufficient counts. \mbox{Figure \ref{Fig5}} (top middle and right panel) shows the jet in the AIA 193~{\AA} channel at \mbox{17:23 UT} and \mbox{17:26 UT} and the  yellow contours (30, 50, 70 and 90$\%$ of the peak flux) are the HXR sources in the \mbox{3-6} and \mbox{6-12 keV} energy bands respectively, integrated over \mbox{17:21-17:27 UT}. The HXR sources are observed to be co-temporal and co-spatial with the footpoint of the jet.


\subsubsection{Radio Emission and PFSS results}
The fast moving nonthermal electrons along the open magnetic field lines emit radio emission at the plasma frequency. This fast drifting emission appears as a pillar structure in the radio dynamic spectrum which is classified as a nonthermal type-III radio burst. These bursts provide evidence for magnetic reconnection and particle acceleration  (\citeads{1995ApJ...447L.135K}, \citeads{2014RAA....14..773R}, \citeads{2012ApJ...754....9G}). We compared the jet timings with observations of dynamic spectra recorded by the WAVES instrument. A nonthermal \mbox{type-III} radio burst was observed at 17:25 UT by the \mbox{RAD 1} and \mbox{RAD 2} radio receivers in the frequency range from \mbox{13 MHz} to \mbox{220 kHz} as shown in \mbox{Fig. \ref{Fig6}} (top panel).

Further, we also investigated the location of these bursts using the Potential Field Source Surface (PFSS) \citepads{1969SoPh....6..442S} extrapolation of the photospheric magnetic field. This tool is used to visualise the solar coronal magnetic field in the active region. \mbox{Figure \ref{Fig6}} (bottom panel) shows the extrapolated coronal magnetic field at \mbox{18:04 UT} using the PFSS analysis technique. The white lines show the closed magnetic structure associated with the active region and the pink lines indicate the open magnetic field lines. The source region of the jet and the open magnetic field lines share the same location. This indicates that the jet is ejected in the direction of these open field structures. The presence of open field lines confirm the source region of the nonthermal type-III radio burst in the jet.


\subsection{Jet 2 : 2013 March 02}

\subsubsection{Overview and Kinematics}

\begin{figure}[!hbtp]
\begin{center}
\includegraphics[trim=0.5cm 0.1cm 9.3cm 2.5cm,width=0.5\textwidth]{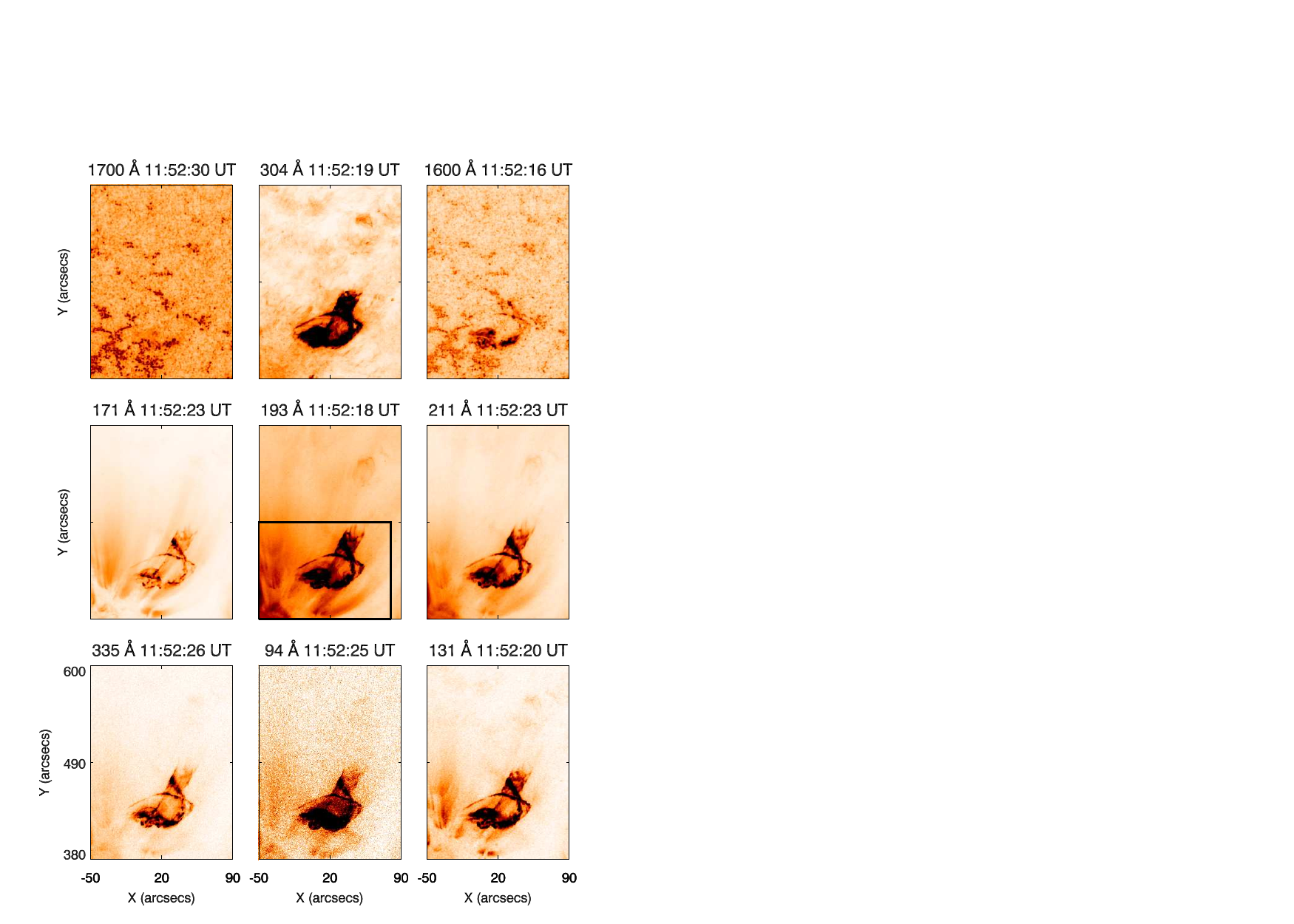}
\caption{The evolution of the jet on 2013 March 02 at \mbox{11:52 UT} in all AIA passbands. The images show the multi-thermal structure of the jet spire originating from the edge of the sunspot (reverse color image). The black over plotted box shows the field-of-view for the region shown in \mbox{figure \ref{Fig10}}. \label{Fig8}}
\end{center}
\end{figure}

The AIA instrument observed a jet on \mbox{2013 March 02} originating from the edge of \mbox{AR 11681} (N17 E41). \mbox{Figure \ref{Fig7}} shows the temporal evolution of the jet in the AIA 193~{\AA} wavelength channel. The jet appeared to evolve from the western periphery of the active region at \mbox{11:49:43} UT in all wavelength channels of AIA as shown in \mbox{Fig. \ref{Fig8}}. The black over plotted box shows the field-of-view for the region shown in \mbox{figure \ref{Fig10}}. A small loop-like structure was observed at the footpoint and as time progressed, it started to grow and erupt as a jet. The first brightening of the loops was seen at \mbox{11:50:18 UT}, before the jet evolution. The jet activity started at \mbox{11:48 UT} and ended at \mbox{12:00 UT}. The continuous untwisting motion of the complex spire was clearly observed in all AIA channels. Another jet was also observed from the same location at \mbox{12:10 UT} which showed a simple spire structure. This observation confirms the recurrent nature of the jet. A B6 X-ray flare was recorded at \mbox{11:48 UT} by GOES. 

\begin{figure}[!hbtp]
\begin{center}
\includegraphics[trim=0.5cm 0.1cm 0.6cm 3.05cm,width=0.5\textwidth]{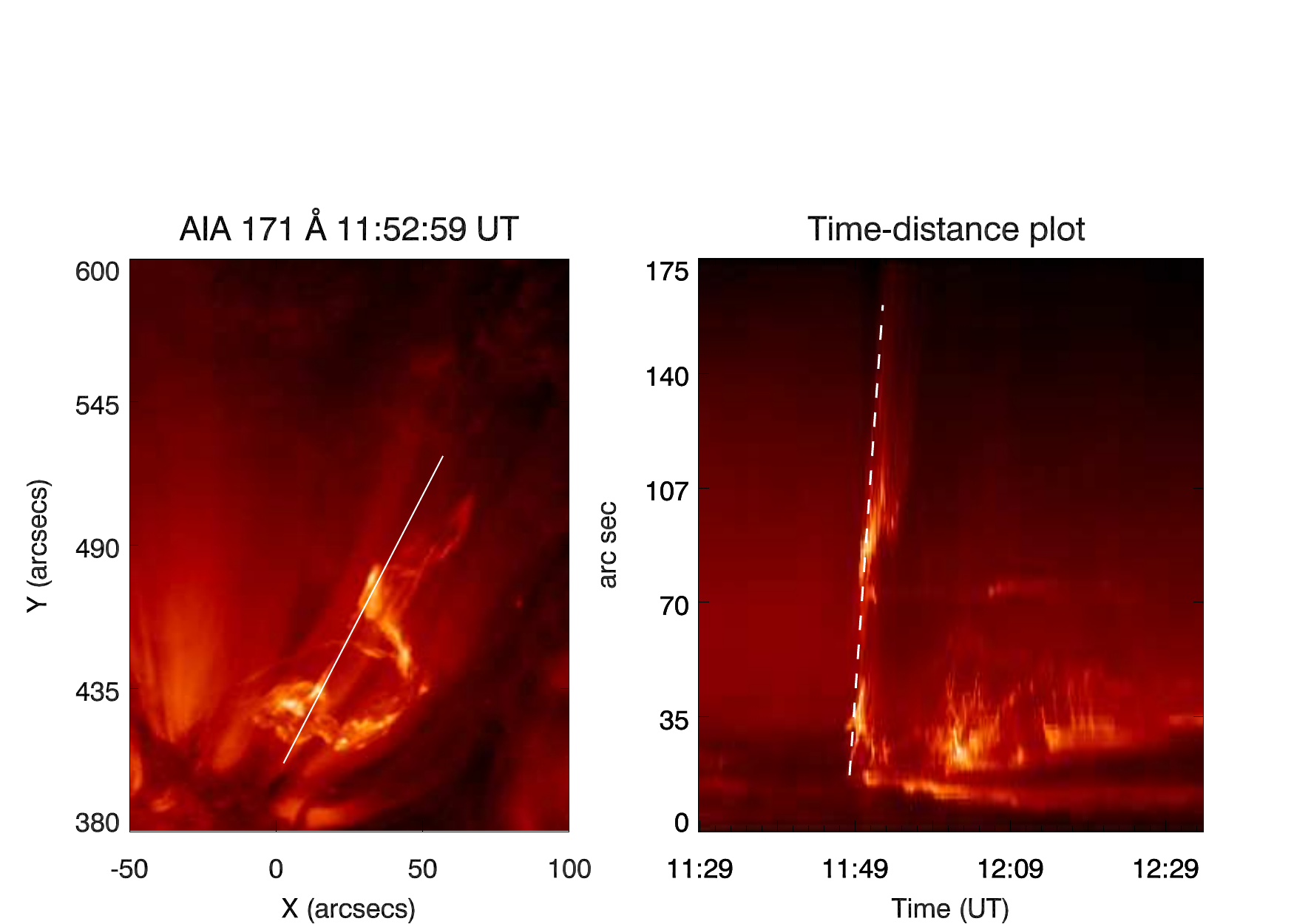}
\caption{Left panel : the jet evolution in the AIA 171~{\AA} channel, The white line shows an artificial slit which is used to produce time-distance plot. Right panel : time-distance plot along the jet spire. The white dashed line is used for the \textit{plane-of-sky} velocity calculation, which is found to be \mbox{316 km/s}.\label{Fig9}}
\end{center}
\end{figure}

We employed the time-distance analysis technique to calculate the \textit{plane-of-sky} jet's velocity. \mbox{Figure \ref{Fig9}} (left panel) shows an image of the jet at \mbox{11:52 UT} in the AIA 171~{\AA} channel and (right panel) shows the time-distance plot of the jet-front. The white  line (left panel) indicates the artificial slit along the jet spire which is used to create a time-distance plot and the white dashed line (right panel) which is used for the velocity calculation, which is found to be \mbox{316 km/s}. The complex, multi-threaded nature of the jet spire and its untwisting nature was well observed. 

\subsubsection{Photospheric Magnetic Field Evolution}
\begin{figure}[!hbtp]
\begin{center}
\includegraphics[trim=0.5cm 6cm 4.5cm 2.5cm,width=0.6\textwidth]{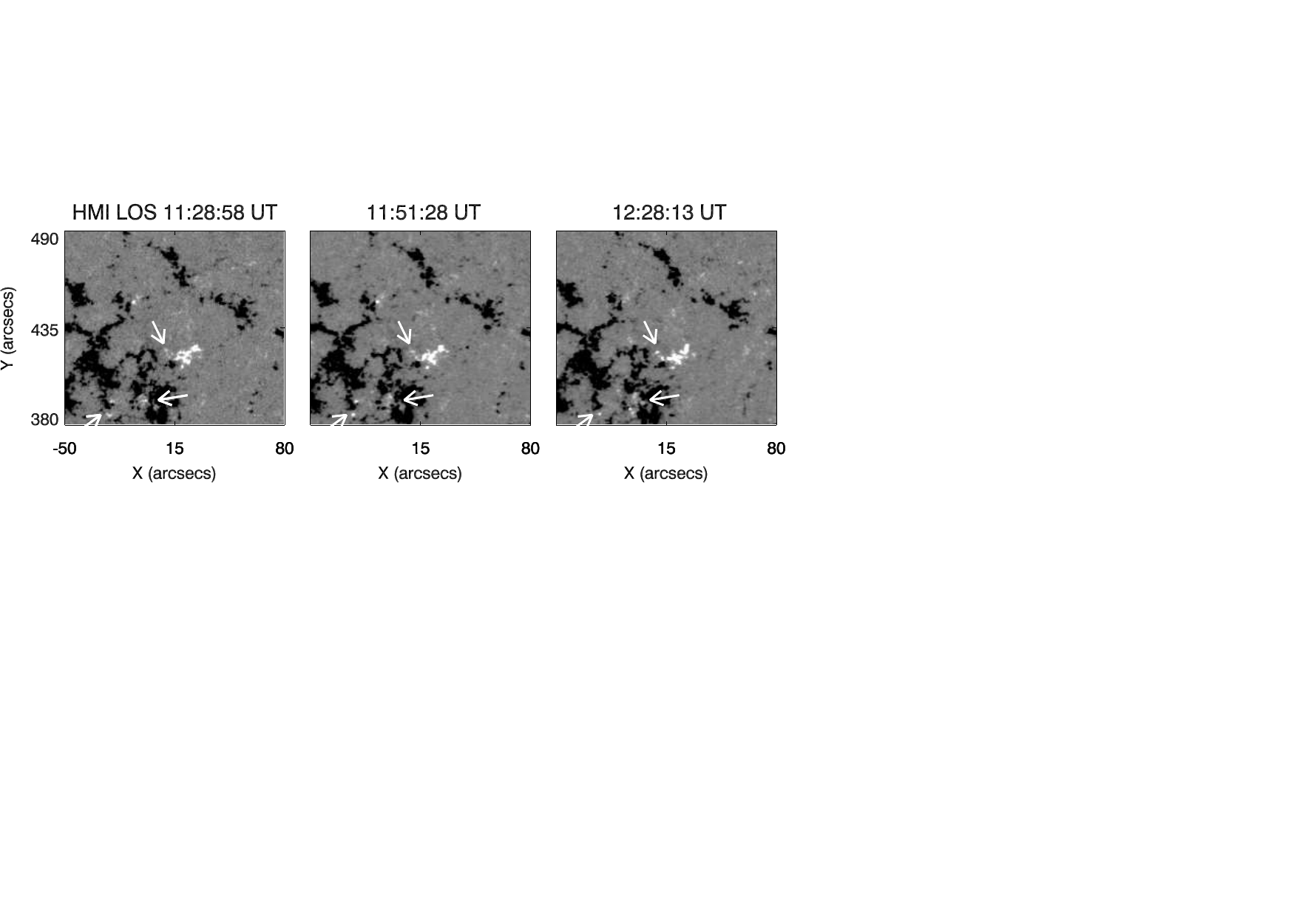}
\caption{The HMI LOS magnetograms during the jet evolution show the positive flux emergence (white) in the negative-polarity (black) and then cancellation. The white arrows show that there is an emergence of positive-polarity during the jet evolution. (See online \mbox{movie2.mp4}) \label{Fig10}}
\end{center}
\end{figure}

\mbox{Figure 10} shows the time evolution of the LOS component of the photospheric magnetic field in the \mbox{+/-100} Gauss range. The active region consists of a negative-polarity region. We clearly observed the positive-flux emergence and cancellation in this region which is shown by the white arrows in \mbox{Fig. \ref{Fig10}}. These regions are co-spatial with the footpoint of the jet. This activity of the magnetic field lasts for half an hour. This observation indicates that the positive flux emergence and cancellation at the footpoint might have contributed to the plasma ejection.


\subsubsection{Radio Emission and PFSS results}

\begin{figure}[!hbtp]
\begin{center}
\includegraphics[trim=0.5cm 0.01cm 3.5cm 1.8cm, width=0.45\textwidth]{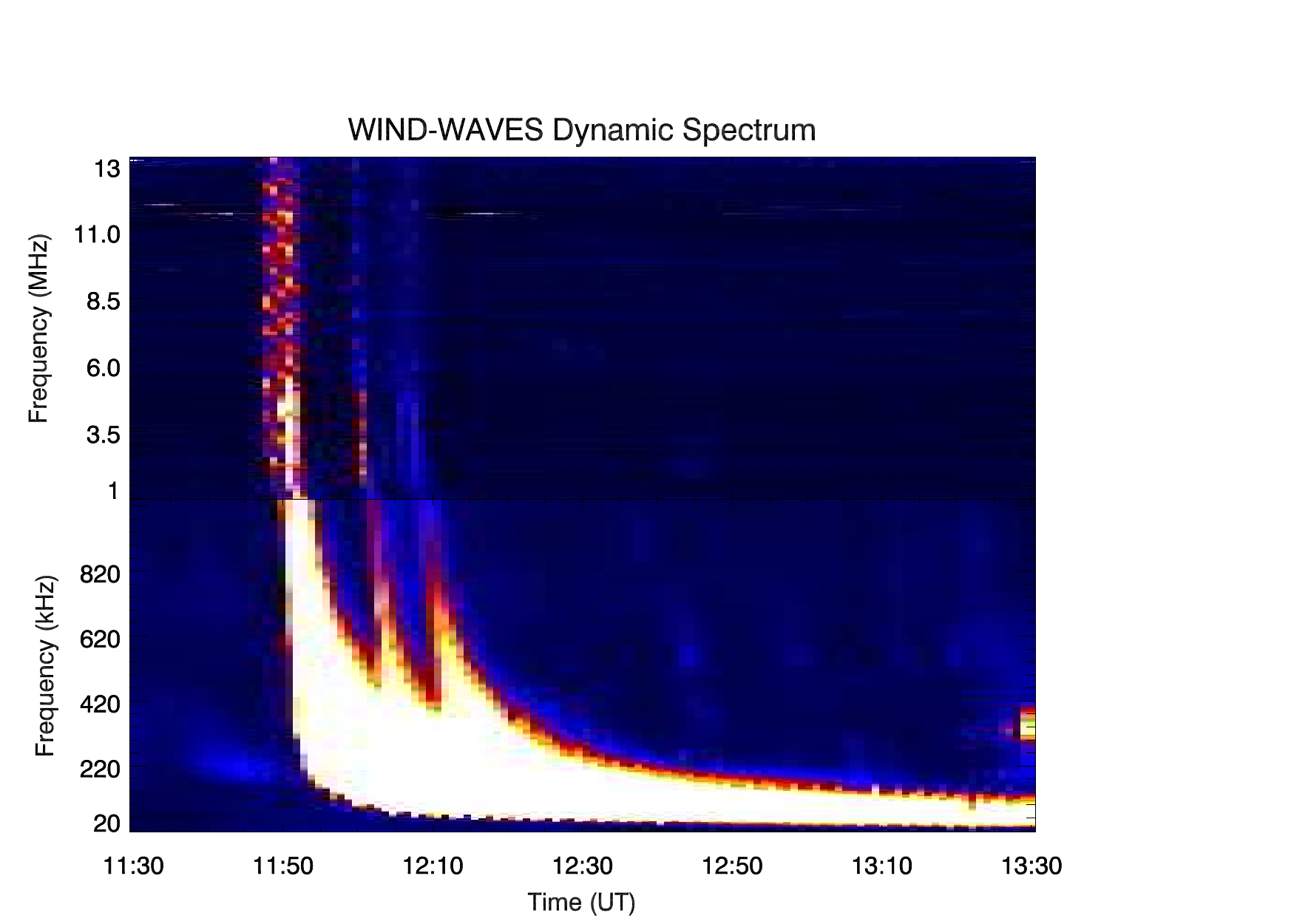}
\includegraphics[trim=1.3cm 0.1cm 1.3cm 0cm, width=0.3\textwidth]{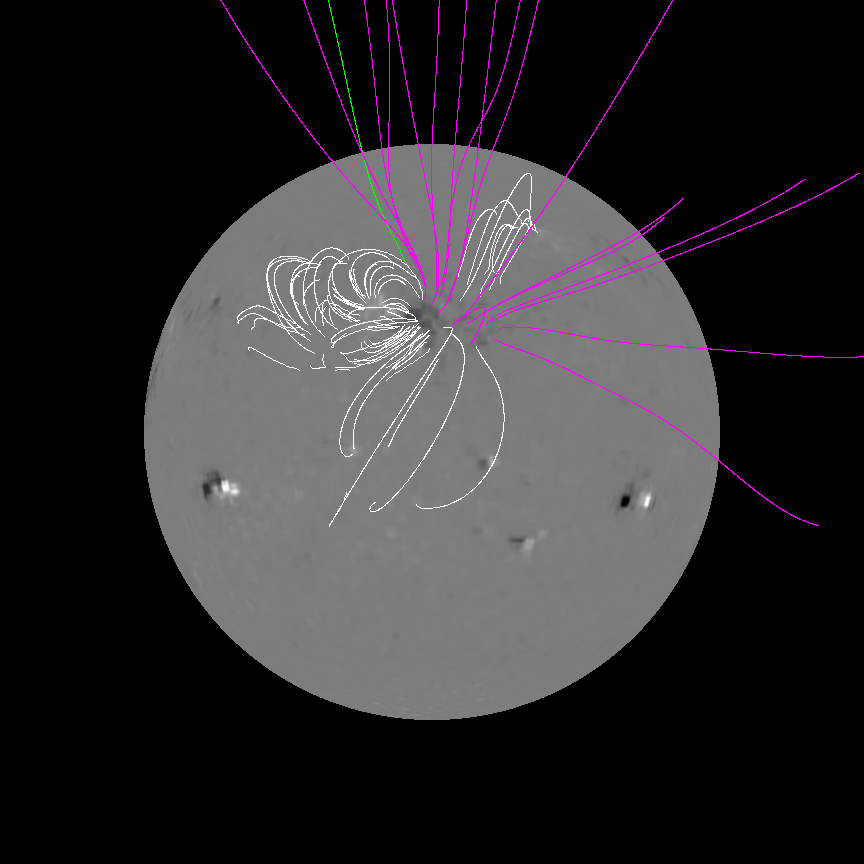}
\caption{Top panel : the dynamic radio spectrum indicates the interplanetary nonthermal type-III radio burst observed by WIND/WAVES at \mbox{11:50 UT}. This burst is co-temporal with the jet evolution. The presence of the nonthermal type-III radio burst indicates the signature of particle acceleration. Bottom panel : the PFSS extrapolation of active region at \mbox{12:04 UT}. The white and pink lines indicate the closed and open magnetic structure in the active region and nearby region respectively. \label{Fig11}}
\end{center}
\end{figure}

The WAVES instrument recorded a nonthermal type-III radio burst during the jet evolution. \mbox{Figure \ref{Fig11}} (top panel) shows the radio dynamic spectrum of a nonthermal type-III radio burst at \mbox{11:50 UT} recorded by the \mbox{RAD 1} \mbox{(20–1040 kHz)} and \mbox{RAD 2} \mbox{(1.075-13.825 MHz)} receivers on the WAVES instrument.

\mbox{Figure \ref{Fig11}} (bottom panel) shows the extrapolated coronal magnetic field at \mbox{12:04 UT} using the PFSS analysis technique. The white lines show the closed magnetic structure associated with the active region and the pink and green lines indicate the open magnetic field lines. The source region of the jet and the open magnetic field lines share the same location which indicates that the jet was ejected in the direction of these open field structures. The presence of open field lines confirm the source region of nonthermal type-III radio burst.


\section{Differential Emission Measure (DEM)}

In this section, we discuss the DEM analysis of the two active region jets described in section 3 and investigate their temperature and electron density structure at the location of the spire and at the footpoint.

\begin{figure}[!hbtp]
\begin{center}
\includegraphics[trim=0.5cm 0.1cm 0.2cm 6.2cm,width=0.5\textwidth]{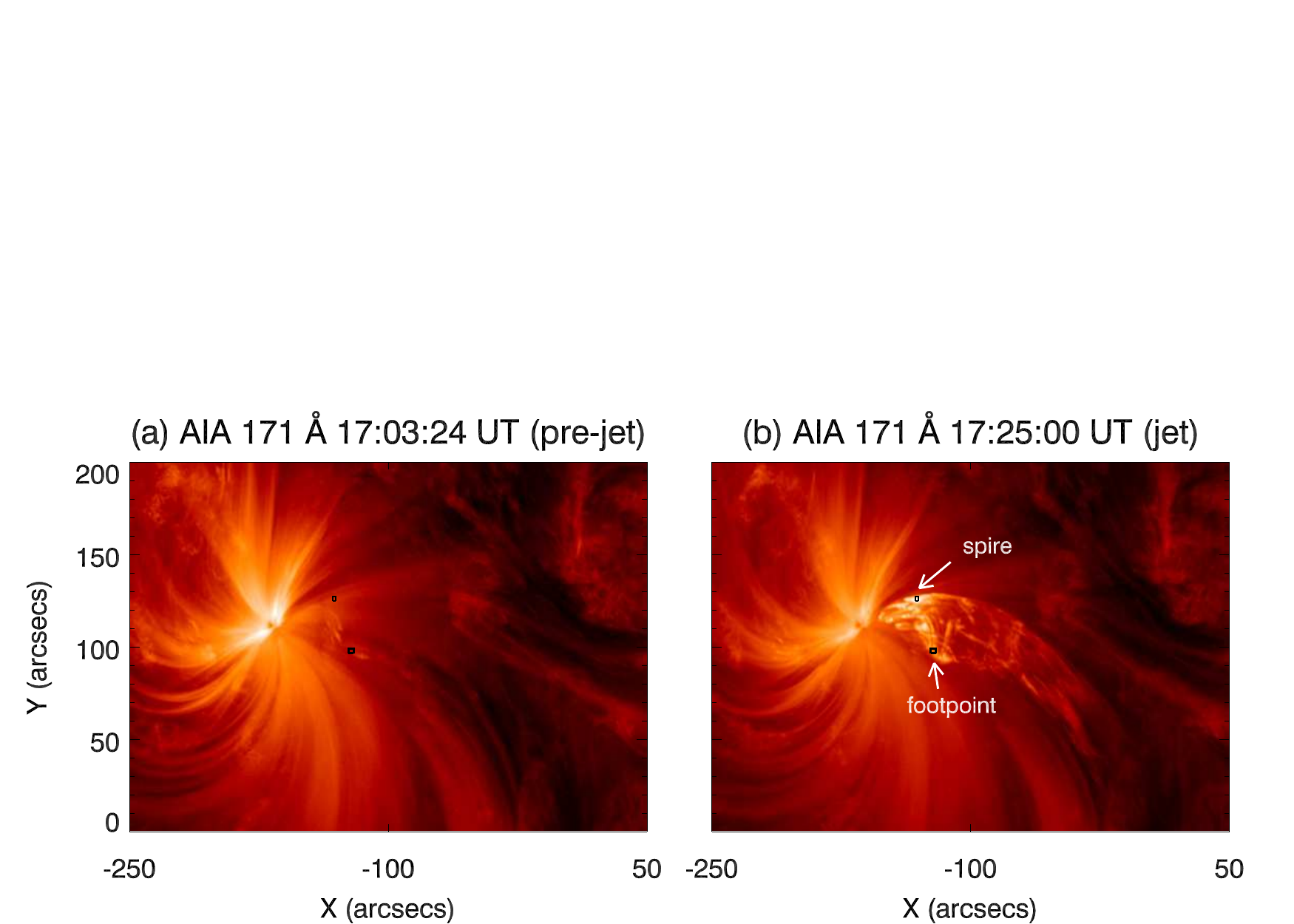}
\includegraphics[trim=1.5cm 0cm 0.1cm 0.1cm,width=0.5\textwidth]{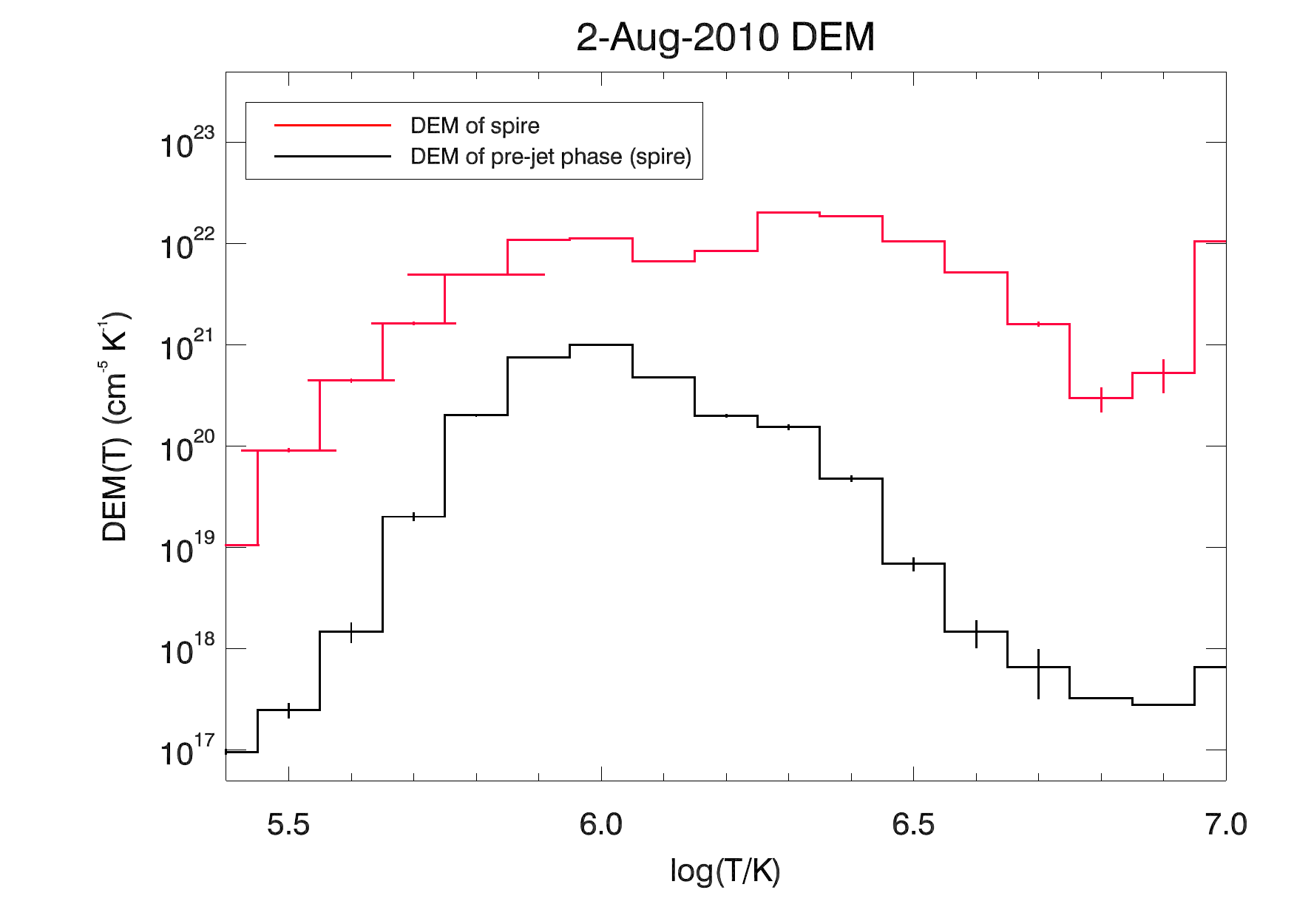}
\includegraphics[trim=1.5cm 0cm 0.1cm 0.1cm,width=0.5\textwidth]{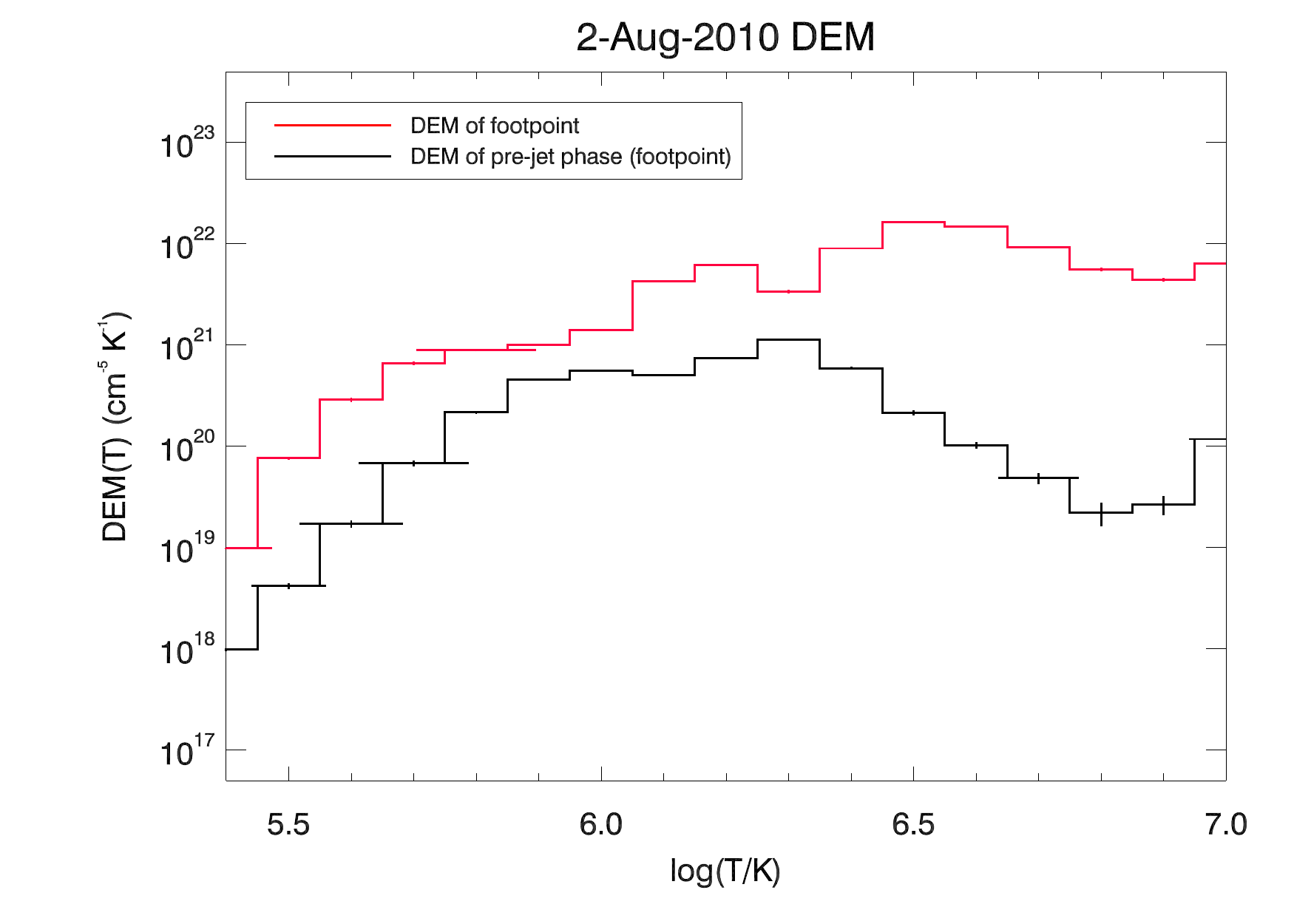}
\caption{Top panel : AIA 193~{\AA} image of (a) pre-jet phase and (b) jet. The two regions in the small boxes are used for the DEM analysis. Middle panel : shows the DEM curves for the spire of the jet (red curve) and at the same region during the pre-jet phase (black curve). Bottom panel : shows the DEM curves for the footpoint of the jet (red curve) and at the same region during the pre-jet phase (black curve). \label{Fig12}}
\end{center}
\end{figure}


Using the six AIA wavelength channels (94, 131, 171, 193, 211 and 335~{\AA}) which are sensitive to a range of coronal temperatures, we performed the LOS Differential Emission Measure (DEM) analysis by using the method of regularized inversion developed by \citetads{2012A&A...539A.146H}. See \citetads{2010A&A...521A..21O}, \citetads{2011A&A...535A..46D}, \citetads{2013A&A...558A..73D} for a detailed description of the EUV filters in AIA and their temperature responses.

The unsaturated AIA images during the pre-jet and jet phase were used for the reconstruction of DEM maps for selected areas. The method returns a regularized DEM as a function of T. For the inversion, we used the zeroth-order regularization in a range of temperature \mbox{log \textit{T} [K] = 5.5} to \mbox{log \textit{T} [K] = 7.0} with 15 temperature bins i.e. $\bigtriangleup$ \mbox{log \textit{T} = 0.1} intervals. The latest version of the AIA filter response function including the new \mbox{CHIANTI v.8} \citepads{2015A&A...582A..56D} and \citetads{1992PhyS...46..202F} coronal abundances were used for the analysis. By assuming a filling factor equal to unity, we calculated an emission measure (EM) and a lower limit of the electron density (\mbox{N$_{e}$}) in the region of the spire and in the region of the footpoint using \mbox{EM $\simeq$ 0.83 $\langle$\textit{N$_{e}$}$\rangle$$^{2}$\textit{$\bigtriangleup$h}} where \textit{$\bigtriangleup$h} is the column depth of the plasma along the \mbox{line-of-sight}. We integrated the DEM values over the temperature range \mbox{log \textit{T} [K] = 5.8} to \mbox{log \textit{T} [K] = 6.7} to calculate the EM values. We obtained the column depth \textit{$\bigtriangleup$h} from the width of the spire (assuming cylindrical geometry) and the footpoint visible in the AIA 171~{\AA} images.

\subsection{2010 August 02}

\mbox{Figure \ref{Fig12}} (top panels (a) \& (b)) show the AIA 171~{\AA} images of the pre-jet phase at \mbox{17:03 UT} and the jet phase at \mbox{17:25 UT} respectively. We used two small regions, one at the footpoint of the jet and the other at the spire of the jet to calculate the LOS DEM. The regions are over plotted as small black boxes in the \mbox{Fig. \ref{Fig12}} (top panel). We identify the same regions in the pre-jet phase at \mbox{17:03 UT} and calculate the DEM curve. The plots (middle panels) show the DEM curves for the spire of the jet (red curve) and at the same location during the pre-jet phase (black curve). This DEM indicates that the temperature at the spire is increased from \mbox{log \textit{T} [K] = 6.0 to log \textit{T} [K] = 6.3} during the jet evolution. We note however, the multi-thermal nature of the emission. The plots (bottom panel) show the DEM curves for the footpoint of the jet (red curve) and at the same region during the pre-jet phase (black curve). It is clear that, the temperature at the region of footpoint is increased from \mbox{log \textit{T} [K] = 6.3 to log \textit{T} [K] = 6.5}. DEM measurements beyond \mbox{log \textit{T} [K] = 7} are not reliable due to the uncertainty in the contribution of emission lines to the AIA 94 and 131~{\AA} channels \citepads{2014SoPh..289.3313Y}.

\begin{figure}[!hbtp]
\begin{center}
\includegraphics[trim=0.5cm 3.3cm 0.2cm 2cm,width=0.85\textwidth]{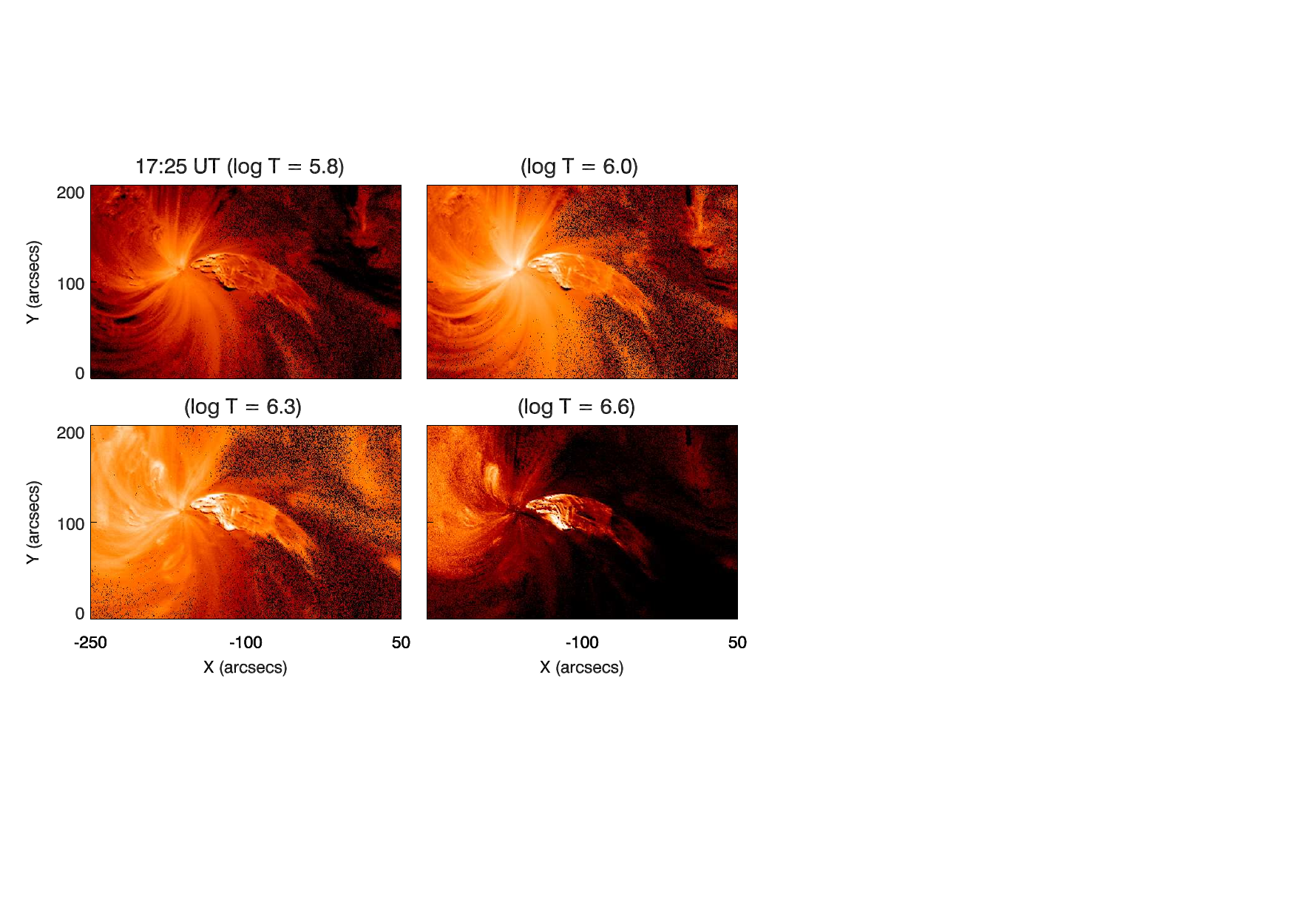}
\caption{The DEM maps of the jet at different temperatures for the jet observed on 2010 August 2. \label{Fig13}}
\end{center}
\end{figure}

\mbox{Figure \ref{Fig13}} shows the DEM maps at different temperatures (at \mbox{log \textit{T} [K] = 5.8, 6.0, 6.3 and 6.6}) at \mbox{17:25 UT}. We investigated the temperature variation of coronal plasma at different locations. We observed that the spire and the footpoint during the jet evolution was at different temperatures and the results indicate that the spire has much more plasma at cooler temperatures than the footpoint. The temperature structure estimated for the small region from the DEM analysis \mbox{(Fig. \ref{Fig12})} shows consistent results with the DEM maps \mbox{(Fig. \ref{Fig13})} at different temperatures.

During the jet-phase, we found an emission measure of \mbox{log EM = 28.6} and a density of \mbox{N$_{e}$ = 1.3$\times$10$^{10}$ cm$^{-3}$} at the location of the spire, and an emission measure of \mbox{log EM = 28.6} and a density of \mbox{N$_{e}$ = 1.1$\times$10$^{10}$ cm$^{-3}$} at the location of the footpoint of the jet. During the pre-jet phase, we found an emission measure of \mbox{log EM = 26.9} and a density of \mbox{N$_{e}$ = 1.8$\times$10$^{9}$ cm$^{-3}$} at the location of the spire, and an emission measure of \mbox{log EM = 27.2} and a density of \mbox{N$_{e}$ = 2.4$\times$10$^{9}$ cm$^{-3}$} at the location of the footpoint of the jet. The results show that, the emission measure was increased by over one order of magnitude and the electron density was increased by factor of ten during the jet evolution at the location of the spire and also at the location of the footpoint of the jet.

\subsection{2013 March 02}

\begin{figure}[!hbtp]
\begin{center}
\includegraphics[trim=0.1cm 1.5cm 3.5cm 1.2cm,width=0.5\textwidth]{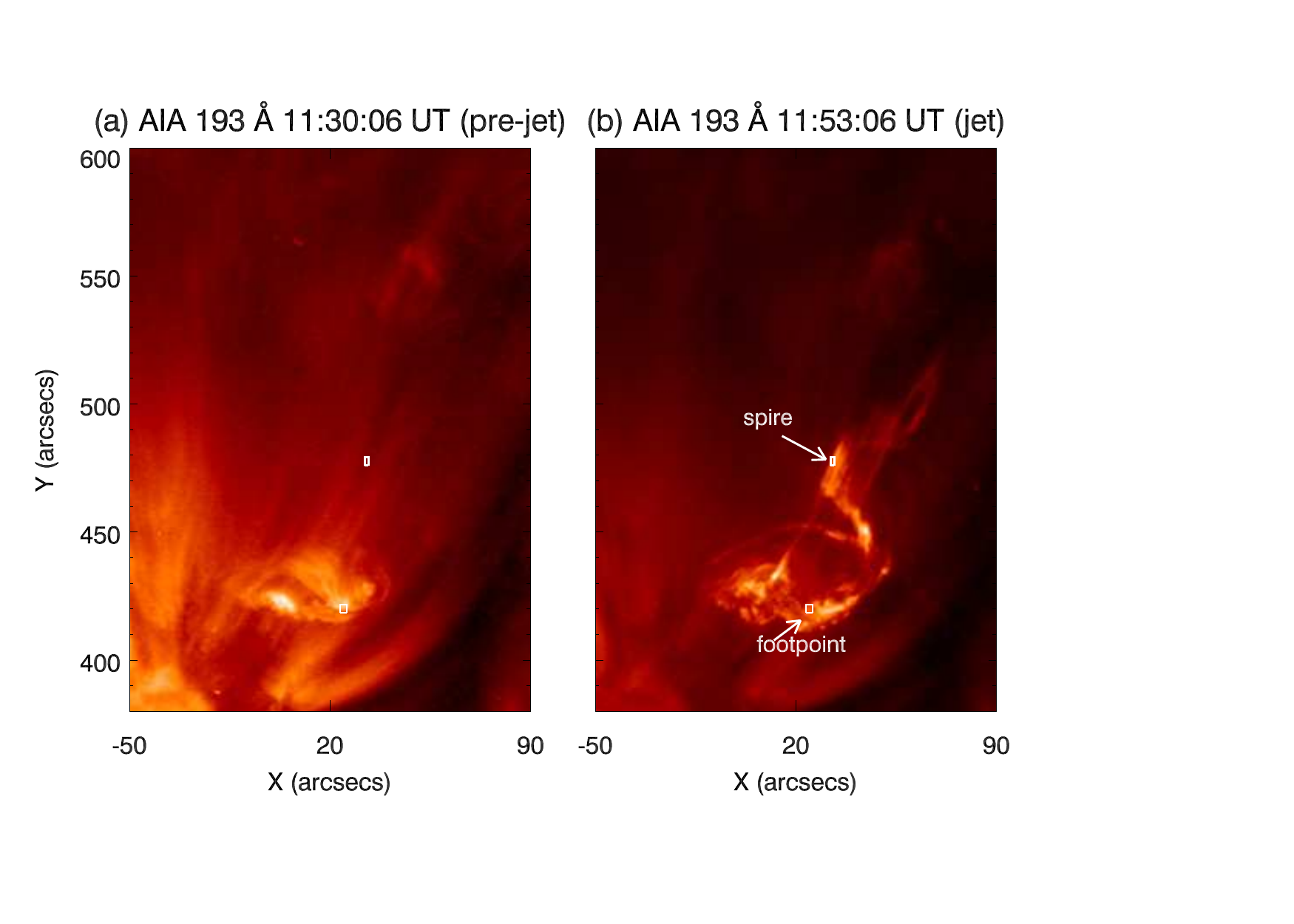}
\includegraphics[trim=1.5cm 0cm 0.1cm 0cm,width=0.5\textwidth]{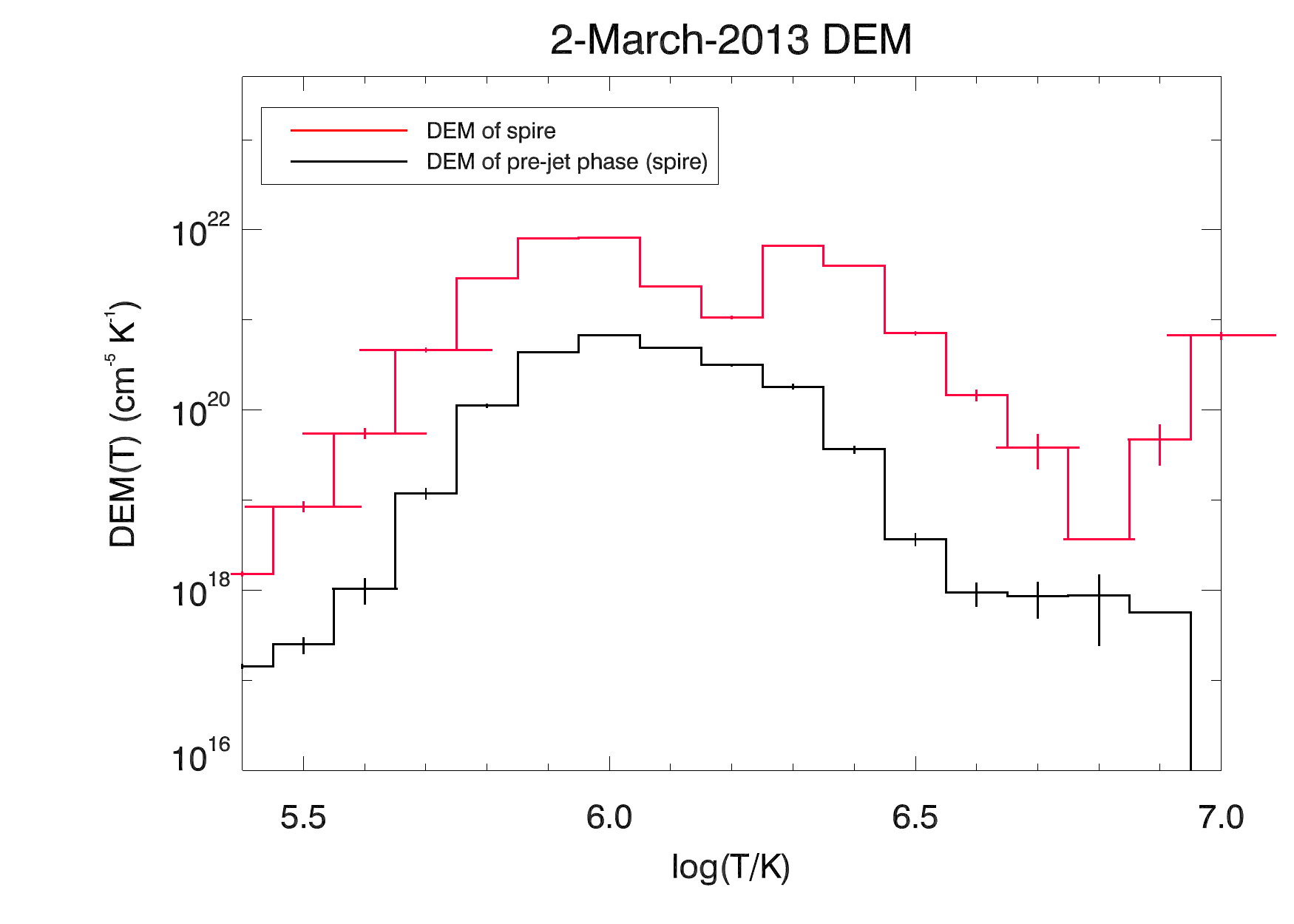}
\includegraphics[trim=1.5cm 0cm 0.1cm 0.1cm,width=0.5\textwidth]{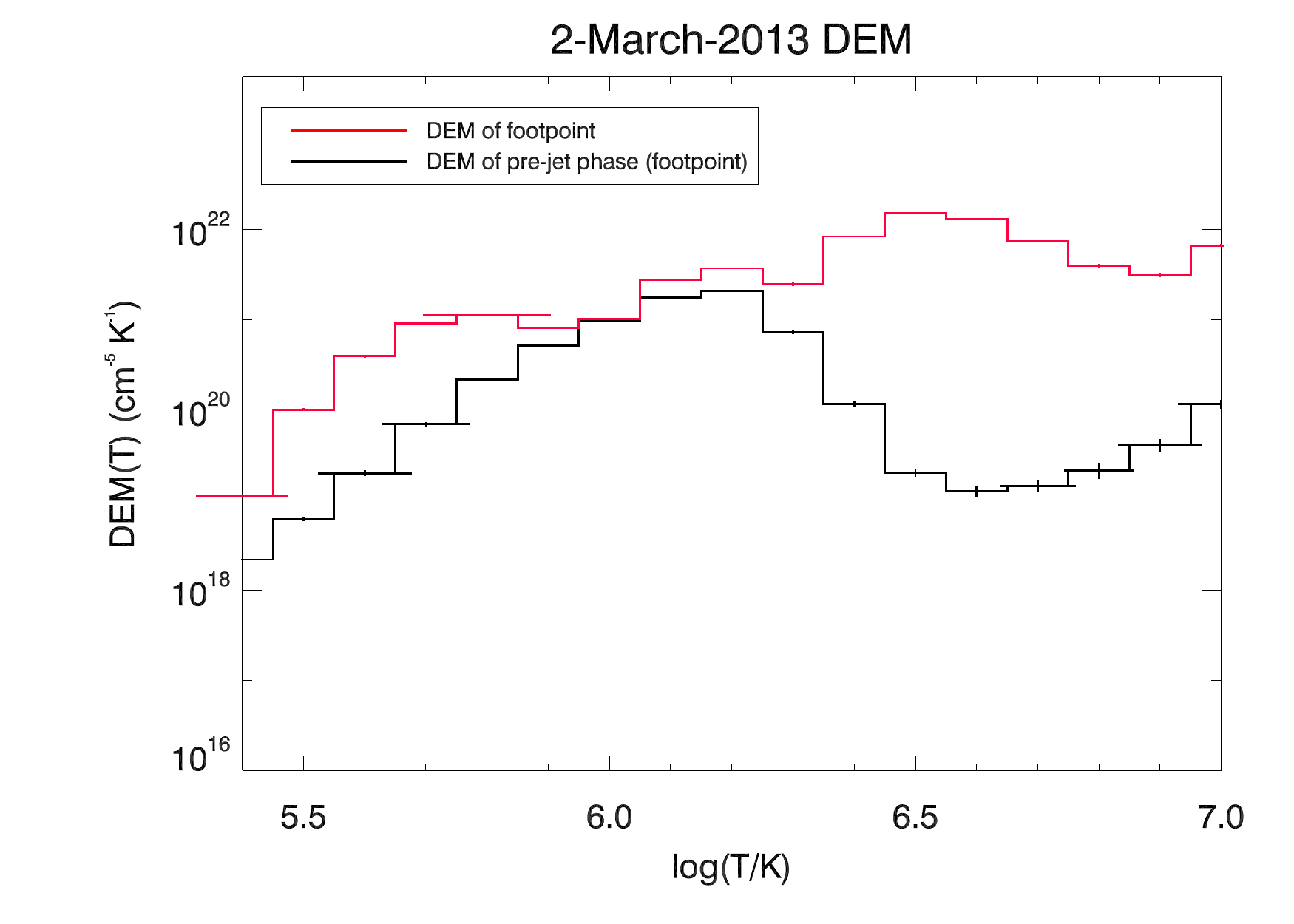}
\caption{Top panel : AIA 193~{\AA} image of (a) pre-jet phase and (b) jet. The two regions in the small boxes are used for the DEM analysis. Middle panel : shows the DEM curves for the spire of the jet (red curve) and at the same region during the pre-jet phase (black curve). Bottom panel : shows the DEM curves for the footpoint of the jet (red curve) and at the same region during the pre-jet phase (black curve).\label{Fig14}}
\end{center}
\end{figure}

We performed a similar DEM analysis for the jet observed on \mbox{2013 March 02}. \mbox{Figure \ref{Fig14}} (top panels (a) \& (b)) shows the AIA 171~{\AA} images of the pre-jet phase at \mbox{11:30 UT} and jet phase at \mbox{11:53 UT} respectively. We used two small regions, one at the footpoint of jet and the other at the spire of the jet for LOS DEM calculation. The regions are over plotted as small black boxes in \mbox{Fig. \ref{Fig14}} (top panel). The plots (middle panel) show the DEM curves for the spire of the jet (red curve) and at the same region during the pre-jet phase (black curve). This indicates that the temperature at the region of spire is increased from \mbox{\textit{T} [K] = 6.0 to \textit{T} [K] = 6.3} during the jet evolution. The plots (bottom panel) show the DEM curves for the footpoint of the jet (red curve) and at the same region during the pre-jet phase (black curve). It is clear that the temperature at the footpoint is increased from \mbox{log \textit{T} [K] = 6.2 to log \textit{T} [K] = 6.5}. 

\begin{figure}[!hbtp]
\begin{center}
\includegraphics[trim=0.5cm 0cm 7.5cm 5.7cm,width=0.85\textwidth]{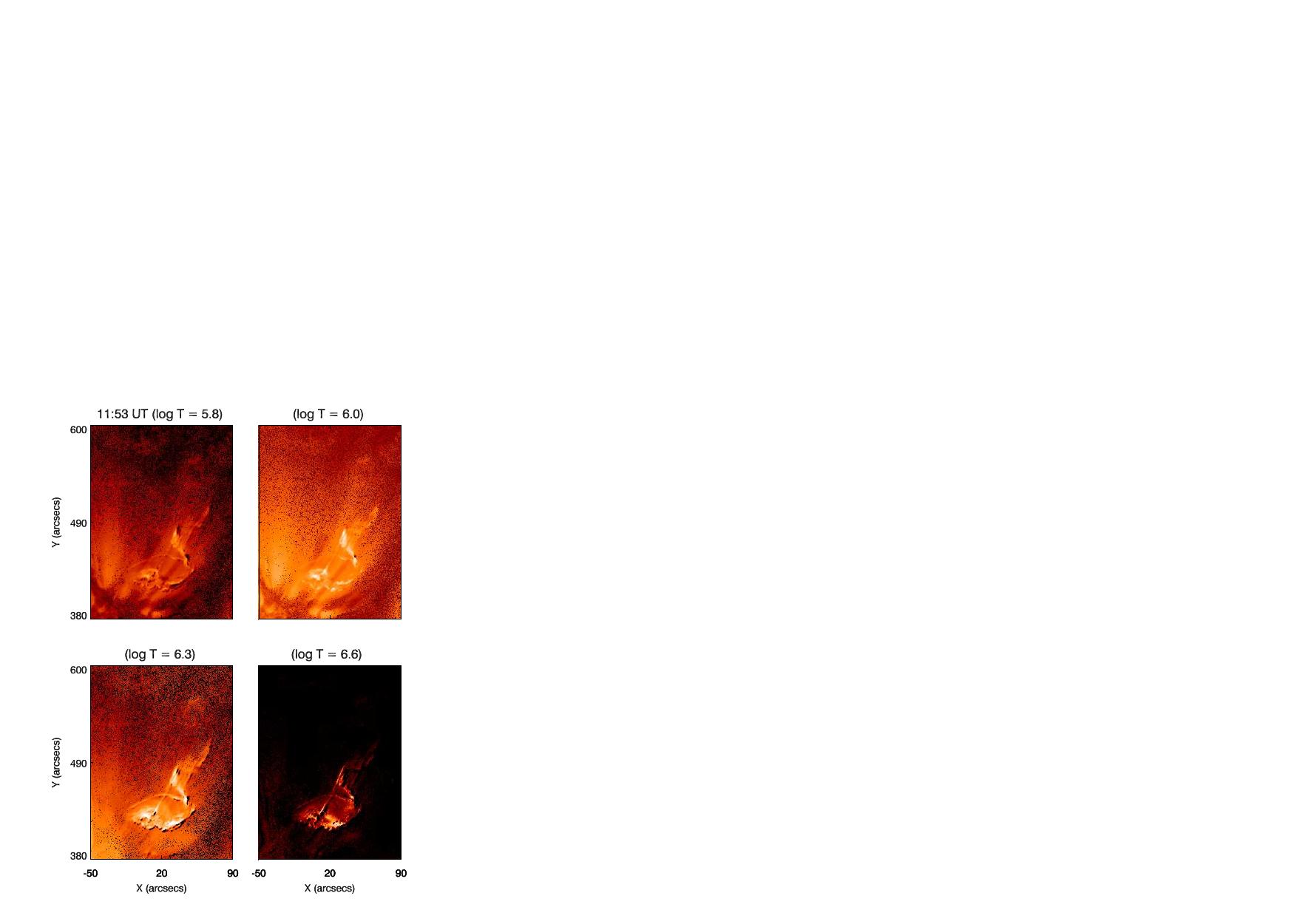}
\caption{The DEM maps of the jet at different temperatures for the jet observed on 2013 March 2.\label{Fig15}}
\end{center}
\end{figure}

\mbox{Figure \ref{Fig15}} shows the DEM maps at different temperatures (at \mbox{log \textit{T} [K] = 5.8, 6.0, 6.3 and 6.6}) at \mbox{11:53 UT.} The temperature structure estimated for the small region from the DEM analysis \mbox{(Fig. \ref{Fig14})} shows consistent results with DEM maps \mbox{(Fig. \ref{Fig15})} at different temperatures.

During the jet-phase, we found an emission measure of \mbox{log EM = 28.0} and a density of \mbox{N$_{e}$ = 8.6$\times$10$^{9}$ cm$^{-3}$} at the location of the spire, and an emission measure of \mbox{log EM = 28.1} and a density of \mbox{N$_{e}$ = 8.4$\times$10$^{9}$ cm$^{-3}$} at the location of the footpoint of the jet. During the pre-jet phase, we found an emission measure of \mbox{log EM = 26.8} and a density of \mbox{N$_{e}$ = 2.1$\times$10$^{9}$ cm$^{-3}$} at the location of the spire, and an emission measure of \mbox{log EM = 27.3} and a density of \mbox{N$_{e}$ = 3.2$\times$10$^{9}$ cm$^{-3}$} at the location of the footpoint of the jet. The results show that, the emission measure was increased by over one order of magnitude and the electron density was increased by factor of two during the jet evolution at the location of the spire. The emission measure was increased by a factor of ten and the electron density was increased by factor of three at the location of the footpoint of the jet.

\section{Discussion and Summary}

In this paper, we present a comprehensive study of \mbox{multiwavelength} observations of 20 active region jets observed during \mbox{August 2010} - \mbox{June 2013}. We have used the observations recorded by SDO/AIA, SDO/HMI, RHESSI, WIND and GONG network. We have investigated the relationship of jets with H$\alpha$ surges, nonthermal type-III radio burst, HXR emission and photospheric magnetic field configuration in the source region of all these jets. In addition, we have measured the general physical properties of the jets such as lifetime and velocity. We have discussed two jet events in detail in section 3 and 4 and the properties of other jets are discussed in the appendix. Using the various filters of AIA, we have studied the temperature structure of these jets in their spires as well as in their footpoints and have provided a lower limit on the electron densities. To the best of our knowledge, this is first comprehensive study of active region jets and their thermodynamic nature. Below we summarise the most important results.

\begin{figure}[!hbtp]
\begin{center}
\includegraphics[trim=0.5cm 0.1cm 1.2cm 7cm,width=0.5\textwidth]{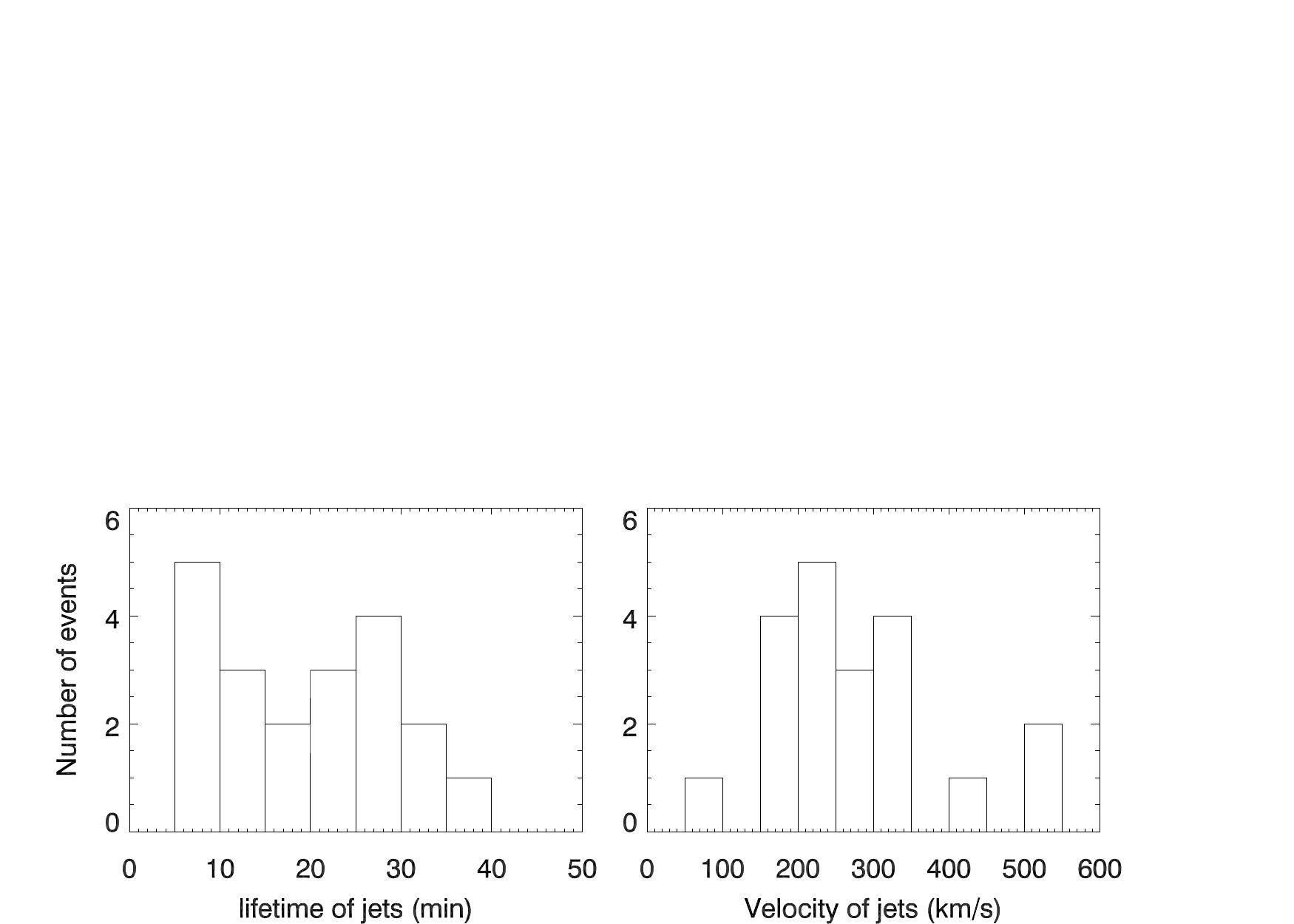}
\caption{Left panel : Distribution of the lifetime of jets observed in the AIA 193~{\AA} channel. Right panel : Distribution of \textit{plane-of-sky} velocities of jets calculated from the time-distance plots.\label{Fig16}}
\end{center}
\end{figure}

Most of the jets originated from the western periphery of active regions close to sunspots. The lifetime (as noted from the AIA 193~{\AA} channel) ranged from \mbox{5 to 39 minutes} with an average of 18 minutes. 25$\%$ of the jets have their lifetime from \mbox{5 to 10 minutes} and 20$\%$ of them have from \mbox{25 to 30 minutes} \mbox{(See Fig.\ref{Fig16})}. The velocities range from \mbox{ 87 to 532 km/s} with an average of \mbox{271 km/s}. 25$\%$ of the jets have velocities from \mbox{200 to 250 km/s} and other 20$\%$ have velocities from \mbox{300 to 350 km/s} \mbox{(See Fig. \ref{Fig16})}. These results are consistent with and are within the range of values obtained by \citetads{1996PASJ...48..123S} from their statistical study of 68 AR jets observed by SXT onboard Yohkoh. We investigated the cool-temperature component of the jets using H$\alpha$ data from ground-based observatories. All of them were found to be associated with H$\alpha$ surges at chromospheric temperatures.

The signature of an outward beam of mildly relativistic electrons from the jet source region to the outer corona and interplanetary space appears as a nonthermal type-III radio burst in the radio dynamic spectrum. To understand the relationship between nonthermal type-III radio bursts and jets, we compared the jet timings with bursts timings using the dynamic radio spectrum obtained from WAVES/WIND. 85$\%$ of the nonthermal type-III radio bursts are found be co-temporal with jets. We also confirm the position of these bursts by performing the PFSS analysis. This analysis showed the presence of open magnetic field lines in the source region which indicates that jets were ejected in the direction of these open magnetic field structure. These observations show the signature of particle acceleration. The RHESSI observations showed the presence of HXR emission during the rising phase of jets. 55$\%$ of the jets are found to be producing HXR emissions. 10 out of 20 events showed that the jets originated in regions of flux cancellation and 6 in regions of flux emergence. 4 events showed a flux emergence and then cancellation during the jet evolution. 

We have performed a detailed analysis of the temperature distribution for the jets by using the computationally fast technique of regularized inversion developed by \citetads{2012A&A...539A.146H}. We found that for most of the spires of the jets, the DEM peaked at around  \mbox{log \textit{T} [K] = 6.2/6.3 ($\sim$2 MK)} (details in Table \ref{table1}). The detailed analysis of two events showed that the temperarture at the spire of the jet 1 and 2 is increased from \mbox{log \textit{T} [K] = 6.0 to log \textit{T} [K] = 6.3}, and the temperarture at the region of footpoint is increased from \mbox{log \textit{T} [K] = 6.3 to log \textit{T} [K] = 6.5} in the case of \mbox{jet 1} and from \mbox{log \textit{T} [K] = 6.2 to log \textit{T} [K] = 6.5} in the case of \mbox{jet 2}. The DEM values also showed consistency with DEM maps (\mbox{figure \ref{Fig13} and \ref{Fig15}}) at different temperartures. These results show that the spires have plasma at cooler temperartures than at the footpoints. The observations of flux cancellation, the association with HXR emission and emission of nonthermal type-III radio burst, suggest that the initiation and therefore heating is taking place at the base of the jets and this is in agreement with the jet model proposed by \citetads{2001ApJ...550.1051S}.  

By assuming a filling factor equal to unity along with coronal abundances \citepads{1992PhyS...46..202F}, we calculated the emission measure and a lower limit of the electron density. We observed that, there is an increase in the emission measure and electron density values during the jet evolution. In the case of \mbox{jet 1}, the emission measure was increased by over one order of magnitude and the electron density was increased by factor of ten during the jet evolution at the location of the spire and also at the location of the footpoint of the jet. In the case of \mbox{jet 2}, the emission measure was increased by over one order of magnitude and the electron density was increased by factor of two during the jet evolution at the location of the spire. The emission measure was increased by a factor of ten and the electron density was increased by factor of three at the location of the footpoint of the jet.

These results are consistent with the results obtained by \citetads{2007PASJ...59S.751C} and \citetads{2013ApJ...763...24K} from their study of individual AR jets. Although, they did not distinguish between the spire and footpoint region of the jet. The electron temperatures and DEMs derived in this study are also in agreement with the results obtained from the Fe XII line ratio diagnostics of a spectroscopic study of the recurrent AR jet (\mbox{Mulay et al.} in preparation).

The results described in this paper provide important and substantial constraints on the thermodynamics of the jet plasma, that will be helpful to the theoretical modelling of jets.

\begin{acknowledgements}
We thank the referee for constructive and detailed comments which helped to improve this paper. The part of the work was done when one of the authors (SMM) was a Junior Research Fellow at Inter-University Centre of Astronomy and Astrophysics (IUCAA), India. SMM and DT acknowledge support from DST under the Fast Track Scheme (SERB/F/3369/2012/2013). SMM also ackowledges support from the Cambridge Trust, University of Cambridge, UK. HEM and GDZ acknowledge the support of STFC. AIA data are courtesy of SDO (NASA) and the AIA consortium. RHESSI work is supported by NASA contract NAS 5-98033. The authors thank the open data policy of WIND/WAVES instrument team. This work utilizes data obtained by the Global Oscillation Network Group (GONG) Program, managed by the National Solar Observatory, which is operated by AURA, Inc. under a cooperative agreement with the National Science Foundation. The data were acquired by instruments operated by the Big Bear Solar Observatory, High Altitude Observatory, Learmonth Solar Observatory, Udaipur Solar Observatory, Instituto de Astrofísica de Canarias, and Cerro Tololo Interamerican Observatory.

\end{acknowledgements}


\bibliographystyle{aa-note} 
 \bibliography{jet_v1}    


\begin{table*}[!btp]
\renewcommand\thetable{1} 
\centering
\caption{\label{tab2} Statistically measured physical parameters of 20 EUV active region jets}
\resizebox{19cm}{!} {
\begin{tabular}{lcccccccccccc}
		
\hline
\hline

{Date}	     	&{Time (UT)} 	&{Lifetime}	&{Active}	  &{Sunspot}	  	&{GOES X-ray Flare}	&{Type III}  	&{RHESSI}   	&{Jet Velocity} &{H$\alpha$}	&{HMI}  &{DEM Peak Temperature}	\\

		&{in 193~{\AA}}	&{(in min)}	&{Region}  	  &{Configuration}	&{Start Time (UT)}	&{(UT)}	     	&{(keV)}	&{(km/s)}       &		&    	&{log \textit{T} [K]}		\\
		
\hline
\hline \space
  
2010 Aug. 02	&17:10 - 17:35  &25		&11092 (N13 E07)  &$\alpha$		&B 1.8 (17:21)  	&17:25		&R1, R2      	&236	      &a	&FE	&6.3			\\

2010 Sep. 17    &00:15 - 00:31	&16		&11106 (S20 W09)  &$\beta$      		&B 5.7 (00:14)  	&00:16		&---	       	&194	      &b	&FEC 	&6.0			\\ 

2011 Feb. 14    &12:51 - 13:02	&11		&11158 (S20 W17)  &$\beta\gamma$ 		&---		  	&12:51		&---		&208	      &d	&FC 	&6.0			\\ 

2011 Feb. 16    &13:33 - 13:42	&09		&11158 (S21 W41)  &$\beta\gamma\delta$ 	&---	      		&13:33		&R1, R2, R3     &207	      &d	&FEC 	&6.0			\\ 
		
2011 Mar. 01    &12:53 - 13:18	&25		&11165 (S21 W00)  &$\beta$             	&---        		&12:53		&---     	&87	      &d	&FC 	&6.3			\\ 

2011 Mar. 07    &21:33 - 22:12	&39		&11166 (N11 E13)  &$\beta\gamma$ 		&---	 		&21:48		&---      	&203	      &d	&FE 	&6.3			\\ 

2011 Oct. 17    &19:48 - 20:04  &16		&11314 (N28 W32)  &$\beta\beta$          &---		 	&19:49		&---		&520	      &b	&FC     &6.3                    \\

2011 Dec. 11    &03:17 - 03:26	&09		&11374 (S17 E27)  &$\alpha$ 		&B 8.6 (03:20)		&03:17		&R1, R2      	&165	      &a 	&FE 	&6.2			\\ 
	
2011 Dec. 11    &12:15 - 12:49  &34		&11374 (S17 E27)  &$\alpha$ 		&---	 		&12:15		&R1, R2		&278	      &d	&FC 	&6.3			\\ 
	
2011 Dec. 11    &23:14 - 23:34  &20		&11374 (S17 E27)  &$\alpha$ 		&---	 		&23:14		&R1, R2      	&165	      &a	&FC 	&6.3			\\
		
2012 Mar. 05    &21:51 - 22:00 	&09		&11429 (N18 E41)  &$\beta\gamma\delta$ 	&---			&21:51		&---     	&532	      &b	&FC     &6.0			\\ 
	
2012 Oct. 10	&14:21 - 14:45	&24		&11585 (S20 W43)  &$\beta$       		&C 2.1 (14:25) 		&14:31		&R2   		&259	      &d 	&FC 	&6.2			\\ 

2013 Mar. 02	&11:49 - 11:57	&08		&11681 (N17 W08)  &$\alpha\gamma$       	&B 6.0 (11:48) 		&11:51		&---   		&316	      &d	&FEC 	&6.0			\\ 

2013 Mar. 02	&12:04 - 12:25	&21		&11681 (N17 W08)  &$\alpha\gamma$      	&B 6.0 (11:48) 		&12:04		&---   		&210	      &d	&FE 	&6.3			\\ 
	
2013 Apr. 28	&20:59 - 21:11  &12		&11731 (N09 E23)  &$\beta\gamma$ 		&C 1.5 (20:59)	 	&---		&---      	&320	      &b	&FEC 	&6.2			\\ 

2013 Apr. 28	&22:51 - 22:59  &08		&11731 (N09 E23)  &$\beta\gamma$ 		&---			&---		&---   		&168	      &b	&FE 	&6.2			\\ 

2013 May 04	&23:15 - 23:49 	&34		&11734 (S19 W04)  &$\beta\gamma\delta$    &C 1.3 (23:25) 		&23:15		&R1, R2, R3, R4	&283	      &b	&FC 	&6.3			\\ 
		
2013 May 25	&08:42 - 08:55	&13		&11748 (N12 W83)  &$\beta\gamma\delta$ 	&---	 		&---	       	&R1, R2		&322	      &d	&FE 	&6.2			\\ 
	
2013 Jun. 17	&08:41 - 09:06	&25		&11770 (S13 E13)  &$\alpha$ 		&--- 			&08:52		&R1, R2, R3     &409	      &c	&FC 	&6.2			\\
		
2013 Jun. 18   	&15:13 - 15:39	&26		&11770 (S14 E02)  &$\alpha$		&---			&15:13		&R1, R2, R3, R4 &338	      &b	&FC 	&6.2			\\ 
		
\hline
\hline

\end{tabular}
}
\tablefoot{ \textbf{a} - H$\alpha$ Surge visible in Solar Magnetic Activity Research Telescope (SMART) Hida Observatory, \textbf{b} - H$\alpha$ Surge visible in Big Bear Solar Observatory (BBSO), \textbf{c} - H$\alpha$ Surge visible in Kanzelh\"{o}he Observatory, \textbf{d} - H$\alpha$ Surge visible in Global Oscillation Network Group (GONG) data, \textbf{R1} - RHESSI 3-6 keV, \textbf{R2} - RHESSI 6-12 keV, \textbf{R3} - RHESSI 12-25 keV, \textbf{R4} - RHESSI 25-50 keV, \textbf{FE} - Flux Emergence, \textbf{FEC} - Flux Emergence and then Cancellation, \textbf{FC} - Flux Cancellation. } \label{table1}
\end{table*}

 \newpage
\begin{appendix}

\section{Information about all jets}
In this section, we discuss the information about all AR jet events used in this multiwavelength analysis. Each of the figures for the jets show three panels that display the \mbox{AIA 171~{\AA}} image of a jet on the left, and in the middle and right panels, two HMI magnetograms during the jet evolution. A region in the small box overplotted on AIA image is used for the DEM analysis.

\begin{figure}[!hbtp]
\begin{center}
\includegraphics[trim=1cm 0.1cm 2.6cm 7.5cm,width=0.55\textwidth]{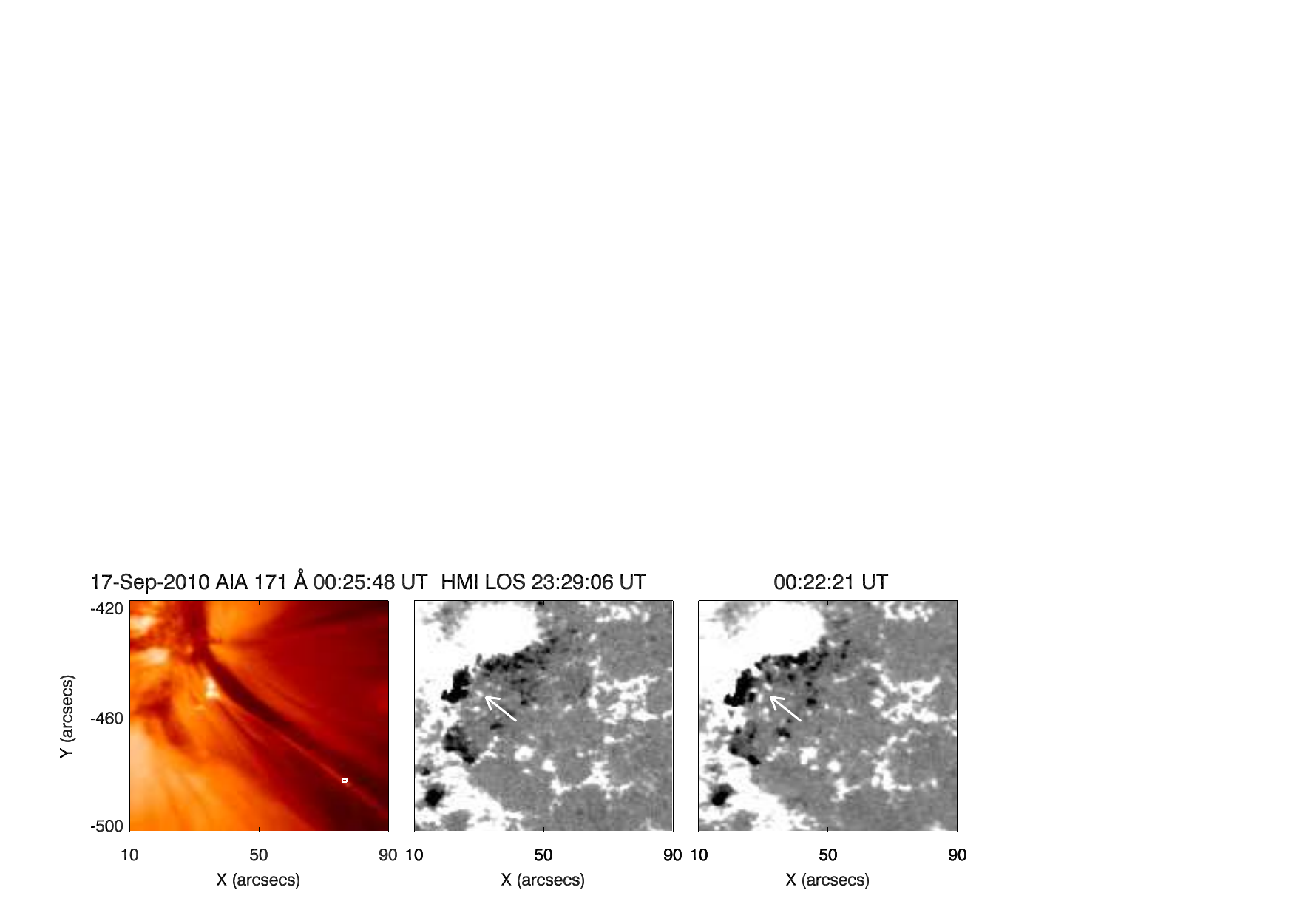}
\caption{Jet 2 : 17 September 2010 - Left panel : AIA 171~{\AA} image of the jet observed on 17 September 2010. The jet appeared to evolve from the edge of the active region 11106 (S20 W09) which was associated with a positive-polarity sunspot with anemone magnetic topology. A small-scale brightening was observed at the jet footpoint before the jet evolution, and the jet evolved as a simple spire. The jet started its activity at 00:15 UT and lasted until 00:31 UT. Middle panel and right panel : The LOS HMI magnetogram image at 23:29 UT before the jet evolution and at 00:22 UT during the jet respectively. The emergence of positive-polarity near the negative-polarity region is shown by white arrows. Further with time, the small negative flux region cancelled with emerging positive-polarity. This magnetic activity lasted for a few hours and it was co-temporal and co-spatial with the jet activity.\label{fig17}}
\end{center}
\end{figure}

\begin{figure}[!hbtp]
\begin{center}
\includegraphics[trim=1cm 0.1cm 2.6cm 8cm,width=0.55\textwidth]{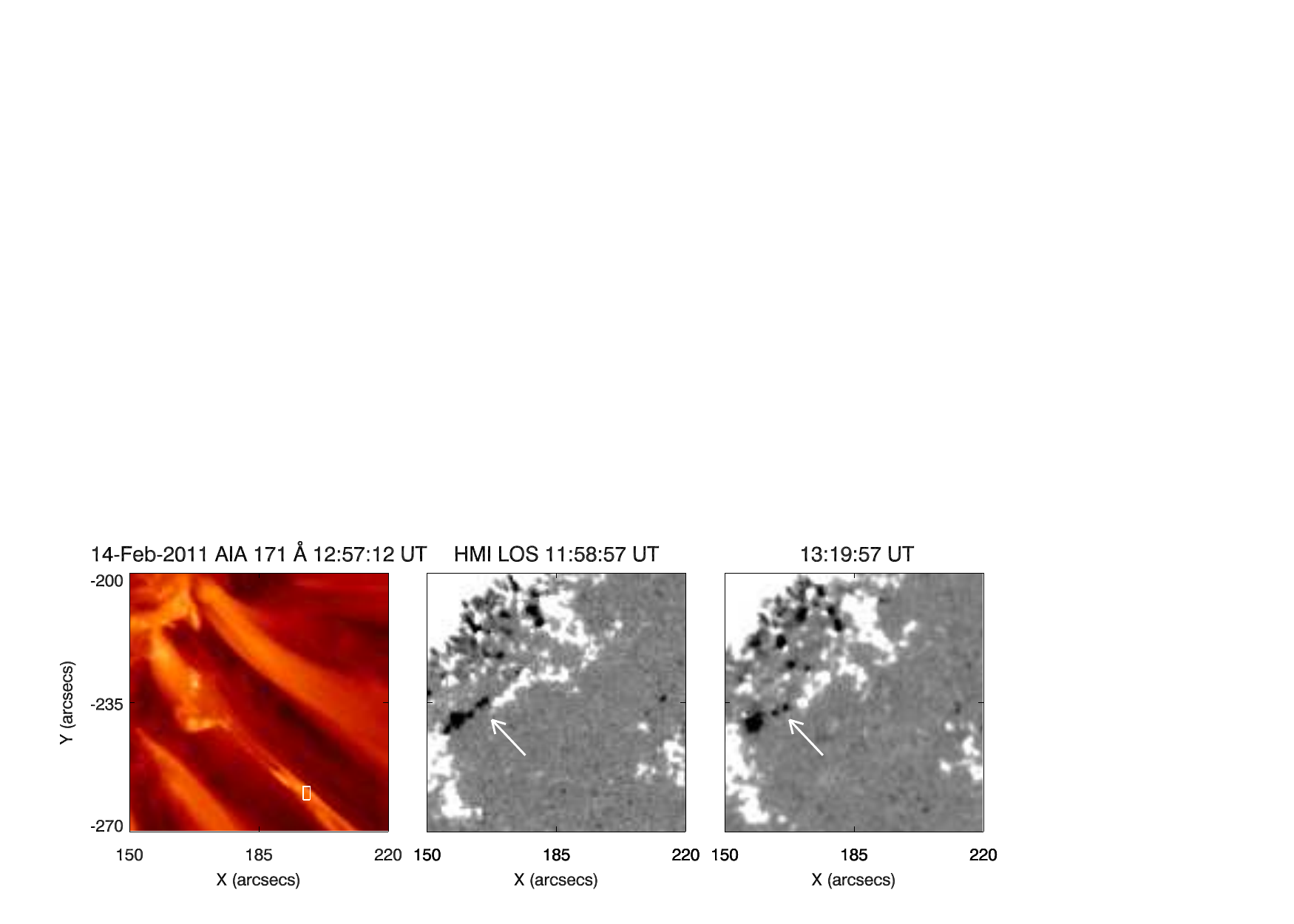}
\caption{Jet 3 : 14 February 2011 - Left panel : AIA 171~{\AA} image of the jet observed on 14 February 2011. The jet appeared to evolve from the edge of the active region 11158 (S20 W17) which was associated with a positive-polarity sunspot. A small-scale brightening was observed at 12:58 UT at the jet footpoint before the jet evolution, and the jet evolved as a simple spire. The jet started its activity at 12:58 UT and lasted until 13:02 UT. Middle panel and right panel : The LOS HMI magnetogram image at 11:58 UT before the jet evolution and at 13:19 UT after the jet evolution respectively. Negative flux cancellation was observed near the positive-polarity region shown by white arrows. This magnetic activity lasted for a few hours and it was co-temporal and co-spatial with the jet activity.\label{fig18}}
\end{center}
\end{figure}

\begin{figure}[!hbtp]
\begin{center}
\includegraphics[trim=1cm 0.1cm 2.6cm 7.5cm,width=0.55\textwidth]{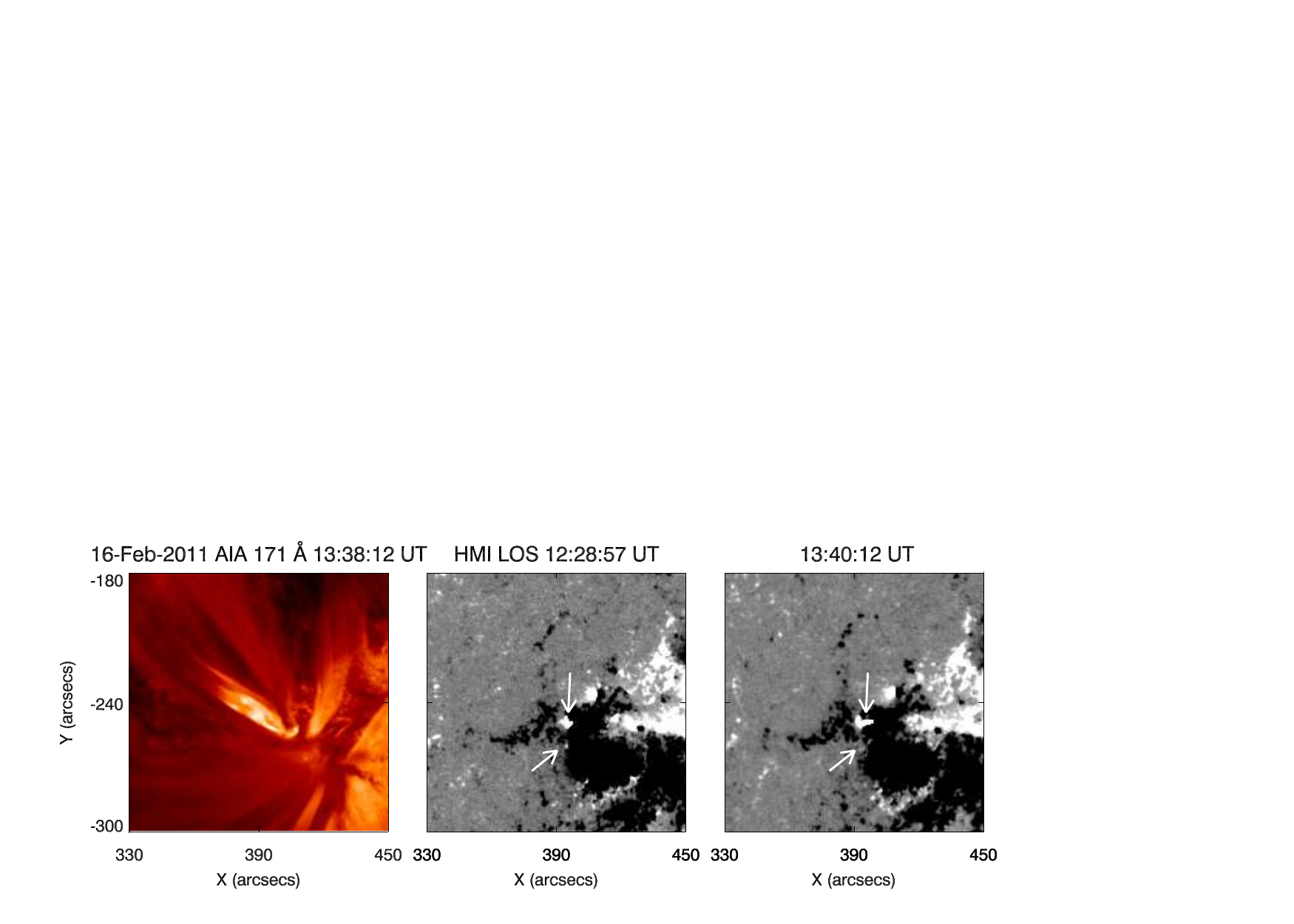}
\caption{Jet 4 : 16 February 2011 - Left panel : AIA 171~{\AA} image of the jet observed on 16 February 2011. The recurrent jet appeared to evolve from the eastern edge of the active region 11158 (S20 W41) which was associated with a negative-polarity sunspot. A small-scale brightening was observed at 13:32 UT at the jet footpoint before the jet evolution, and the jet evolved as complex, multi-threaded spire. The observations showed the continuous untwisting motion of the jet spire with radial motion. The jet started its activity at 13:24 UT and lasted until 13:42 UT. Middle panel and right panel : The LOS HMI magnetogram image at 12:28 UT before the jet evolution and at 13:40 UT during the jet evolution respectively. The positive flux emergence and cancellation was observed near the negative-polarity sunspot region shown by white arrows. This magnetic activity lasted for a few hours and it was co-temporal and co-spatial with the jet activity.\label{fig19}}
\end{center}
\end{figure}

\begin{figure}[!hbtp]
\begin{center}
\includegraphics[trim=1cm 0.1cm 2.6cm 7.5cm,width=0.55\textwidth]{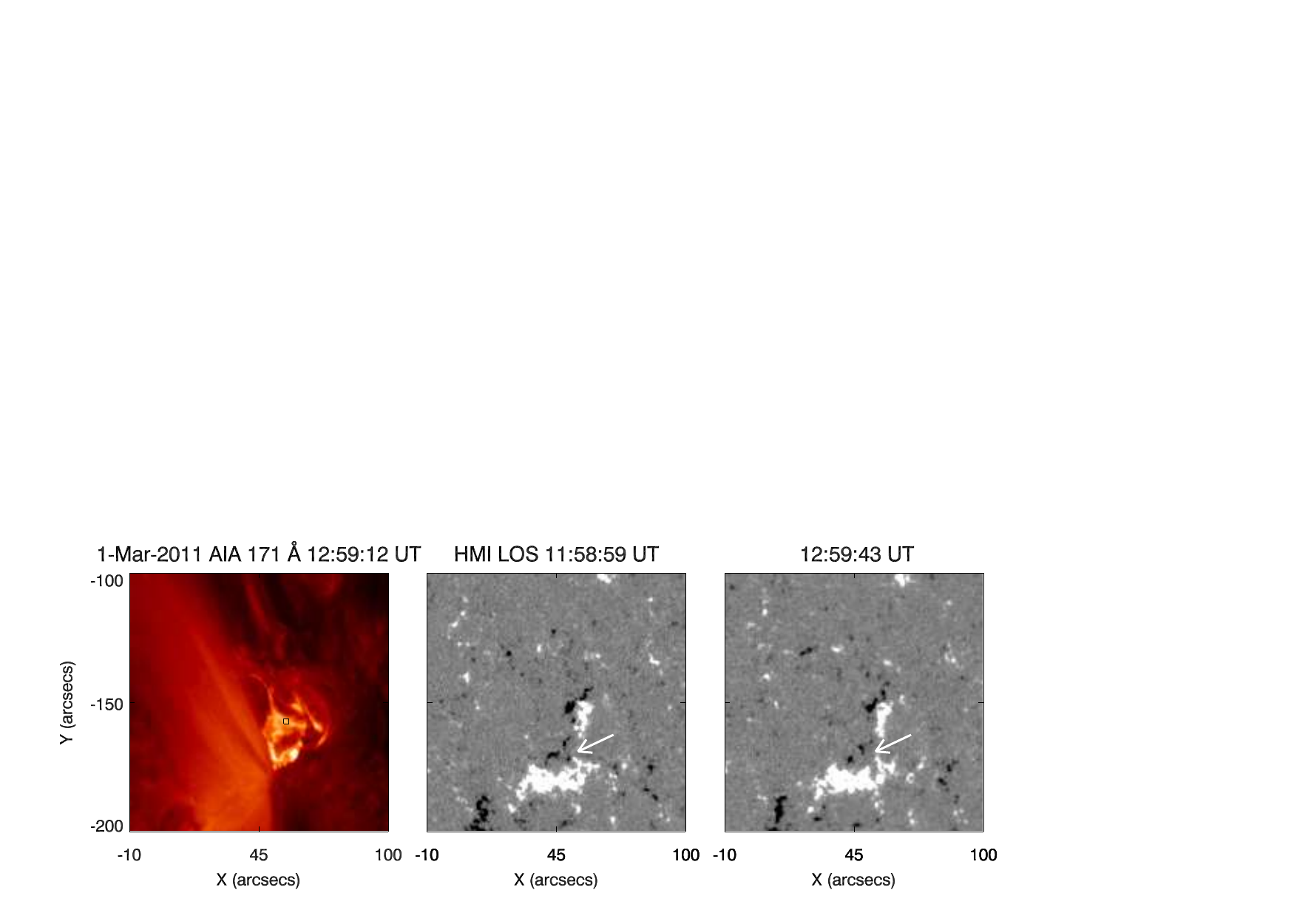}
\caption{Jet 5 : 1 March 2011 - Left panel : AIA 171~{\AA} image of the jet observed on 1 March 2011. The jet appeared to evolve from the edge of the active region 11165 (S21 W00). A small-scale brightening was observed at 12:53 UT at the jet footpoint before the jet evolution, and the jet evolved as complex, multi-threaded spire. The jet started its activity at 12:53 UT and lasted until 13:18 UT. Middle panel and right panel : The LOS HMI magnetogram image at 11:58 UT before the jet evolution and at 12:59 UT during the jet evolution respectively. The negative flux cancellation with nearby positive-polarity region was observed and it is shown by white arrows. This magnetic activity lasted for a few hours and it was co-temporal and co-spatial with the jet activity.\label{fig20}}
\end{center}
\end{figure}

 \newpage
\begin{figure}[!hbtp]
\begin{center}
\includegraphics[trim=1cm 0.1cm 2.6cm 7.5cm,width=0.55\textwidth]{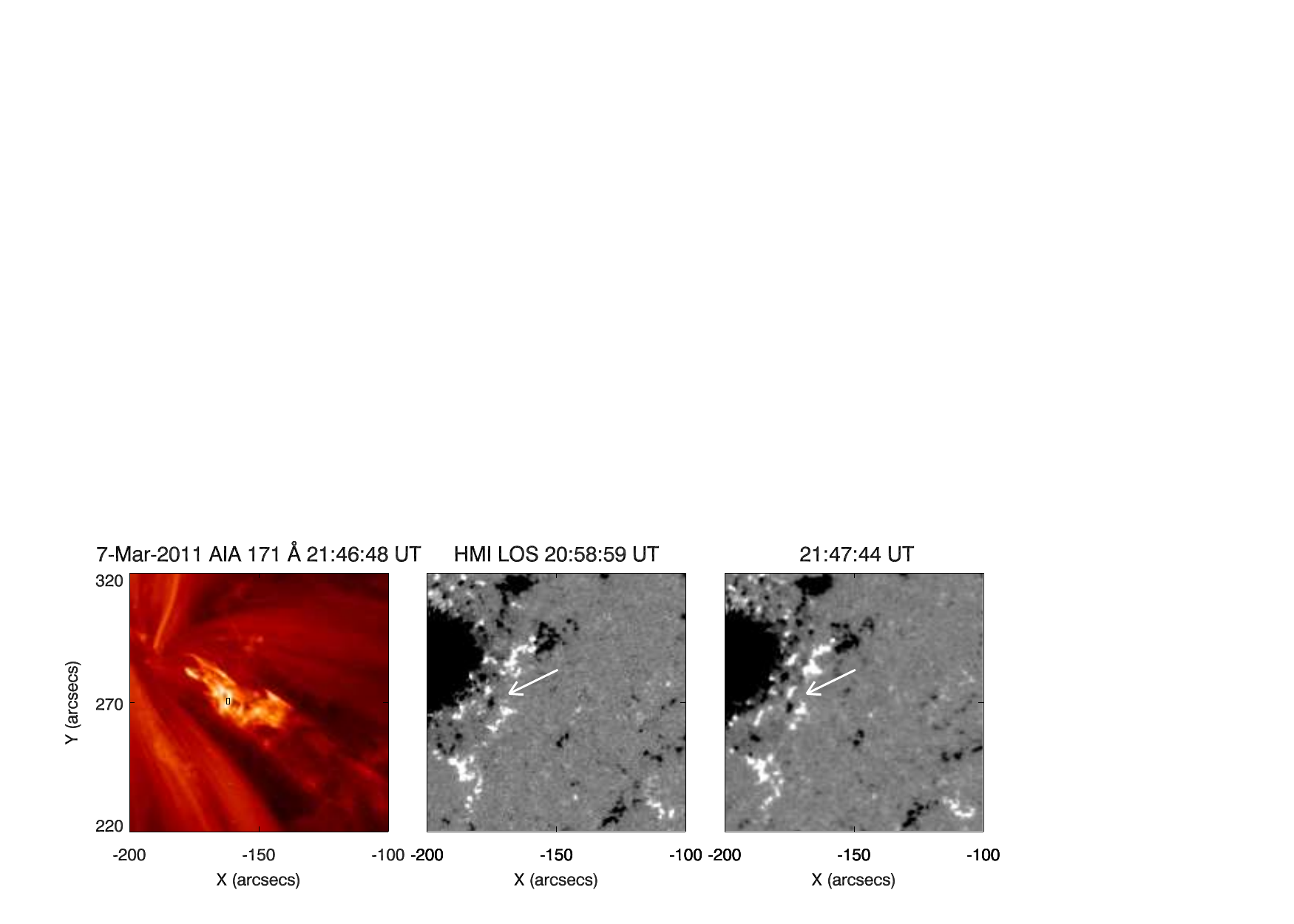}
\caption{Jet 6 : 7 March 2011 - Left panel : AIA 171~{\AA} image of the jet observed on 7 March 2011. The jet appeared to evolve from the edge of the active region 11166 (N11 E13) which was associated with a negative-polarity sunspot with anemone magnetic topology. A small-scale brightening was observed at 21:33 UT at the jet footpoint before the jet evolution, and the jet evolved as complex, multi-threaded spire. The observations showed the continuous untwisting motion of jet spire with radial motion. The jet started its activity at 21:33 UT and lasted until 22:12 UT. Middle panel and Right panel : The LOS HMI magnetogram image at 20:58 UT before the jet evolution and at 21:47 UT during the jet evolution respectively. The positive flux emergence at the edge of the negative-polarity region was observed and it is shown by white arrows. \label{fig21}}
\end{center}
\end{figure}

\begin{figure}[!hbtp]
\begin{center}
\includegraphics[trim=0.5cm 0.1cm 2.5cm 8cm,width=0.55\textwidth]{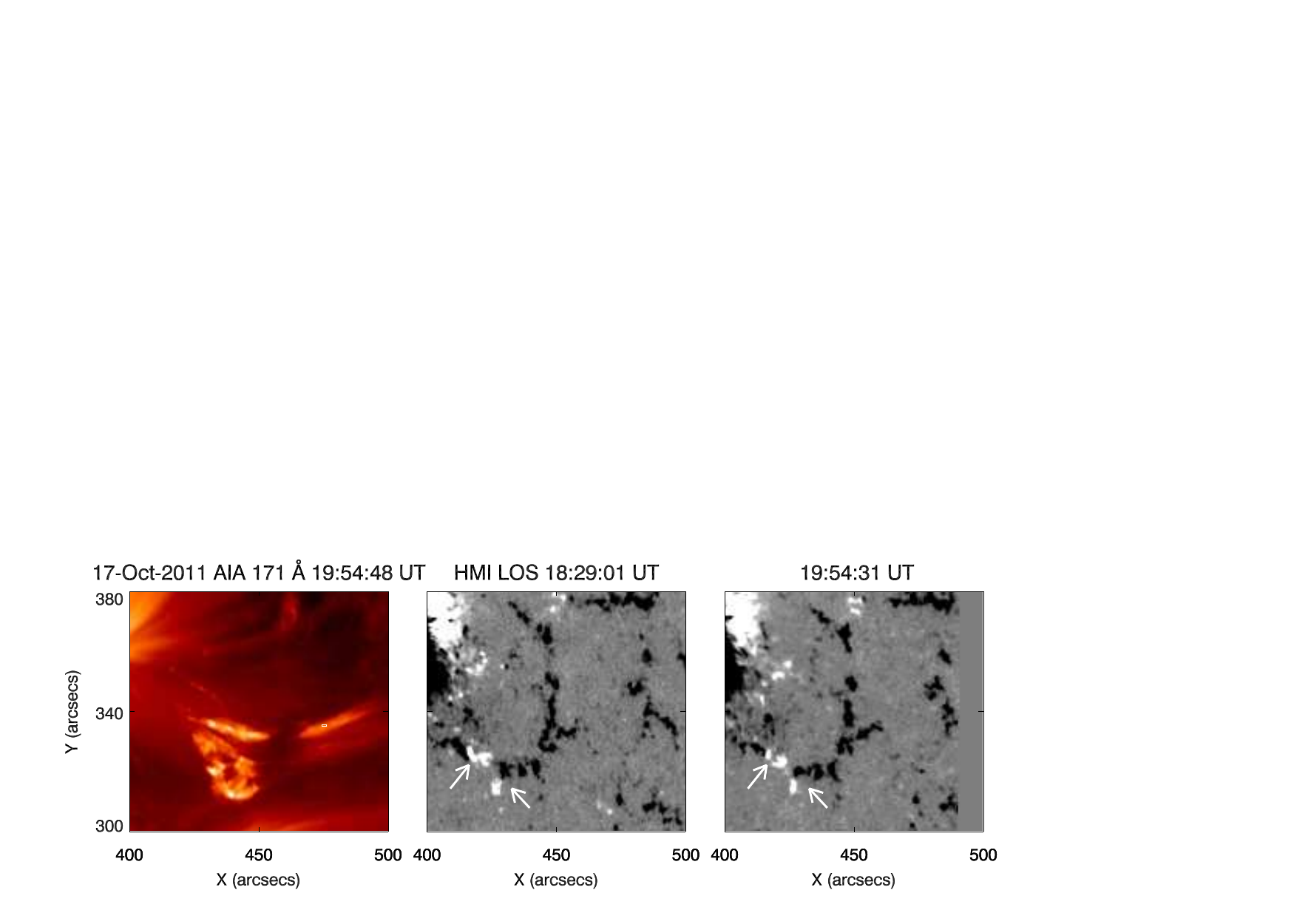}
\caption{Jet 7 : 17 October 2011 - Left panel : AIA 171~{\AA} image of the jet observed on 17 October 2011. The jet appeared to evolve from the edge of the active region 11314 (N28 W32) which was associated with a negative-polarity sunspot. A small-scale brightening was observed at 19:49 UT at the jet footpoint before the jet evolution, and the jet evolved as a complex, multi-threaded spire with inverted Y topology. The jet started its activity at 19:48 UT and lasted until 20:04 UT. Another similar jet was also observed at 20:07 UT in a nearby region. Middle panel and right panel : The LOS HMI magnetogram image at 18:29 UT before the jet evolution and at 18:54 UT during the jet evolution respectively. Flux cancellation was observed and it is shown by white arrows. This magnetic activity lasted for a few hours and it was co-temporal and co-spatial with the jet activity. \label{fig22}}
\end{center}
\end{figure}

\newpage
\begin{figure}[!hbtp]
\begin{center}
\includegraphics[trim=0.5cm 0.1cm 2.5cm 8cm,width=0.55\textwidth]{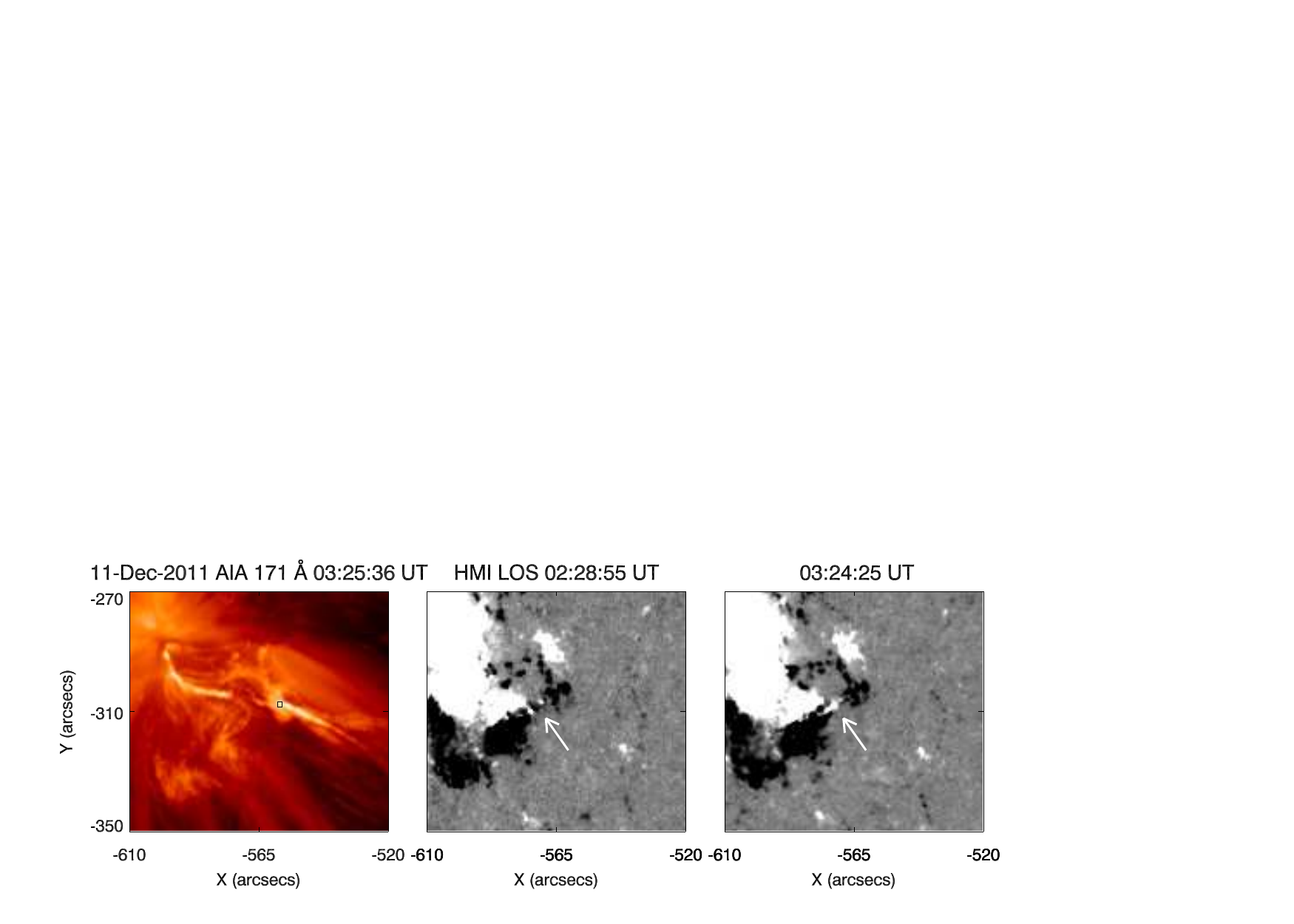}
\caption{Jet 8 : 11 December 2011 - Left panel : AIA 171~{\AA} image of the jet observed on 11 December 2011. The jet appeared to evolve from the edge of the active region 11374 (S17 E27) which was associated with a positive-polarity sunspot with anemone magnetic topology. A small-scale brightening was observed at a 03:21 UT at the jet footpoint before the jet evolution. The jet evolved as complex, multi-threaded spire. The jet started its activity at 03:21 UT and lasted until 03:26 UT. Middle panel and right panel : The LOS HMI magnetogram image at 02:28 UT before the jet evolution and at 03:24 UT during the jet evolution respectively. The negative-polarity cancellation followed by positive flux emergence at the edge of the negative-polarity region was observed and it is shown by white arrows. This magnetic activity lasted for a few hours and it was co-temporal and co-spatial with the jet activity. \label{fig23}}
\end{center}
\end{figure}

\begin{figure}[!hbtp]
\begin{center}
\includegraphics[trim=0.5cm 0.1cm 2.5cm 8cm,width=0.55\textwidth]{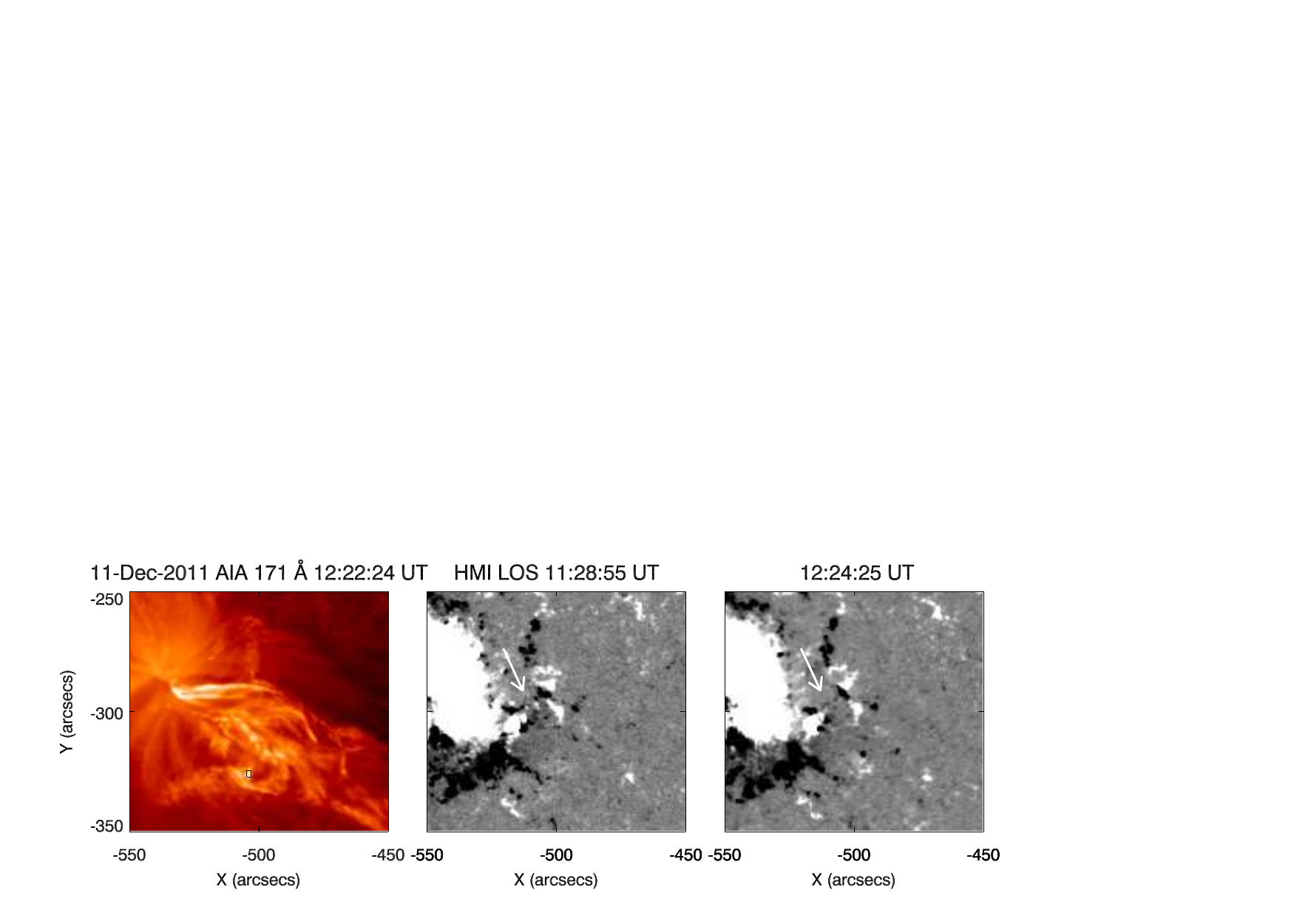}
\caption{Jet 9 : 11 December 2011 - Left panel : AIA 171~{\AA} image of the jet observed on 11 December 2011. The jet appeared to evolve from the edge of the active region 11374 (S17 E27) which was associated with a positive-polarity sunspot with anemone magnetic topology. A small-scale brightening was observed at 12:15 UT at the jet footpoint before the jet evolution, and the jet evolved as complex, multi-threaded spire. The observations showed the continuous untwisting motion of jet spire. The jet started its activity at 12:15 UT and lasted until 12:49 UT. Middle panel and right panel : The LOS HMI magnetogram image at 11:28 UT before the jet evolution and at 12:24 UT during the jet evolution respectively. The positive-polarity cancellation at the edge of the negative-polarity region was observed and it is shown by white arrows. This magnetic activity lasted for a few hours and it was co-temporal and co-spatial with the jet activity. \label{fig24}}
\end{center}
\end{figure}

\newpage
\begin{figure}[!hbtp]
\begin{center}
\includegraphics[trim=0.5cm 0.1cm 2.5cm 8cm,width=0.55\textwidth]{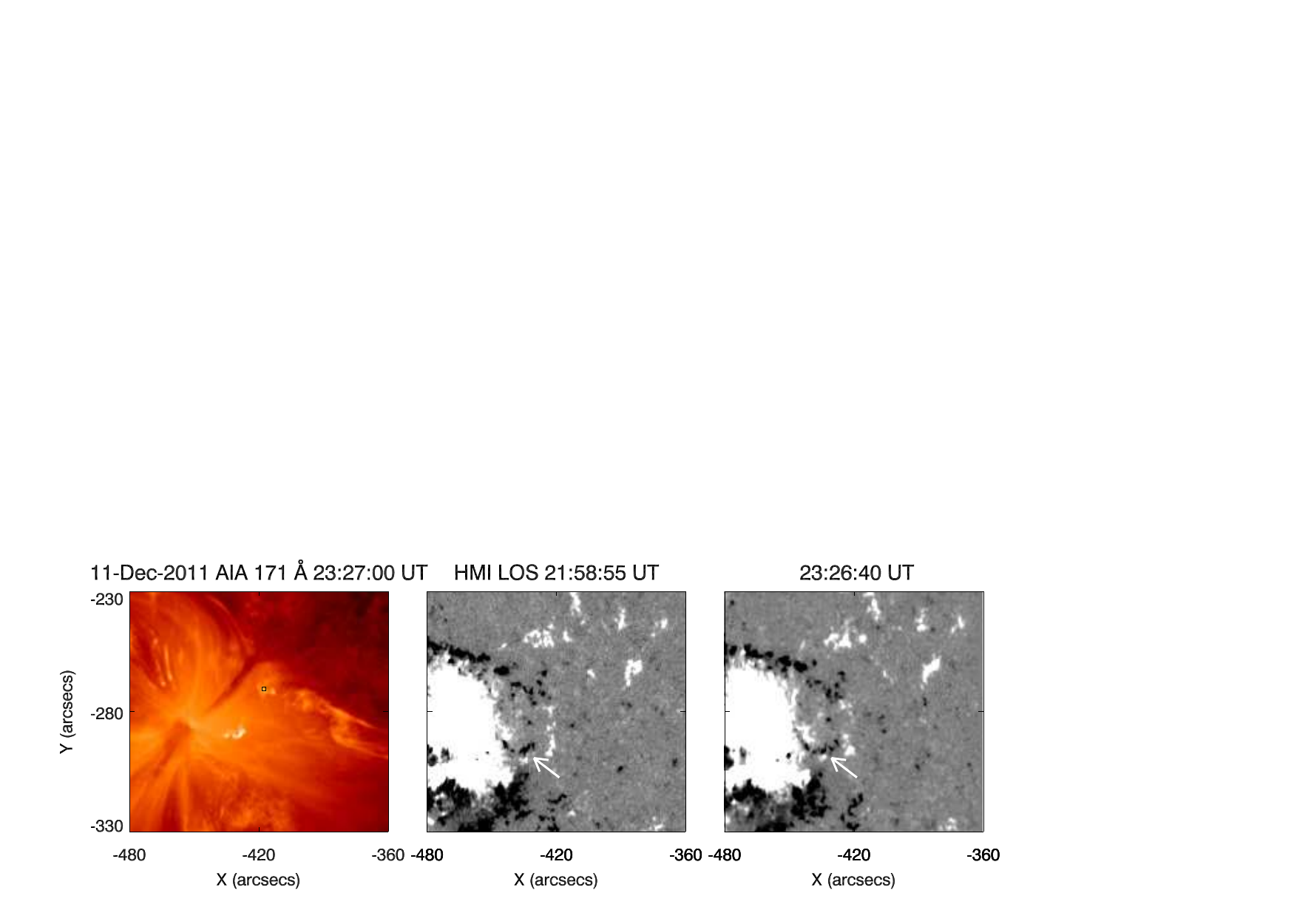}
\caption{Jet 10 : 11 December 2011 - Left panel : AIA 171~{\AA} image of the jet observed on 11 December 2012. The jet appeared to evolve from the edge of the active region 11374 (S17 E27) which was associated with a positive-polarity sunspot with anemone magnetic topology. A small-scale brightening was observed at 23:14 UT at the jet footpoint before the jet evolution, and the jet evolved as complex, multi-threaded spire. The observations showed the continuous untwisting motion of jet spire. The jet started its activity at 23:14 UT and lasted until 23:34 UT. Middle panel and right panel : The LOS HMI magnetogram image at 21:58 UT before the jet evolution and at 12:24 UT during the jet evolution respectively. The positive-polarity cancellation at the edge of the negative-polarity region was observed and it is shown by white arrows. This magnetic activity lasted for a few hours and it was co-temporal and co-spatial with the jet activity. \label{fig25}}
\end{center}
\end{figure}

\begin{figure}[!hbtp]
\begin{center}
\includegraphics[trim=0.5cm 0.1cm 2.5cm 8cm,width=0.55\textwidth]{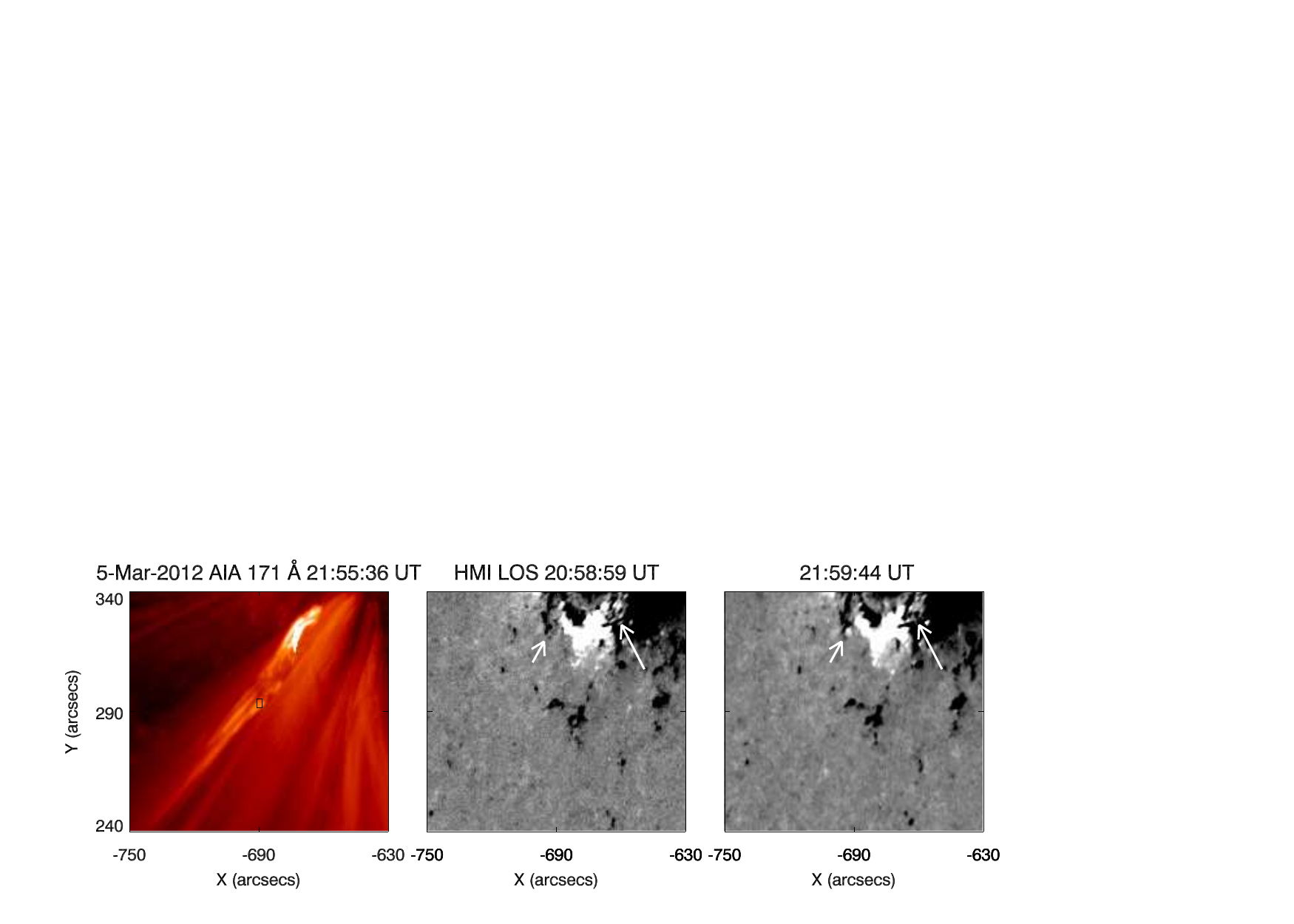}
\caption{Jet 11 : 5 March 2012 - Left panel : AIA 171~{\AA} image of the jet observed on 5 March 2012. The jet appeared to evolve from the edge of the active region 11429 (N18 E41). A small-scale brightening was observed at 21:50 UT at the jet footpoint before the jet evolution, and the jet evolved as multi-threaded spire. A small loop is observed at the footpoint. The jet started its activity at 21:51 UT and lasted until 22:00 UT. Middle panel and right panel : The LOS HMI magnetogram image at 20:58 UT before the jet evolution and at 21:59 UT during the jet evolution respectively. The positive-polarity cancellation with nearby negative-polarity region was observed and it is shown by white arrows. This magnetic activity lasted for a few hours and it was co-temporal and co-spatial with the jet activity. \label{fig26}}
\end{center}
\end{figure}

\newpage
\begin{figure}[!hbtp]
\begin{center}
\includegraphics[trim=0.5cm 0.1cm 2.5cm 8cm,width=0.55\textwidth]{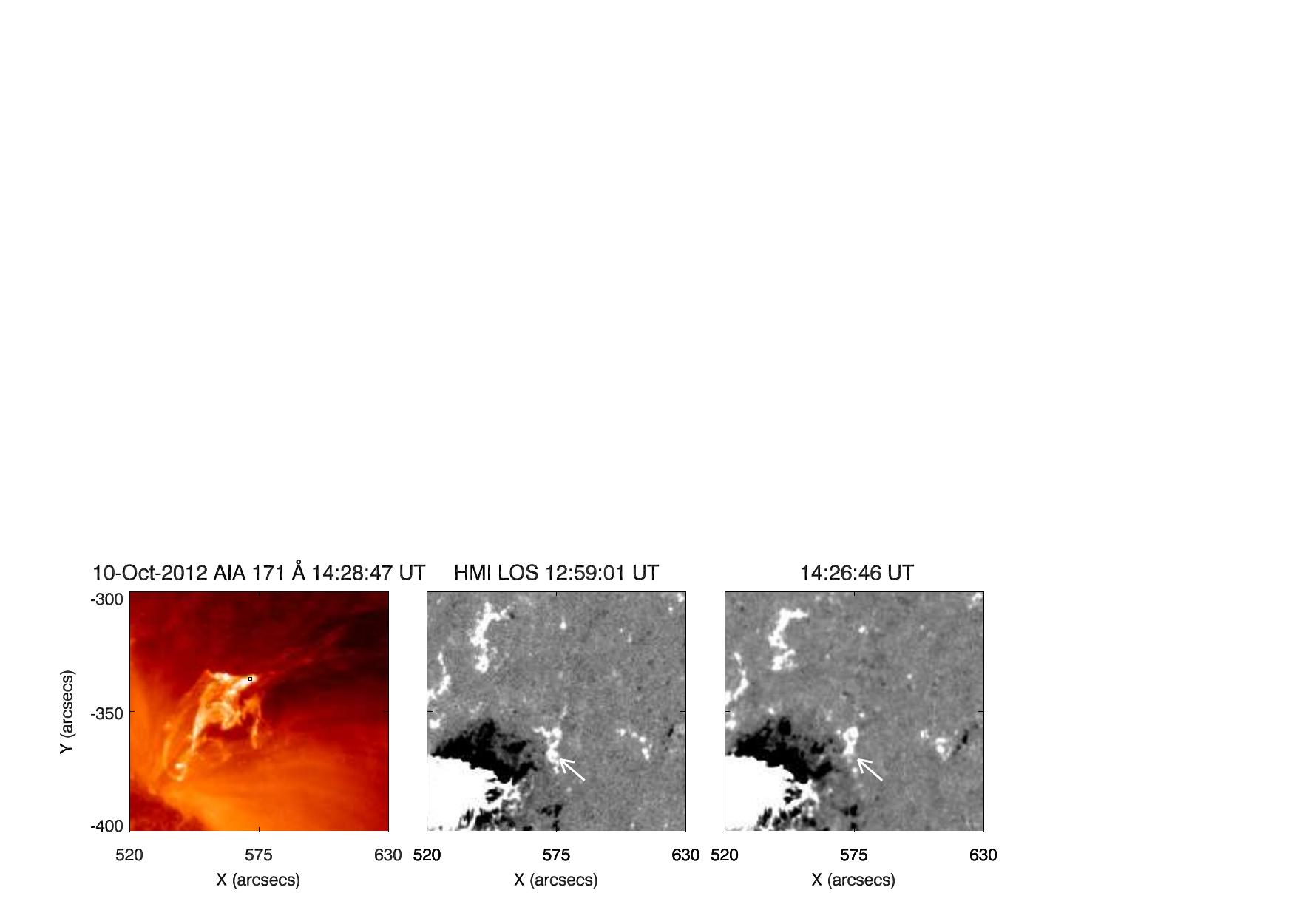}
\caption{Jet 12 : 10 October 2012 - Left panel : AIA 171~{\AA} image of the jet observed on 10 October 2012. The jet appeared to evolve from the edge of the active region 11585 (S20 W43). A small-scale brightening was observed at 14:26 UT at the jet footpoint before the jet evolution, and the jet evolved as multi-threaded spire. A small burst is observed at the footpoint. The jet started its activity at 14:21 UT and lasted until 14:45 UT. Middle panel and right panel : The LOS HMI magnetogram image at 12:59 UT before the jet evolution and at 14:26 UT during the jet evolution respectively. The positive-polarity cancellation with nearby negative-polarity region was observed and it is shown by white arrows. This magnetic activity lasted for a few hours and it was co-temporal and co-spatial with the jet activity. \label{fig27}}
\end{center}
\end{figure}

 \begin{figure}[!hbtp]
 \begin{center}
 \includegraphics[trim=0.5cm 0.1cm 2.5cm 8cm,width=0.55\textwidth]{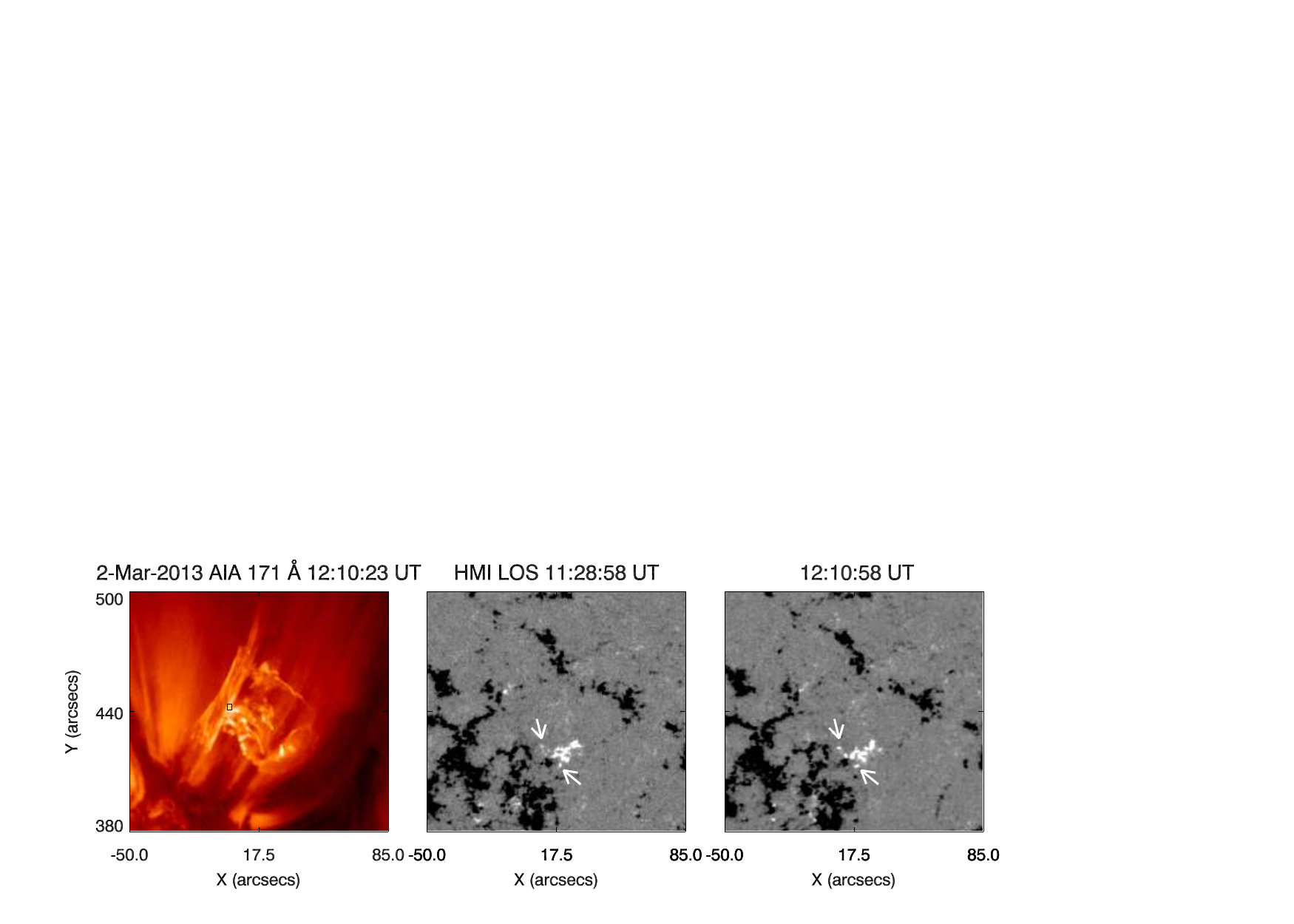}
 \caption{Jet 14 : 2 March 2013 - Left panel : AIA 171~{\AA} image of the jet observed on 2 March 2013. The jet appeared to evolve from the edge of the active region 11681 (N17 W08). A small-scale brightening was observed at 12:03 UT at the jet footpoint before the jet evolution, and the jet evolved as simple spire. The observations showed the continuous untwisting motion of jet spire. The jet started its activity at 12:04 UT and lasted until 12:25 UT. Middle panel and right panel : The LOS HMI magnetogram image at 11:28 UT before the jet evolution and at 12:10 UT during the jet evolution respectively. The positive-polarity emergence at nearby negative-polarity region was observed and it is shown by white arrows. This magnetic activity lasted for a few hours and it was co-temporal and co-spatial with the jet activity. \label{fig28}}
 \end{center}
 \end{figure}

\newpage
\begin{figure}[!hbtp]
\begin{center}
\includegraphics[trim=0.5cm 0.1cm 2.5cm 8cm,width=0.55\textwidth]{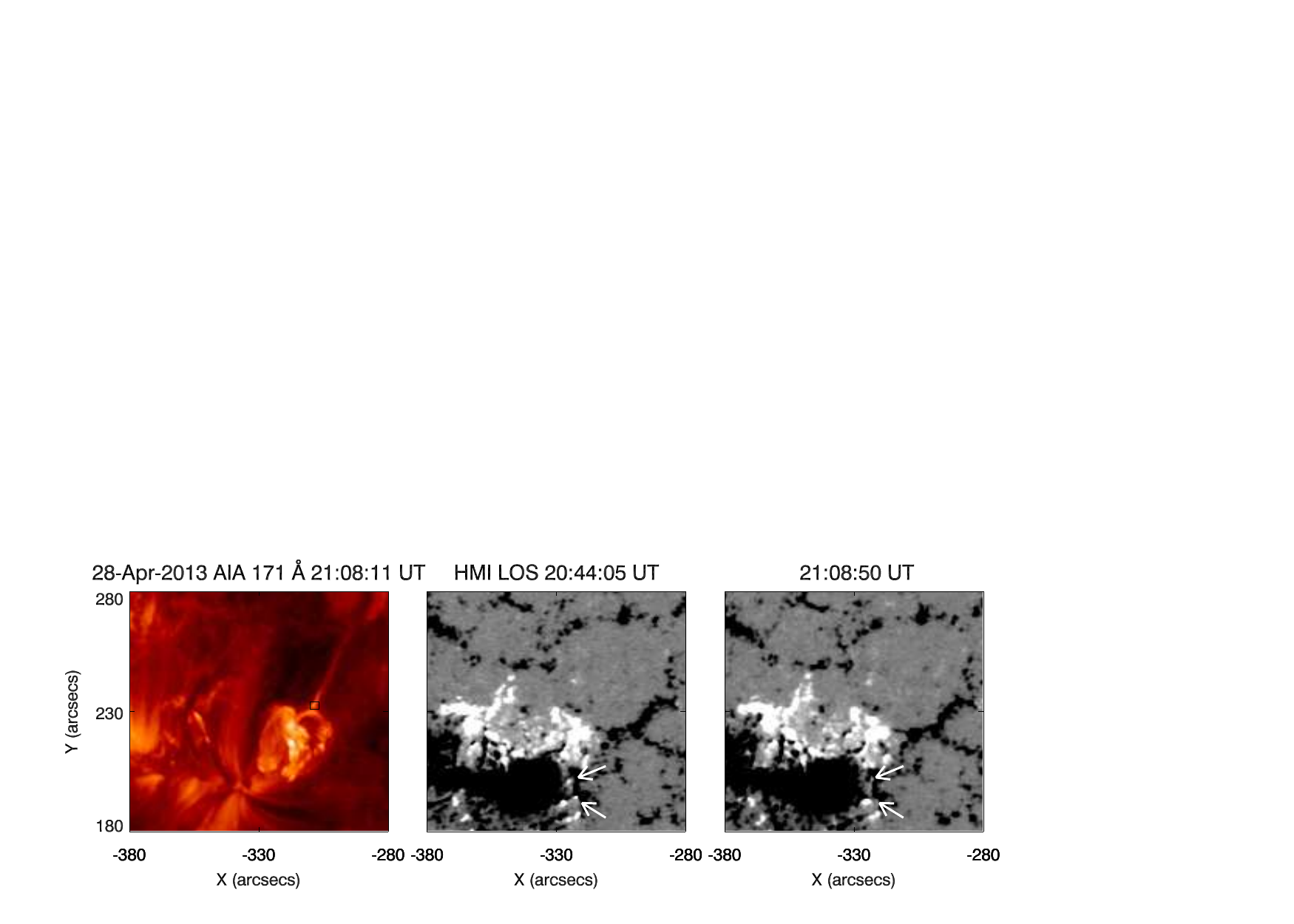}
\caption{Jet 15 : 28 April 2013 - Left panel : AIA 171~{\AA} image of the jet observed on 28 April 2013. The jet appeared to evolve from the edge of the active region 11681 (N17 W08). A small-scale brightening was observed at 20:59 UT at the jet footpoint before the jet evolution, and the jet evolved as complex, multi-threaded spire. The observations showed the continuous untwisting motion of jet spire. The jet started its activity at 20:59 UT and lasted until 21:11 UT. Middle panel and right panel : The LOS HMI magnetogram image at 20:44 UT before the jet evolution and at 21:08 UT during the jet evolution respectively. The positive-polarity emergence and then cancellation observed near the negative-polarity region and it is shown by white arrows. This magnetic activity lasted for a few hours and it was co-temporal and co-spatial with the jet activity. \label{fig29}}
\end{center}
\end{figure}

\begin{figure}[!hbtp]
\begin{center}
\includegraphics[trim=0.5cm 0.1cm 2.5cm 8cm,width=0.55\textwidth]{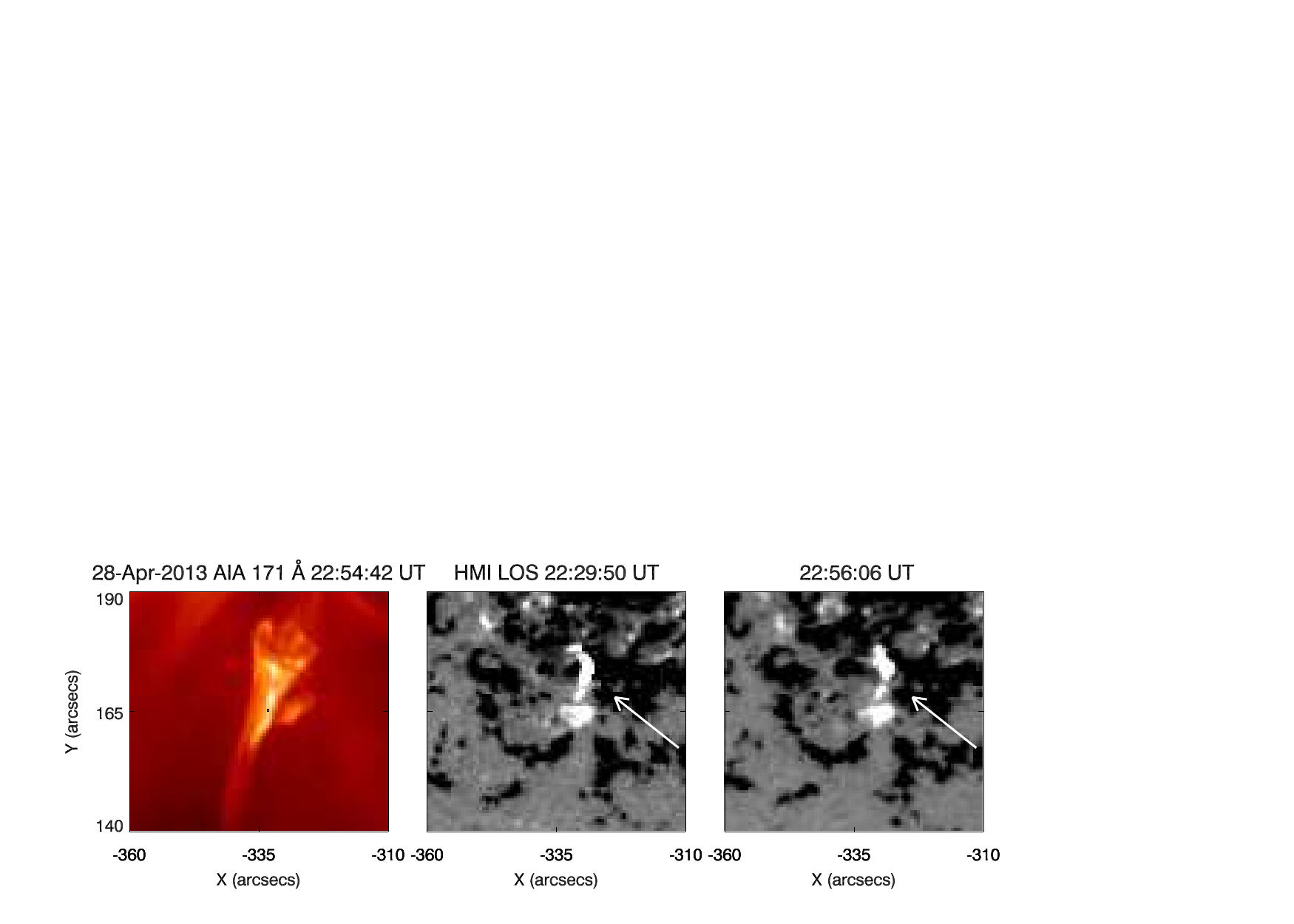}
\caption{Jet 16 : 28 April 2013 - Left panel : AIA 171~{\AA} image of the jet observed on 28 April 2013. The jet appeared to evolve from the edge of the active region 11681 (N17 W08) which was associated with negative-polarity sunspot. A small-scale brightening was observed at 22:51 UT at the jet footpoint before the jet evolution, and the jet evolved as multi-threaded spire with inverted Y topology of magnetic field. The jet started its activity at 22:51 UT and lasted until 22:59 UT. Middle panel and right panel : The LOS HMI magnetogram image at 22:29 UT before the jet evolution and at 22:56 UT during the jet evolution respectively. The positive-polarity emergence at nearby negative-polarity region was observed and it is shown by white arrows. This magnetic activity lasted for a few hours and it was co-temporal and co-spatial with the jet activity. \label{fig30}}
\end{center}
\end{figure}

\newpage
\begin{figure}[!hbtp]
\begin{center}
\includegraphics[trim=0.5cm 0.1cm 2.5cm 8cm,width=0.55\textwidth]{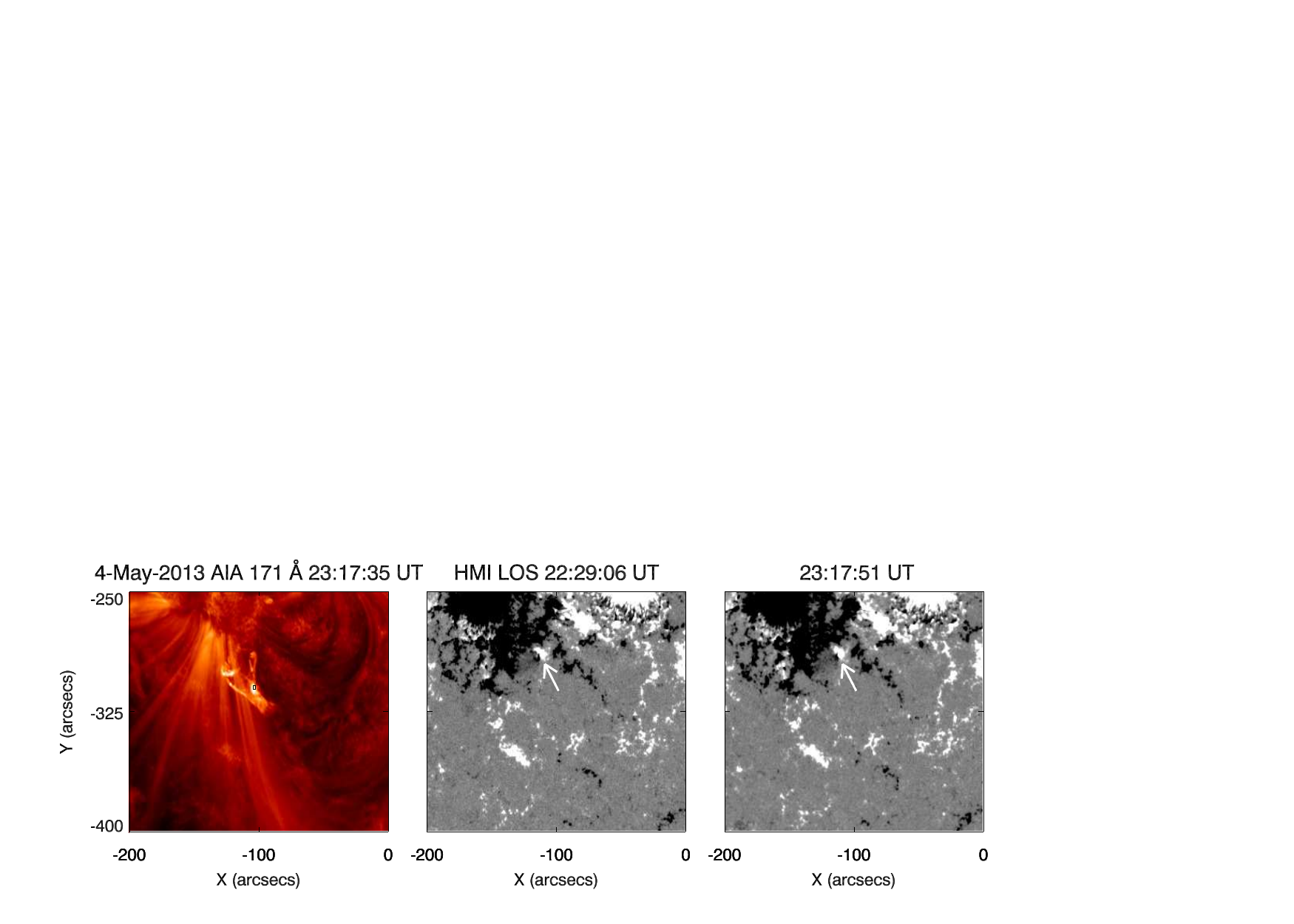}
\caption{Jet 17 : 4 May 2013 - Left panel : AIA 171~{\AA} image of the jet observed on 4 May 2013. The jet appeared to evolve from the edge of the active region 11734 (S19 W04). The jet was associated with a closed loop structure and complex, multi-threaded jet was originated from one of the footpoints of loop. It has been observed that the jet plasma couldn't escape fully in the solar atmosphere, a part of it flows along the closed loop structure. The jet started its activity at 23:15 UT and lasted until 23:49 UT. Middle panel and right panel : The LOS HMI magnetogram image at 22:29 UT before the jet evolution and at 23:17 UT during the jet evolution respectively. The negative-polarity cancellation at nearby positive-polarity region was observed and it is shown by white arrows. This magnetic activity lasted for a few hours and it was co-temporal and co-spatial with the jet activity. \label{fig31}}
\end{center}
\end{figure}

\begin{figure}[!hbtp]
\begin{center}
\includegraphics[trim=0.5cm 0.1cm 2.5cm 8cm,width=0.55\textwidth]{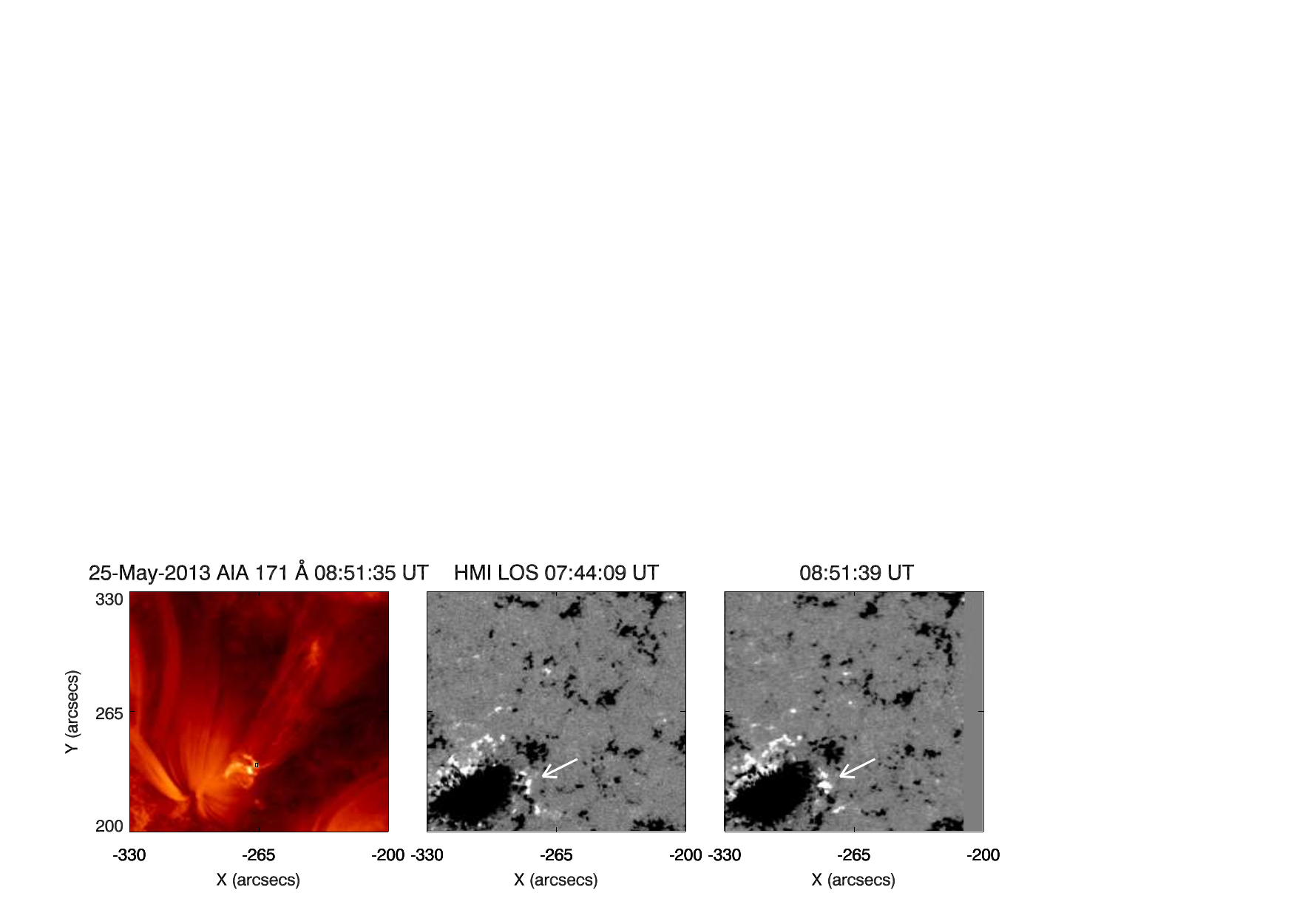}
\caption{Jet 18 : 25 May 2013 - Left panel : AIA 171~{\AA} image of the jet observed on 25 May 2013. The jet appeared to evolve from the edge of the active region 11748 (N12 W83). The jet was associated with a small loop  at the footpoint and complex, multi-threaded jet was originated from it showing the untwisting nature. The jet started its activity at 08:45 UT and lasted until 08:53 UT. Middle panel and right panel : The LOS HMI magnetogram image at 07:44 UT before the jet evolution and at 08:51 UT during the jet evolution respectively. The positive-polarity emergence at the edge of the negative-polarity sunspot was observed and it is shown by white arrows. This magnetic activity lasted for a few hours and it was co-temporal and co-spatial with the jet activity. \label{fig32}}
\end{center}
\end{figure}

\newpage
\begin{figure}[!hbtp]
\begin{center}
\includegraphics[trim=0.5cm 0.1cm 2.5cm 8cm,width=0.55\textwidth]{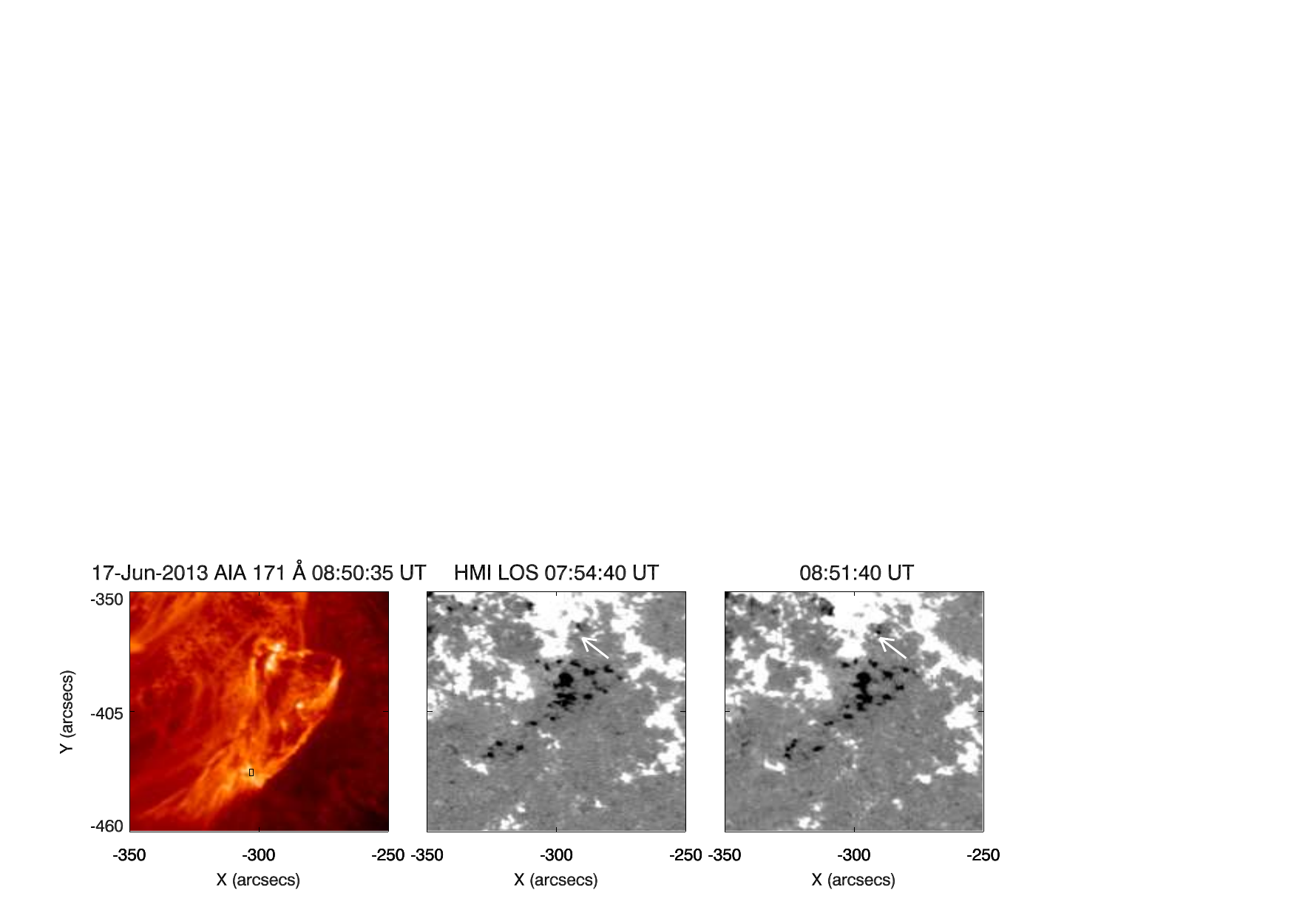}
\caption{Jet 19 : 17 Jun 2013 - Left panel : AIA 171~{\AA} image of the jet observed on 17 June 2013. The jet appeared to evolve from the edge of the active region 11770 (S13 E13). The jet was associated with a closed loop structure and complex, multi-threaded jet was originated from one of the footpoints of loop showing its untwisting nature. It has been observed that the jet plasma couldn't escape fully in the solar atmosphere, a part of it flows along the closed loop structure. The jet started its activity at 08:40 UT and lasted until 09:06 UT. Middle panel and right panel : The LOS HMI magnetogram image at 07:54 UT before the jet evolution and at 08:51 UT during the jet evolution respectively. The positive-polarity cancellation at the nearby negative-polarity region was observed and it is shown by white arrows. This magnetic activity lasted for a few hours and it was co-temporal and co-spatial with the jet activity.  \label{fig33}}
\end{center}
\end{figure}

\begin{figure}[!hbtp]
\begin{center}
\includegraphics[trim=0.5cm 0.1cm 2.5cm 8cm,width=0.55\textwidth]{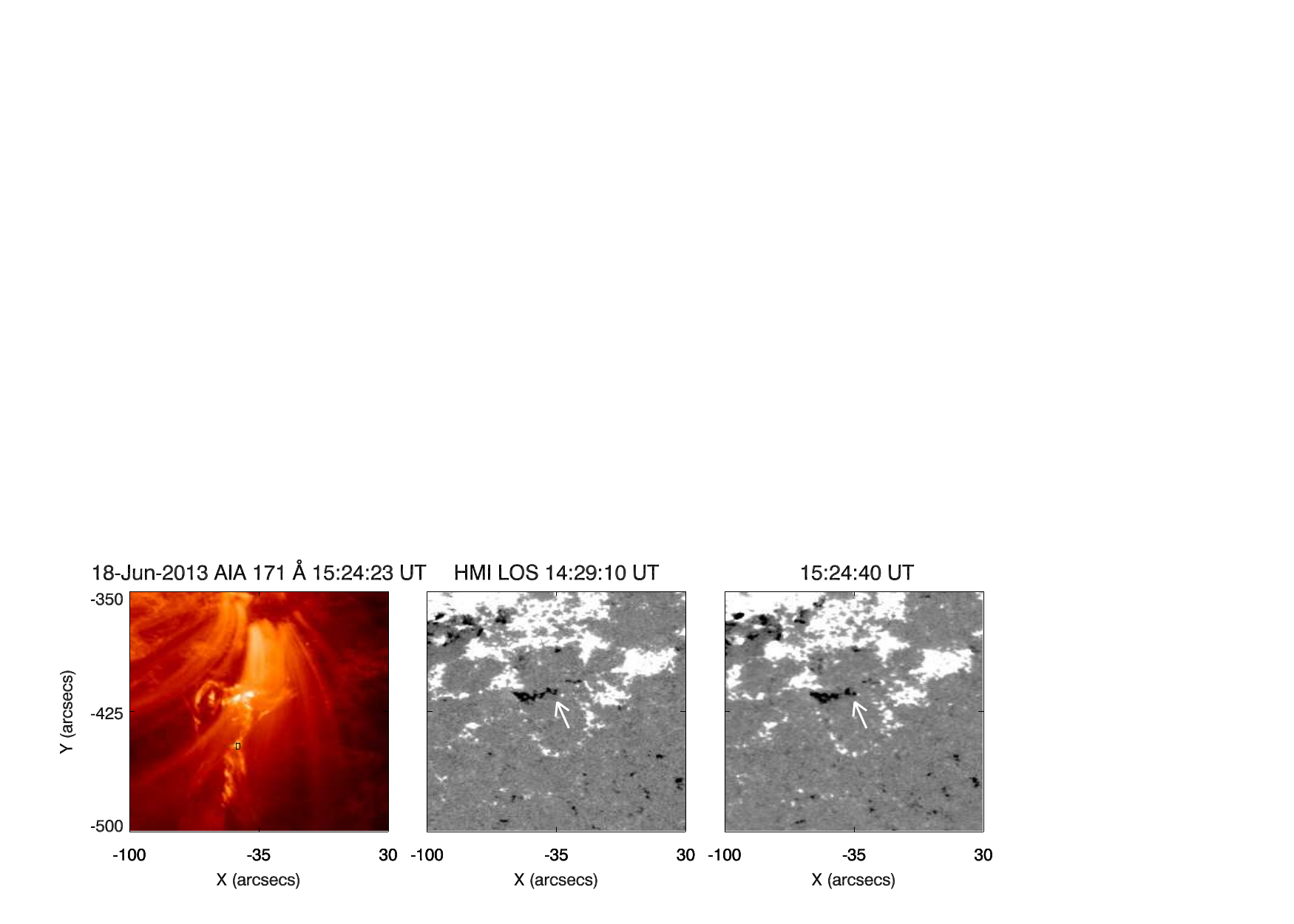}
\caption{Jet 20 : 18 Jun 2013 - Left panel : AIA 171~{\AA} image of the jet observed on 18 June 2013. The jet appeared to evolve from the edge of the active region 11770 (S13 E13). The complex, multi-threaded jet was originated showing its untwisting nature. The jet started its activity at 15:13 UT and lasted until 15:39 UT. Middle panel and right panel : The LOS HMI magnetogram image at 14:29 UT before the jet evolution and at 15:24 UT during the jet evolution respectively. The positive-polarity cancellation at the nearby negative-polarity region was observed and it is shown by white arrows. This magnetic activity lasted for a few hours and it was co-temporal and co-spatial with the jet activity. \label{fig34}}
\end{center}
\end{figure}


\end{appendix}

\end{document}